\pdfoutput=1
\documentclass[12pt,english]{article}
\usepackage[T1]{fontenc}
\usepackage[latin9]{inputenc}
\usepackage{geometry}
\usepackage{amssymb}
\usepackage[authoryear]{natbib}
\defcitealias{AndrewsBarwick2012}{D. Andrews \& Barwick (2012)}
\defcitealias{AndrewsSoares2010}{D. Andrews \& Soares (2010)}
\defcitealias{AndrewsGuggenberger2009}{D. Andrews \& Guggenberger (2009)}
\defcitealias{AndrewsShi2013}{D. Andrews \& Shi (2013)}
\defcitealias{Andrewsetal2017}{D. Andrews et al. (2019)}

\let\counterwithin\relax

\newcommand{\argmax}{\mbox{argmax}}
\newcommand\ind{\protect\mathpalette{\protect\independenT}{\perp}}
\def\independenT#1#2{\mathrel{\rlap{$#1#2$}\mkern2mu{#1#2}}}

\usepackage{color}
\usepackage{amsmath}
\usepackage[normalem]{ulem}
\usepackage{enumitem}
\usepackage{algorithm}
\usepackage[noend]{algpseudocode}
\usepackage{subfig}
\usepackage{array,booktabs,ragged2e}
\usepackage[flushleft]{threeparttable}
\usepackage{adjustbox}
\usepackage{xr-hyper}
\usepackage{hyperref}
\usepackage{chngcntr}
\makeatletter

\usepackage{epsf}
\usepackage{graphicx}

\newtheorem{lemma}{Lemma}\newtheorem{definition}{Definition}
\newtheorem{assumption}{Assumption}
\newtheorem{proposition}{Proposition}

\usepackage[subtle]{savetrees}
\usepackage[small,compact]{titlesec}
 
\usepackage{babel}

\usepackage{multibib}
\newcites{supp}{Supplement References}

\usepackage{setspace}
\newcommand{\var}[1]{Var\left(#1\right)}
\newcommand{\reals}{\mathbb{R}}
\newcolumntype{R}[1]{>{\RaggedLeft\arraybackslash}p{#1}}

\makeatother

\usepackage{babel}
\begin{document}

\title{Inference for Linear Conditional\\ Moment Inequalities\footnote{We thank Tim Armstrong, Gary Chamberlain, Ivan Canay, Jiafeng Chen, Kirill Evdokimov, Jerry Hausman, Bulat Gafarov, Hiroaki Kaido, Adam McCloskey, Francesca Molinari, Whitney Newey, Ashesh Rambachan, Bas Sanders, Jesse Shapiro, Brit Sharoni, Xiaoxia Shi, Joerg Stoye, Chris Walker, and participants at several seminars for helpful comments, and thank Thomas Wollmann for helpful discussion of his application. We are grateful to Xiaoxia Shi and Matt Thirkettle for sharing code and providing advice on its implementation. Andrews gratefully acknowledges financial support from the NSF under Grant 1654234.  Roth gratefully acknowledges financial support from an NSF Graduate Research Fellowship under Grant DGE1144152. Andrews: iandrews@fas.harvard.edu.  Roth: jonathan\_roth@brown.edu.  Pakes: apakes@fas.harvard.edu}}

\author{Isaiah Andrews  ~~~ Jonathan Roth ~~~ Ariel Pakes}
\maketitle
\begin{abstract}
	We show that moment inequalities in a wide variety of economic applications have a particular linear conditional structure. We use this structure to construct uniformly valid confidence sets that remain computationally tractable even in settings with nuisance parameters. We first introduce least favorable critical values which deliver non-conservative tests if all moments are binding. Next, we introduce a novel conditional inference approach which ensures a strong form of insensitivity to slack moments. Our recommended approach is a hybrid technique which combines desirable aspects of the least favorable and conditional methods. The hybrid approach performs well in simulations calibrated to \citet{Wollmann}, with favorable power and computational time comparisons relative to existing alternatives. 
\end{abstract}

\section{Introduction}

Moment inequalities are a useful tool in a wide range of fields in empirical economics. As described in recent reviews by \cite{HoRosen} and \cite{Molinari2019}, moment inequalities can be used to exploit the most direct implications of utility or profit maximization for inference in both single-agent settings and games. They can also be used to weaken parametric, behavioral, measurement, and selection assumptions in a range of problems.
Inference using moment inequalities raises practical challenges, however, particularly when there are nuisance parameters (e.g. coefficients on control variables) that are not of direct interest. 

A first challenge is obtaining tests that are computationally tractable. Many available moment inequality methods rely on test inversion over a grid for the full parameter vector (including the nuisance parameters), but the computational costs of such approaches grow exponentially in the dimension of the parameter vector. This has necessitated the development of alternative approaches that either profile out (i.e. optimize over) the nuisance parameters in the computation of the test statistic \citep[e.g.,][]{BugniCanayShi2017} or use computational shortcuts to form projection confidence sets without computing the test for all values of the nuisance parameter \citep[e.g.,][]{Kaidoetal2016}. Nevertheless, computation can still be challenging when the dimension of the nuisance parameters is moderate or large. 

A second challenge is obtaining tests with good power. When there are nuisance parameters, tests for the parameter of interest can be obtained via projection, but this can lead to conservative tests with poor power \citep[see][]{BugniCanayShi2017, Kaidoetal2016}. Moreover, the power of many existing procedures can be negatively affected by the inclusion of non-binding moments, yet it may not be clear \textit{ex ante} which of the moments implied by economic theory will be binding. This has prompted a variety of approaches to eliminate or reduce the sensitivity of moment inequality tests to slack moments including work by \citetalias{AndrewsSoares2010}, \citetalias{AndrewsBarwick2012}, \citet{Romanoetal2014}, \citet{Chernozhukovetal2015}, \citet{BugniCanayShi2017}, and \citet{Bellonietal2018}, among many others.

In this paper, we show that a variety of applications of moment inequalities have a particular structure that can be exploited to address these challenges. Specifically, we study settings with moment inequalities of the form $E[ Y_i(\beta_0) - X_i(\beta_0) \delta | Z_i ] \leq 0$, where $\beta_0$ is the parameter of interest, $\delta$ is a nuisance parameter, and $X_i(\beta_0)$ is a function of $Z_i$. That is, we study conditional moment inequalities that (a) are linear in the nuisance parameters $\delta$, and (b) have conditional variance (given the instruments $Z_i$) that does not depend on the nuisance parameters. In Section \ref{sec: Linear Conditional MI}, we highlight several recent applications of moment inequalities that have this structure, including interval-valued regression and revealed preference models in industrial organization. 

Under this linear conditional structure, the profiled studentized max statistic can be represented as a linear program, and can thus be computed efficiently even when the dimension of the nuisance parameters is large. Linear conditional structure is also helpful for deriving tractable critical values, since it implies that the asymptotic variance of the moments (conditional on the instruments) does not depend on the value of the nuisance parameters. These features allow us to construct profiling-based confidence sets that rely on test inversion only for the target parameter and not for the nuisance parameters, and thus are computationally tractable even when the dimension of the nuisance parameters is large. We exploit this linear conditional structure to develop two tests that have different desirable properties, as well as a third hybrid approach that combines the two and is our preferred approach. 
	
Our first approach is based on the least-favorable (LF) asymptotic distribution of our test statistic. We show that the distribution of the test statistic is increasing (in the sense of first-order stochastic dominance) in the mean of the moments, and thus the least-favorable distribution under the null corresponds with the case where the mean of all of the moments is zero.\footnote{This presumes that the set of data-generating processes considered allows for the possibility that all moments bind simultaneously. If not, then the distribution used for our critical value is an upper bound on the least-favorable distribution under the null.} It is then straightforward to calculate a critical value under the least-favorable distribution via simulation. The LF test has exact asymptotic size when all of the moments are simultaneously binding in population, and thus avoids conservativeness from projection in this case. A downside of the LF test, however, is that its power can be negatively affected by the inclusion of slack moments. 

To address sensitivity to slack moments, we introduce a second test based on a novel conditioning argument. We condition on the Lagrange multipliers in the optimization to compute the test statistic, which intuitively correspond with the set of binding moments in sample after profiling out the nuisance parameters. We show that the set of values of the moments for which a particular Lagrange multiplier is optimal is a polyhedron, and we then derive critical values using results from \citet{Leeetal2016} on polyhedral conditioning events. We prove that the resulting conditional test is insensitive to slack moments in the strong sense that, as a subset of the moments becomes arbitrarily slack, the conditional test converges to the test that drops these moments ex-ante. A downside of the conditional test, however, is that it may have poor power in settings where multiple moments are approximately equally violated. Finally, given the different relative strengths of the LF and conditional approaches, we introduce a hybrid approach that combines the LF and conditional approaches, while avoiding the conservativeness of Bonferroni approaches.

The critical values for all of our tests are based on a normal approximation to the distribution of the moments conditional on the instruments. If this normal approximation holds exactly with known variance, our proposed tests control size in finite samples. In Section \ref{sec: Asymptotics} we provide regularity conditions under which size control in this finite sample normal model translates to uniform asymptotic size control over a large class of data-generating distributions. A desirable feature of our proposed tests is that they they achieve uniform asymptotic size control without having to specify a sequence of tuning parameters that converges at a certain rate. Nevertheless, our tests do require the researcher to make some choices. To use the hybrid test, the researcher must specify the size of the ``first-stage'' least favorable test $\kappa$, although this choice only affects the power of the test and not its asymptotic validity.\footnote{We recommend using $\kappa=\alpha/10$, and implement this choice in our simulations, following the recommendation for the two-step procedure in \citet{Romanoetal2014}.\label{FN: kappa}} Additionally, although conditional moment inequalities can imply an infinite number of unconditional moments, our tests only exploit the implications of $k$ unconditional moments that must be specified by the researcher. We provide heuristic guidance on the choice of the $k$ moments in Section \ref{sec: choice of moments}.

To explore the numerical performance of our methods, we apply our techniques in simulations calibrated to \cite{Wollmann}'s study of the US auto bailout.  We consider designs with up to ten nuisance parameters, and find that our proposed tests remain computationally tractable and have good size control in all specifications. The power of the hybrid test is similar to or better than that of the LF and conditional tests in all specifications, and we thus recommend the hybrid approach among our proposed procedures. We also find that the hybrid test has power dominating that of the projection-based tests of \citetalias{AndrewsSoares2010} and \citet{Kaidoetal2016} in all specifications for which we are able to compute these tests, and computation time for the hybrid can be over 10 times faster than for either of the projection-based approaches. The hybrid approach is also competitive with the sCC and sRCC tests proposed in concurrent work by \citet{cox_simple_2020}, although neither approach dominates the other across all specifications in terms of power or computational speed. 

\paragraph{Related Literature.} \citet{cox_simple_2020} consider the class of linear conditional moment inequalities introduced in this paper and propose tests based on a profiled quasi-likelihood ratio (QLR) statistic, whereas our tests are based on the profiled studentized max statistic.  \citet{cox_simple_2020} and the present paper independently developed conditional testing approaches, but due to the difference in test statistics, the conditioning events and resulting tests are different.  As discussed in Section \ref{sec: Monte Carlo}, we find in our Monte Carlo simulations that our preferred test (the hybrid) has non-nested power with those proposed by \citet{cox_simple_2020}, which accords with the intuition that tests based on the max and QLR statistics direct power towards different parts of the parameter space. 

Subvector inference for moment inequalities with linear parameters is also considered in \citet{cho_simple_2021, Garfarov2016} and \citet{Flynn2019}. The setting in these papers differs from ours in that they consider unconditional moment inequalities, whereas we consider conditional moments; our paper also differs in that we allow the target parameters to potentially enter the moments non-linearly. One advantage of our approach relative to these previous papers is that we do not require a linear independence constraint qualification (LICQ) assumption, which restricts what moments can bind in population; see Section \ref{sec: Asymptotics} for further discussion.\footnote{\citet{cho_simple_2021} show that LICQ can be guaranteed to hold by adding a stochastic perturbation to the moments, at the expense of obtaining inference on an outer set of the sharp identified set.} Another related paper is \citet{kaido_asymptotically_2014}, who consider efficient estimation and inference for the support function in settings with convex moment inequalities, which nests the problem of subvector estimation/inference in moment inequality models where all parameters enter linearly. Their approach, however, relies on a Slater constraint qualification that, for example, rules out moment equalities cast as inequalities. Our approach is thus complementary, since we do not require such a constraint qualification but also do not provide any formal efficiency results.

Our approach uses a profiled maximum statistic, and thus is also related to other profiling-based methods for moment inequalities. The profiling-based approach in \citet{BugniCanayShi2017} differs from ours in that it accommodates unconditional moment inequalities and does not require that the parameters enter the moments linearly. However, the linear structure that we consider enables highly-tractable computation since the profiled test statistic is computed with a linear program, and also enables us to develop tests that are uniformly asymptotically valid without relying on drifting sequences of tuning parameters. \citet{Bellonietal2018} build on the approach of \citet{BugniCanayShi2017} to develop methods for subvector inference with high-dimensional unconditional moments. \citet{Fangetal2021} propose a test based on the solution to a linear program that is applicable for a large class of problems that nests a high-dimensional version of the conditional linear inequalities considered in this paper, although at the cost of either introducing a sample-size dependent tuning parameter or obtaining a conservative test. Alternative approaches to subvector inference in moment inequality models include projection-based methods \citep[e.g.,][]{kaido_calibrated_2017}; sub-sampling approaches \citep[e.g.,][]{RomanoShaikh2008}; and quasi-posterior Monte Carlo methods \citep{CCT2018}.\footnote{The approach of \citet{CCT2018} delivers inference on the identified set, rather than on points within the identified set.}  We emphasize that the aforementioned methods do not impose the specific linear conditional structure considered in this paper, and thus are applicable in a much wider class of problems. We provide comparisons to the profiling-based approach of \citet{cox_simple_2020} as well as two projection-based methods in our Monte Carlo simulations.

One important limitation of our approach is that --- while we assume that conditional moment inequalities are satisfied --- we consider tests that exploit only a fixed number ($k$) of the implied unconditional inequalities. This contrasts with papers that consider asymptotics in which the number of moments grows with the sample size, such as \citetalias{AndrewsShi2013} for full-vector inference, and \citet{Chernozhukovetal2015} and \citet{Bellonietal2018} for subvector inference.\footnote{\citet{Flynn2019} considers a continuum of unconditional moment inequalities.} An interesting open question is whether the tests proposed in this paper can be extended to the setting with a diverging number of moments. See Section \ref{sec: Linear Conditional MI} below for additional discussion.

\section{Linear Conditional Moment Inequalities}
\label{sec: Linear Conditional MI}

We assume that we observe independent
and identically distributed data $D_{i}$, $i=1,...,n$ drawn from
an unknown distribution $P \in \mathcal{P}$, for a class $\mathcal{P}$ of distributions. The true values of the parameters $(\beta,\delta)$ are assumed to satisfy the conditional moment inequalities
\begin{equation}
	E_{P_{D|Z}}[Y_i(\beta)-X_i(\beta)\delta|Z_i]\le 0 \mbox{ almost surely},\label{eq: Linear Conditional Moments}
\end{equation}
 
\noindent where $Z_i$ is a subvector of $D_i$, $Y_i(\beta)=y(D_i,\beta)\in \mathbb{R}^k$ and $X_i(\beta)=x(Z_i,\beta)\in \mathbb{R}^{k\times p}$ for known functions $y(\cdot,\cdot)$ and $x(\cdot,\cdot),$ and $P_{D|Z}$ denotes the conditional distribution of $D_i$ given $Z_i$. We are interested in $\beta,$ while $\delta\in\mathbb{R}^p$ is a nuisance parameter. Specifically, we want to test that a given value $\beta_0$ belongs to the identified set for $\beta$, $\widetilde H_0:\beta_0\in B_I(P),$ where 
\begin{equation}\label{eq: ID set}
B_{I}\left(P\right)=\left\{ \beta:\mbox{ there exists }\delta\mbox{ such that }E_{P_{D|Z}}[Y_i(\beta)-X_i(\beta)\delta|Z_i]\le 0\mbox{ almost surely}\right\} 
\end{equation}
is the set of values $\beta$ such that there exists $\delta$ which makes (\ref{eq: Linear Conditional Moments}) hold.  For the remainder of the paper we omit the phrase ``almost surely'' for brevity.
We call restrictions of the form (\ref{eq: Linear Conditional Moments}) \emph{linear conditional moment inequalities}.  They have two key properties: first, the nuisance parameter $\delta$ enters linearly and, second, the Jacobian of the moments with respect to $\delta$, $-X_i(\beta)$, is non-random conditional on $Z_i$. This structure implies that the variance of the moments conditional on $Z_i$ does not depend on $\delta$.

It is helpful to compare (\ref{eq: Linear Conditional Moments}) to the linear regression model
\begin{equation}
Y^*_{i}={X^{*}_{i}}'\delta+\varepsilon_{i}\mbox{ where }E_{P_{D|X^*}}\left[\varepsilon_{i}|X^*_{i}\right]=0\label{eq: Linear Regression Model}
\end{equation}
for $Y_i^*\in\mathbb{R}$ and $X^*_i\in\mathbb{R}^p$. Specifically, (\ref{eq: Linear Conditional Moments}) implies
\begin{equation}
Y_{i}(\beta)=X_{i}(\beta)\delta+\varepsilon_{i}(\beta)\mbox{ where }E_{P_{D|Z}}\left[\varepsilon_{i}(\beta)|Z_{i}\right]\le0,\label{eq: Linear Moment Inequality Model}
\end{equation}
where $Y_{i}(\beta)\in\mathbb{R}^k$ and $X_i(\beta)\in\mathbb{R}^{k\times p}$.
Linear conditional moment inequalities thus generalize the traditional regression model to (a) relax the conditional moment restriction on the errors $\varepsilon_i$ to an inequality, (b) allow the possibility that there are instruments $Z_i$ beyond the regressors $X_i$,  (c) allow a vector-valued outcome, and (d) allow $\beta$ to enter the moments non-linearly.  

\subsection{Examples of Linear Conditional Moment Inequalities}

Linear conditional moment inequalities appear in a variety of economic applications.

\paragraph{Example 1} Linear conditional moment inequalities arise naturally from the linear regression model (\ref{eq: Linear Regression Model}), and its instrumental variables generalization, when we  observe only bounds on the outcome $Y_i^*$. Consider the model 
\begin{equation}\label{eq: IV model}
Y_{i}^*=W_i\beta+V_{i}'\delta+\varepsilon_{i},\mbox{  }E_{P_{D|Z}}\left[\varepsilon_{i}|Z_{i}\right]=0
\end{equation}
where $V_i$ is a function of $Z_i$ while $W_i$ may be endogenous. For instance, $\beta$ may be a causal effect of interest whereas $V_i$ represents a set of control variables.  This is a linear instrumental variables model where the error is mean-independent of the instrument.

As in e.g. \cite{manskitamer2002}, suppose that rather than observing $Y_i^*$ we instead observe bounds $Y_i^L$ and $Y_i^U$ where $Y_i^L\le Y_i^*\le Y_i^U$ with probability one.
 The model (\ref{eq: IV model}) implies that $E_{P_{D|Z}}[Y_i^L-W_i\beta-V_i'\delta|Z_i]\le0$ and $E_{P_{D|Z}}[W_i\beta+V_i'\delta-Y_i^U|Z_i]\le0$, so we obtain conditional moment inequalities.  To cast these inequalities into our framework, suppose we are interested in inference on $\beta,$ and for any vector of non-negative functions of the instruments $f(Z_i)$ let $Y_i(\beta)=(Y_i^L-W_i\beta,W_i\beta-Y_i^U)'\otimes f(Z_i),$ and $X_i=(V_i', -V_i')'\otimes f(Z_i)$, for ``$\otimes$'' the Kroneker product.  This yields the moments $ E_{P_{D|Z}}[Y_i(\beta)-X_i\delta | Z_i] \leq 0,$ as desired.\footnote{Our approach to this application relies on the conditional moment restriction $E_{P_{D|Z}}\left[\varepsilon_{i}|Z_{i}\right]=0$.  As discussed by \cite{PonomarevaTamer2011}, this means that the identified set may be empty if the linear model is incorrect.  For $Z_i=(W_i,V_i')'$, \cite{BeresteanuMolinari2008} assume only that $E_{P}[\varepsilon_{i}Z_i]=0$ and conduct inference on the (necessarily nonempty) set of best linear predictors.  \cite{Bontempsetal2012} study identification and inference, including specification tests, for a class of linear models with unconditional moment restrictions.} $\triangle$

\paragraph{Example 2} \cite{Katz2007} studies the impact of travel time on supermarket choice.
Katz assumes that utility is additively separable in the basket of goods bought ($B_{i}$), the travel time to the supermarket 
chosen ($T_{i,s}$), and the cost of the basket ($\pi(B_{i},s)$).  Normalizing coefficient on cost to one,  agent $i$'s realized utility is
$$
U_i(B_{i},s)=   U_i(B_{i}) + C_s'\delta  - (\beta + \nu_i) T_{i,s} - \pi(B_{i},s),
$$
where $C_s$ are observed characteristics of the supermarket, $T_{i,s}$ is the travel time for $i$ going to $s$, and  $\beta+\nu_i$ is its impact on utility, where $\nu_i$ has mean zero given supermarket characteristics and travel times.

Katz assumes travel times and store characteristics are known to the shopper.
For $\tilde{s}$ a supermarket with $T_{i,\tilde{s}} > T_{i,s}$ that also marketed $B_{i}$, he  divides the difference $U_i(B_{i},s)-U_i(B_{i},\tilde{s})$ by $T_{i,s}-T_{i,\tilde{s}}$
and notes that a combination of
expected utility maximization and revealed preference implies that 
 $
 E_{P_{D|Z}}[Y_i(\beta)-X_i\delta | Z_i] \le 0, 
$
for
$$
Y_i(\beta) \equiv  -\beta-\frac{[\pi(B_{i},s)- \pi(B_{i},\tilde{s})]}{T_{i,s}-T_{i,\tilde{s}}}, \;  X_i \equiv -\frac{C_s'-C_{\tilde{s}}'}{T_{i,s}-T_{i,\tilde{s}}},  \hbox{ and }  Z_i \equiv (T_{i,s},T_{i,\tilde{s}}, C_s',C_{\tilde{s}}')'.
$$
Together with an analogous inequality which uses a store closer to the agent,  Katz obtains both upper and lower bounds for $\beta$. 
$\triangle$

\paragraph{Example 3} \cite{Wollmann} considers the bailout of GM and Chrysler's commercial truck divisions during the 2008 financial crisis and asks what would have happened had they instead been allowed to either fail or merge with another firm.  This example is the basis for our simulations below.

Merger analysis focuses on price differences pre- and post-merger. Wollmann notes that some commercial truck production is modular (it is possible to connect different cab types to different trailers), so some products would likely have been repositioned  after the change in the environment.  To analyze
 product repositioning he requires estimates for the fixed costs of marketing a product.  His estimated demand  and cost systems enable him to estimate counterfactual profits from adding
or deleting products. Assuming firms maximize expected profits, differences in expected profits from adding or subtracting products imply bounds on fixed costs. 

To illustrate, let  $J_{f,t}$ be the set of models
that firm $f$  marketed in year $t$ and let $J_{f,t}\setminus j $ be that set excluding product $j$, while $\Delta \pi(J_{f,t}, J_{f,t}\setminus j)$ is the 
difference in expected profits between marketing $J_{f,t}$  and $J_{f,t}\setminus j$. The fixed cost to firm $f$ of marketing  product $j$ at time $t$ is given by $(\delta_{c,f} + \delta_g g_j)$ if the product was not marketed previously ($j\not\in J_{f,t-1}$), and $\beta (\delta_{c,f} + \delta_g g_j)$ if it was previously marketed.  Here $\delta_{c,f}$ is a firm-specific intercept, $g_j$ is the weight of product $j$, $\delta_g$ is the cost of adding additional weight (assumed common across firms), and $\beta$ captures the cost savings of marketing a pre-existing product. We can write the fixed cost as $X_{j,f,t}^*(\beta)\delta$, where $X_{j,f,t}^*(\beta)$ contains a firm indicator and the product's weight, possibly multiplied by $\beta$ depending on whether $j \in J_{j,f,t-1}$. For $Z_{f,t}$ a set of variables known to the firm when marketing decisions were made, including the variables used to form $X_{j,f,t}^*(\beta)$,
\begin{equation}
 E_{P_{D|Z}}[Y_{j,f,t}  - X_{j,f,t}(\beta) \delta |Z_{f,t}] \geq 0 \mbox{ for all }j, \label{eq: Wollmann moment example} 
\end{equation}
by the firm's equilibrium conditions, where 
$$ 
Y_{j,f,t} \equiv \Delta \pi(J_{f,t}, J_{f,t} \setminus j) \cdot 1\{ j \in J_{f,t}, j \in J_{f, t-1}\}, \; \hspace{0.5cm} X_{j,f,t}(\beta) \equiv  X_{f,j,t}^*(\beta) \cdot  1\{ j \in J_{f,t}, j \in J_{f, t-1}\}
$$
and $1\{A\}$ is an indicator for the event $A$.
Additional inequalities can be added for marketing a product that was not marketed in the prior period, for withdrawing products, and for  combining the withdrawal
of one product with adding another.   $\triangle$

\smallskip

\cite{cox_simple_2020} note that moment inequalities in \cite{Eizenberg2014} and \cite{Gandhietal2019} also have linear conditional structure.  Further recent examples appear in \cite{HoPakes2014}, \cite{Moralesetal2017},  \citet{rambachan_more_2022}, and \citet{rambachan_identifying_2021}.

\subsection{Simplifications from Linear Conditional Structure \label{subsec: simplifications from conditional structure}}

In addition to arising frequently in applications, the structure of linear conditional moment inequalities can be exploited to develop simple and computationally tractable tests of (\ref{eq: Linear Conditional Moments}). We begin by describing an asymptotic framework frequently used to test moment inequalities, and some challenges it generates. We then describe how  linear conditional structure can be used to circumvent some of these issues. We focus on the intuition here, deferring formal results to the following sections.

\paragraph{Unconditional asymptotics} Conditional moment inequalities are often tested indirectly. In particular, (\ref{eq: Linear Conditional Moments}) implies that
$
E_P[ Y_i(\beta) - X_i(\beta) \delta   ] \leq 0.
$
 To test $\widetilde{H}_0:\beta_0\in B_I(P),$ we may therefore test that there exists a value of $\delta$ such that $E_P[ Y_i(\beta_0) - X_i(\beta_0) \delta] \leq 0$.
Letting $Y_{n,0}=\frac{1}{\sqrt{n}}\sum_i Y_i(\beta_0)$ and $X_{n,0} = \frac{1}{\sqrt{n}}\sum_i X_i(\beta_0)$, the central limit theorem implies that for each $\delta$, $Y_{n,0}-X_{n,0}\delta- \mu_{U,n,0}(\delta) \to_d N(0,\Sigma_{U,0}(\delta)),$ for $\mu_{U,n,0}(\delta) = \sqrt{n}E_P[Y_{i}(\beta_0)-X_i(\beta_0)\delta]$ and $\Sigma_{U,0}(\delta)=Var_P(Y_i(\beta_0)-X_i(\beta_0)\delta)$. This suggests the approximation 
\begin{equation}
	Y_{n,0}-X_{n,0}\delta \approx^d N(\mu_{U,n,0}(\delta),\Sigma_{U,0}(\delta)), \label{eq: unconditional normal approximation}
\end{equation}
\noindent where $\approx^d$ denotes approximate equality in distribution.  The normal approximation (\ref{eq: unconditional normal approximation}) may be used to test $H_0^{joint}(\delta_0):\mu_{U,n,0}(\delta_0)\le 0$, which jointly restricts $(\beta,\delta)$.  This allows a projection test of
$\widetilde{H}_0:\beta_0\in B_I(P)$, which rejects if and only if we reject $H_0^{joint}(\delta_0)$ for all $\delta_0$. Simple projection tests can be quite conservative, however, which has motivated approaches based on the joint limiting distribution across different values of $\delta$ \citep[e.g.][]{Kaidoetal2019}.
 
Even if we are happy to use the projection method, projection tests based on (\ref{eq: unconditional normal approximation}) are complicated by the dependence of the variance matrix $\Sigma_{U,0}(\delta_0)$ on the value of $\delta_0$, since critical values for tests of $H_0^{joint}(\delta_0)$ will typically depend on $\delta_0$ as well. When the nuisance parameter $\delta$ has even moderate dimension, calculating the critical value for many values of $\delta_0$ can become computationally burdensome, necessitating careful attention to algorithms to mitigate the computational cost \citep[e.g.,][]{Kaidoetal2019}.  

\paragraph{Conditional asymptotics} Linear conditional structure allows an alternative asymptotic approximation, which avoids complications discussed above by conditioning on the sequence of realized instrument values $\{Z_i\} = \{Z_i\}_{i=1}^{\infty}$. For $\mu_i(\beta,{P_{D|Z}}) = E_{P_{D|Z}}[Y_i(\beta) | Z_i]$ and $\mu_{n,0} = \frac{1}{\sqrt{n}}\sum_i \mu_i(\beta_0,{P_{D|Z}})$, the Lindeberg-Feller central limit theorem implies that under mild conditions $Y_{n,0} - \mu_{n,0} | \{Z_i\}\rightarrow_d N(0,\Sigma_0)$, where $\Sigma_0 = E_{P}[Var_{P_{D|Z}}(Y_i(\beta_0)|Z_i)]$. Since $X_{n,0}$ is non-random conditional on $\{Z_i\}$, this suggests the approximation 
\begin{equation}\label{eq: normal approximation}
	Y_{n,0}-X_{n,0}\delta | \{Z_i\} \approx^d N(\mu_{n,0} -X_{n,0}\delta,\Sigma_0).
\end{equation}
\noindent Importantly, and in contrast to (\ref{eq: unconditional normal approximation}), the variance $\Sigma_0$ in (\ref{eq: normal approximation}) does not depend on the value of $\delta$. This substantially simplifies the problem of constructing tests.  Further, since $X_{n,0}$ is non-stochastic conditional on $\{Z_i\}$, (\ref{eq: normal approximation}) holds jointly across values of $\delta.$

\label{text: conditional approx discussion} To construct tests based on this conditional approximation, observe that if $\widetilde{H}_0: \beta_0 \in B_I(P)$ holds, then there exists (almost surely) a value of $\delta$ such that $\mu_{n,0}- X_{n,0}\delta \leq 0$. The null $\widetilde{H}_0:\beta_0\in B_I(P)$ thus implies the null $H_0: \mu_{n,0} \in \mathcal{M}_{n,0}$, where 
\[\mathcal{M}_{n,0} = \{ \mu\in\mathbb{R}^k : \text{ there exists } \delta \text{ such that }  \mu - X_{n,0} \delta \leq 0\}\]
is  non-stochastic conditional on $\{Z_i\}$.\footnote{\label{FN: def of Mn0}In fact, $\widetilde{H}_0$ implies that $\mu_{n,0} \in \mathcal{M}_{n,0} \cap \mathcal{M}_{n,0,\mathcal{P}_{D|Z}}$, where $\mathcal{P}_{D|Z}$ is the family of conditional distributions implied by $\mathcal{P}$, while $\mathcal{M}_{n,0,\mathcal{P}_{D|Z}} = \left\{ \frac{1}{\sqrt{n}} \sum_i E_{P_{D|Z}}[Y_i(\beta_0)  |Z_i] \,|\, P_{D|Z} \in \mathcal{P}_{D|Z} \right\} $. For tractability, we focus on the implied null that $ \mu_{n,0} \in \mathcal{M}_{n,0} $ rather than $\mu_{n,0} \in \mathcal{M}_{n,0} \cap \mathcal{M}_{n,0,\mathcal{P}_{D|Z}}$. This yields valid but potentially conservative tests if $ X_{n,0}\delta \not\in \mathcal{M}_{n,0,\mathcal{P}_{D|Z}}$ for all $\delta$, i.e. if $\mathcal{P}_{D|Z}$ does not allow all moments to simultaneously bind; see Section \ref{sec: lf tests} for additional discussion.} Equation (\ref{eq: normal approximation}) with $\delta=0$ further implies that $Y_{n,0} | \{Z_i\} \approx^d N(\mu_{n,0}, \Sigma_0)$, so testing $H_0$ reduces, asymptotically, to testing a restriction on the mean of a multivariate normal vector.

\paragraph{Indirect Tests}
While indirect tests of $\widetilde{H}_0:\beta_0\in B_I(P)$ are natural, they can entail a loss of consistency. The original null hypothesis $\widetilde{H}_0:\beta_0\in B_I(P)$ implies that there exists a $\delta$ such that $\frac{1}{\sqrt{n}}\sum_i (E_{P_{D|Z}}[Y_i(\beta_0) | Z_i] - X_i(\beta_0) \delta) \otimes f(Z_i) \leq 0$ for all non-negative functions $f(Z_i)$, whereas $H_0: \mu_{n,0} \in \mathcal{M}_{n,0}$ only tests that this is satisfied for $f(Z_i)=1$.\footnote{Note that if one starts with $(Y_i,X_i)$ satisfying (\ref{eq: Linear Conditional Moments}), then $E_{P_{D|Z}}[\tilde{Y}_i - \tilde{X}_i \delta | Z_i]\leq0$  for $ (\tilde{Y}_i, \tilde{X}_i) = (Y_i,X_i) \otimes f(Z_i)$ and any non-negative finite instrument function $f(Z_i)$. Thus, a key restriction imposed in our framework is that the researcher chooses a finite set of instruments with which to interact the initial moments.} 
Indeed, conditional moment inequalities based on continuously distributed instruments $Z_i$ generate an infinite number of unconditional inequalities, as discussed in e.g. \citetalias{AndrewsShi2013}, \citet{Armstrong2014},  \citet{Chernozhukovetal2015}, and \citet{Chetverikov2013}. 
As a result, the tests we develop do not in general yield consistent tests when the instruments are continuously distributed. This contrasts with the aforementioned papers, which develop consistent tests by checking an (asymptotically) infinite number of moment restrictions.  

Inference based on a finite, researcher-selected set of inequalities nonetheless appears widespread in applications, and is the approach adopted in all the empirical applications discussed above save \cite{Gandhietal2019}.  This raises the question of how to select the finite set of moments (i.e, which restrictions to include in $Y_i$), which we discuss informally in Section \ref{sec: choice of moments} below. Whether one can go further, either characterizing an optimal selection of moments or combining our results with those in the previous literature on conditional moment inequalities to ensure consistent inference in settings with continuously distributed $Z_i$, is an interesting question for future work.

\section{Inference Procedures in the Normal Model}\label{sec:Inference}
We now introduce our tests. Motivated by the asymptotic approximation (\ref{eq: normal approximation}), we begin with tests of $H_0:\mu_{n,0}\in\mathcal{M}_{n,0}$ in the exact normal model
\begin{equation}
	Y_{n,0} \sim N(\mu_{n,0}, \Sigma_0) \text{ for known } \Sigma_0. \label{eqn: finite sample normal model}
\end{equation} 
\noindent The next section presents sufficient conditions for feasible versions of our tests, based on non-normal data and estimates of $\Sigma_0$, to uniformly control asymptotic size.

\subsection{Test Statistic\label{sec:test_statistic}}

Given $Y_{n,0} \sim N(\mu_{n,0}, \Sigma_0)$ for known $\Sigma_0$, we construct tests for the hypothesis $H_0:\mu_{n,0} \in\mathcal{M}_{n,0}$, that is, that there exists some $\delta$ such that $\mu_{n,0}-X_{n,0}\delta \le 0$. We eliminate the nuisance parameter $\delta$ by using the profiled max statistic,
$$
\hat{\eta}_{n,0} = \min_\delta \hspace{0.1cm} \max_{j} \,\, \left\{  e_j'\left( Y_{n,0}-X_{n,0}\delta\right)/\sigma_{0,j} \right\} \label{eq: max statistic}
$$
for $e_j$ the $j$th standard basis vector and $\sigma_{0,j}=\sqrt{e_j'\Sigma_{0}e_j}$.\footnote{We define $\frac{c}{0}=\infty$ for all $c>0$.} Our test statistic thus profiles the maximum-criterion statistic ($S_3$ in the notation of \citetalias{AndrewsSoares2010}). By a profiled test statistic, we mean one that optimizes over the nuisance parameter $\delta$ to find the value that makes the test statistic as small as possible. Specifically, note that $\max_{j} \left\{ e_j'\left( Y_{n,0}-X_{n,0}\delta\right)/\sigma_{0,j} \right\}$
calculates the maximum studentized violation of the sample moments at a given $\delta$, so $\hat{\eta}_{n,0}$ corresponds to the maximum violation at the value of $\delta$ that makes this violation the smallest.  One could profile test statistics other than the max statistic --- e.g. \cite{cox_simple_2020} study profiled QLR statistics and \citet{BugniCanayShi2017} study profiled modified method of moments (MMM) statistics (among others) --- but it will be helpful for our analysis that the profiled max statistic admits an equivalent representation as the solution to the linear program, 
\begin{equation}
	\hat{\eta}_{n,0} = \min_{\eta,\delta} \hspace{.1cm} \eta 
	\mbox{ subject to }Y_{n,0}-X_{n,0}\delta\le\eta\cdot \sigma_{0},
	\label{eq: Projection Linear Program}
\end{equation}
for $\sigma_{0}=(\sigma_{0,1},...,\sigma_{0,k})'$.  This allows for tractable computation of $\hat{\eta}_{n,0}$ even when the dimension of $\delta$ is large, and the linear structure plays a key role in the construction of our tests. 

\subsubsection{Dual representation of the test statistic \label{subsec: dual}}
To derive critical values, we will make use of the dual representation of the linear program (\ref{eq: Projection Linear Program}). Standard results in linear programming (e.g., Chapter 7.4 of \citet{Schrijver1986}) imply that when $\hat{\eta}_{n,0} > -\infty$ it is the solution of the dual linear program,\footnote{Observe that $\hat{\eta}_{n,0}$ is equal to $-\infty$ if and only if $\min_\delta \max_j e_j' X_{n,0} \delta =-\infty$, in which case $H_0$ is satisfied regardless of the value of $\mu_{n,0}$, so the testing problem is trivial.  Finiteness of $\hat{\eta}_{n,0}$ implies that $X_{n,0}$ does not have full row rank, for instance because $k>p$. \label{FN: infinite eta}}
\begin{equation}
	\hat{\eta}_{n,0} = \max_{\gamma} \hspace{.1cm} \gamma'Y_{n,0} \text{ s.t. } \gamma \geq 0, \gamma' X_{n,0} = 0, \gamma'\sigma_0 = 1. \label{eqn: dual problem}
\end{equation}
\noindent Moreover, the maximum is obtained at one of the finite set of vertices of the feasible set. Intuitively, the set of feasible values $F(X_{n,0}, \sigma_0) = \{  \gamma \geq 0 |  \gamma' X_{n,0} = 0, \gamma'\sigma_0 = 1 \}$ is a polyhedron, i.e. a convex set with flat sides, and a vertex corresponds with a ``corner'' of this set. More formally, as described in e.g. \citet[][Section 8.5]{Schrijver1986}, $\gamma\in  F(X_{n,0}, \sigma_0)$ is a vertex if it can be realized as a unique solution to (\ref{eqn: dual problem}) for some value of $Y_{n,0}$:
\begin{definition}
The set of vertices $V(X_{n,0}, \sigma_0)$ of $F(X_{n,0}, \sigma_0)$  is  \[V(X_{n,0}, \sigma_0)=\left\{\gamma\in F(X_{n,0}, \sigma_0) : \exists y\in\mathbb{R}^k \text{ such that }\gamma'y>\tilde\gamma'y\text{ for all }\tilde\gamma\in F(X_{n,0}, \sigma_0)\setminus\{\gamma\}\right\}.\] 
\end{definition}
\noindent As a simple example, if $\Sigma =I$ and $X_{n,0}=0$, then $V(X_{n,0}, \sigma_0)$ is the set of standard basis vectors in $\mathbb{R}^k$. In Lemma \ref{lem: vertex characterization} in the appendix, we give an alternative characterization of the set of vertices, which shows that $\gamma \in F(X_{n,0}, \sigma_0)$ is a vertex if and only if $\gamma$ is the solution to the system of equations defined by a full-rank subset of the constraints in (\ref{eqn: dual problem}). Since there are a finite number of constraints in (\ref{eqn: dual problem}), this immediately implies that $V(X_{n,0},\sigma_0)$ is finite. It is neither necessary nor recommended to enumerate all of the elements of $V(X_{n,0},\sigma_0)$ to compute our test statistic and critical values (see Section \ref{sec: Practical Implementation} for details on computation), but this representation will be useful for explaining our approach. 

The dual representation for $\hat\eta_{n,0}$ implies that in the finite sample normal model the test statistic $\hat{\eta}_{n,0}$ is the maximum of a multivariate normal vector, $\hat{\eta}_{n,0}=\max_{\gamma\in V(X_{n,0}, \sigma_0)}\gamma'Y_{n,0} = \max\{\gamma_{(1)}'Y_{n,0},...,\gamma_{(J)}'Y_{n,0} \}$, for $\gamma_{(1)},...,\gamma_{(J)}$ the elements of $V(X_{n,0} , \sigma_{0})$. Our critical values will then be based on properties of the maximum of a correlated Gaussian vector. 

\subsection{Least Favorable Tests \label{sec: lf tests}}

Our first test is based on the ``least-favorable'' value of $\mu_{n,0}$ under the null hypothesis $H_0$. Recall that $\hat{\eta}_{n,0} =  \max_{\gamma \in V(X_{n,0}, \sigma_0)} \gamma' Y_{n,0} $. Hence 
$$\hat{\eta}_{n,0} = \max_{\gamma \in V(X_{n,0}, \sigma_0)} \left\{\gamma' \mu_{n,0}  + \gamma' (Y_{n,0} -\mu_{n,0})\right\} \leq \max_{\gamma \in V(X_{n,0}, \sigma_0)} \gamma' \mu_{n,0}  +  \max_{\gamma \in V(X_{n,0}, \sigma_0)} \gamma' (Y_{n,0} -\mu_{n,0}).$$
 Under $H_0$, however, there exists $\delta$ such that $\mu_{n,0} - X_{n,0} \delta \leq 0$. Since every $\gamma \in V(X_{n,0}, \sigma_{0})$ is feasible in (\ref{eqn: dual problem}) by construction, we also have that $\gamma'X_{n,0} = 0$ and $\gamma \geq 0$ for all $\gamma \in V(X_{n,0}, \sigma_0)$. It follows that under the null,  $ \gamma'\mu_{n,0} =  \gamma'(\mu_{n,0} - X_{n,0} \delta) \leq 0$ for all $\gamma \in V(X_{n,0}, \sigma_{0})$. Combined with the previous display, this implies that  under $H_0,$
\begin{equation}
\hat{\eta}_{n,0} \leq  \max_{\gamma \in V(X_{n,0}, \sigma_0)} \gamma' (Y_{n,0} -\mu_{n,0}). \label{eqn: inequality for lf}
\end{equation}
Since $Y_{n,0} -\mu_{n,0} \sim N(0,\Sigma_0)$, we define the least-favorable critical value $c_{\alpha,LF}=c_{\alpha,LF}(X_{n,0}, \sigma_{0})$ as the $1-\alpha$ quantile of $ \max_{\gamma \in V(X_{n,0}, \sigma_0)} \gamma' \xi$ for $\xi \sim N(0,\Sigma_0)$ and consider the test that rejects when $\hat{\eta}_{n,0}$ exceeds this critical value, $\phi_{LF} = 1\left\{ \hat{\eta}_{n,0} > c_{\alpha,LF} \right\}$. It follows immediately from the inequality (\ref{eqn: inequality for lf}) that under the finite sample normal model $E[\phi_{LF}] \leq \alpha$ whenever $H_0: \mu_{n,0} \in \mathcal{M}_{n,0}$ holds. Moreover, the inequality (\ref{eqn: inequality for lf}) reduces to an equality if $\gamma' \mu_{n,0} = 0$ for all $\gamma \in V(X_{n,0}, \sigma_0)$, as for example occurs if $\mu_{n,0} =0$ or more generally if $\mu_{n,0}=X_{n,0}\delta$ for some $\delta$, in which case $E[\phi_{LF}] = \alpha$. Thus, the LF test has exact size in the finite sample normal model if it is possible for all moments to bind simultaneously. We note, however, that this may not be possible for some data-generating processes (e.g., if certain pairs of moments correspond to upper and lower bounds that cannot simultaneously bind), in which case the least favorable test may have size strictly less than $\alpha$.\footnote{\label{FN: modification of LF}
In such cases, where $0\not\in\mathcal{M}_{n,0}\cap\mathcal{M}_{n,0,\mathcal{P}_{D|Z}}$ for $\mathcal{M}_{n,0,\mathcal{P}_{D|Z}}$ as defined in footnote \ref{FN: def of Mn0},
tests based on the critical value $ c_{\alpha,LF}+\psi$ for $\psi=\max_{\mu_{n,0}\in\mathcal{M}_{n,0}\cap\mathcal{M}_{n,0,\mathcal{P}_{D|Z}}}\max_{\gamma\in V(X_{n,0},\sigma_0)}\gamma'\mu_{n,0}$ will also control size.  These tests have (weakly) improved power since $\psi\le 0$ by definition.  The adjustment factor $\psi$ depends on the class of conditional data generating processes $\mathcal{P}_{D|Z}$ considered, however, so we focus on results using $c_{\alpha,LF}$ for simplicity.
}

\paragraph{Sensitivity to slack moments} An undesirable feature of the LF test is that it may be sensitive to the inclusion of slack moments. That is, the power of the test may be negatively affected if one includes in $Y_{n,0}$ moments that are very far from binding (i.e. elements $j$ with $\mu_{n,0,j}\ll 0$). The reason is that the critical value $c_{\alpha,LF}$ is based on the distribution of the test statistic when $\mu_{n,0}=0$, and thus generally increases when adding additional moments, even though the test statistic $\hat{\eta}_{n,0}$ will generally not be affected by the inclusion of very slack moments. Motivated by this fact,  \citetalias{AndrewsSoares2010}, \citetalias{AndrewsBarwick2012},  \cite{Romanoetal2014}, and related papers propose techniques that use information from the data to either select moments or shift the mean of the distribution from which the critical values are calculated.  This yields tests with higher power in cases where many of the moments are slack.  Unfortunately, applying these existing methods in our setting breaks the linear structure, and hence the computational advantages from using linear programming, which motivates us to introduce an alternative approach.

\subsection{Conditional Test}\label{subsec: conditional test}
We next introduce a test that is less sensitive to the inclusion of slack moments than the LF test while also exploiting the linear conditional structure in our context. This test is based on the distribution of $\hat{\eta}_{n,0}$ conditional on the 
identity of the optimal vertex in the dual problem, $\hat\gamma=\arg\max_{\gamma \in V(X_{n,0},\sigma_0)}\gamma'Y_{n,0}$.\footnote{$\hat\gamma$ depends on $n$ and $\beta_0$, but we leave this dependence implicit for simplicity of notation.}   For simplicity of exposition, we begin by assuming that $\hat\gamma$ is unique, in the sense that $\arg\max_{\gamma \in V(X_{n,0},\sigma_0)}\gamma'Y_{n,0}$ is a singleton; we will discuss the case of a non-unique dual below.\footnote{Our asymptotic results in the next section impose a sufficient condition for uniqueness to hold with probability one asymptotically.} 
If $\hat\gamma'\Sigma_0\hat\gamma=0$ then we define the conditional test to reject if and only if $\hat\eta_{n,0}>0.$ For the remainder of this section, we thus assume that $\hat\gamma'\Sigma_0\hat\gamma>0.$
For any $\gamma \in V(X_{n,0},\sigma_0),$ note that $\hat\gamma=\gamma$ only if $\gamma'Y_{n,0}\ge \tilde\gamma'Y_{n,0}$ for all $\tilde\gamma\in V(X_{n,0},\sigma_0)$.  Hence, $\hat\gamma=\gamma$ is optimal only if $Y_{n,0}$ lies in the polyhedron $\{y \,|\, (\gamma - \tilde\gamma)' y \geq 0, \, \forall \tilde\gamma \in V(X_{n,0}, \sigma_0)\}$. This representation allows us to characterize the distribution of $\hat{\eta}_{n,0}$ conditional on $\hat\gamma=\gamma$ using Lemma 5.1 in \citet{Leeetal2016}, which characterizes the behavior of Gaussian random variables conditional on polyhedral events. 

\begin{lemma} \label{lem:conditional distribution of etahat}
Let $S_{n,0,{\gamma}}=\left(I-\frac{\Sigma_0{\gamma}{\gamma}'}{{\gamma}'\Sigma_0{\gamma}}\right)Y_{n,0}$. Then under (\ref{eqn: finite sample normal model}),
\begin{equation}\hat{\eta}_{n,0} \,|\, \left\{ \hat\gamma=\gamma, S_{n,0,{\gamma}} = s \right\} \sim TN( \gamma'\mu_{n,0}, \gamma' \Sigma_0 \gamma, [\mathcal{V}^{lo}_{n,0}, \mathcal{V}^{up}_{n,0}]  ), \label{eqn: condl distribution of eta}
\end{equation}
\noindent where $TN(\mu,\sigma^2, [a,b])$ denotes the $N(\mu,\sigma^2)$ distribution truncated to $[a,b]$,  
\begin{equation}
	\mathcal{V}_{n,0}^{lo}=\max_{\tiny 
		\begin{array}{c}\tilde\gamma\in V(X_{n,0},\sigma_0):\\
			\gamma'\Sigma_0\gamma>\gamma'\Sigma_0\tilde\gamma
		\end{array}
	}\frac{\gamma'\Sigma_0\gamma\cdot\tilde\gamma's}{\gamma'\Sigma_0\gamma-\gamma'\Sigma_0\tilde\gamma},
	~
	\mathcal{V}_{n,0}^{up}=\min_{ \tiny 
		\begin{array}{c}\tilde\gamma\in V(X_{n,0},\sigma_0):\\
			\gamma'\Sigma_0\gamma<\gamma'\Sigma_0\tilde\gamma
		\end{array}
	}
	\frac{\gamma'\Sigma_0\gamma\cdot\tilde\gamma's}{\gamma'\Sigma_0\gamma-\gamma'\Sigma_0\tilde\gamma},\label{eq: V^lo, V^up}
\end{equation}
and we define $\mathcal{V}_{n,0}^{lo}=-\infty$ and $ \mathcal{V}_{n,0}^{up}=\infty$, respectively, when we optimize over the empty set.
\end{lemma}

Recall that under $H_0$, $\gamma'\mu_{n,0} \leq 0$ for all $\gamma \in V(X_{n,0}, \sigma_0)$. Additionally, Lemma A.1 in \citet{Leeetal2016} shows that the $TN(\mu,\sigma^2;[a,b])$ distribution is increasing in $\mu$ in the sense of first order stochastic dominance. It follows that the distribution on the right-hand side of (\ref{eqn: condl distribution of eta}) is weakly dominated by the $TN(0,\hat\gamma' \Sigma_0 \hat\gamma, [\mathcal{V}^{lo}_{n,0}, \mathcal{V}^{up}_{n,0}])$ distribution under the null. We therefore base our test on this distribution. Letting $\bar{c}_{\alpha,C}$ be the $1-\alpha$ quantile of the $TN(0,\hat\gamma' \Sigma_0 \hat\gamma, [\mathcal{V}^{lo}_{n,0}, \mathcal{V}^{up}_{n,0}])$ distribution, we define the conditional critical value as $c_{\alpha,C} = c_{\alpha,C}(Y_{n,0},X_{n,0},\Sigma_0) = \max\{ \bar{c}_{\alpha,C}, 0\}$ and reject if $\hat{\eta}_{n,0}$ exceeds it, $\phi_C= 1\left\{ \hat{\eta}_{n,0} > c_{\alpha,C} \right\}$.\footnote{The censoring of the critical value at 0 is unnecessary for size control in the finite-sample normal model, but simplifies asymptotic arguments. It is also substantively reasonable as it prevents the test from rejecting when all of the moment inequalities are satisfied in sample ($\hat{\eta}_{n,0}\leq0$).} It follows immediately that $\phi_C$ controls size conditionally in the finite sample normal model, with $E[\phi_{C} | \hat\gamma =\gamma, S_{n,0,\gamma} ] \leq \alpha$ whenever $\mu_{n,0}\in \mathcal{M}_{n,0}$.\footnote{As for the least favorable test, if $X_{n,0}\delta\not\in\mathcal{M}_{n,0}\cap\mathcal{M}_{n,0,\mathcal{P}_{D|Z}}$ for all $\delta,$ we can potentially use smaller critical values, replacing $\bar{c}_{\alpha,C}$ with the $1-\alpha$ quantile of a $TN(\psi_{\hat\gamma},\hat\gamma' \Sigma_0 \hat\gamma, [\mathcal{V}^{lo}_{n,0}, \mathcal{V}^{up}_{n,0}])$ distribution for $\psi_{\hat\gamma}=\max_{\mu_{n,0}\in\mathcal{M}_{n,0}\cap\mathcal{M}_{n,0,\mathcal{P}_{D|Z}}}\hat\gamma'\mu_{n,0}$. As before, $\psi_{\hat\gamma}$ will depend on the specification of $\mathcal{P}_{D|Z}$, and we focus on tests based on $\bar{c}_{\alpha,C}$ for simplicity.} Unconditional size control follows by the law of iterated expectations.

\paragraph{Example (uncorrelated moments)} Consider the case where $Y_{n,0}\sim N(\mu_{n,0},I)$, and $X_{n,0}=0$, so that there is no nuisance parameter $\delta$. Then $V(X_{n,0},\sigma_0)$ is simply the set of standard basis vectors, so $\hat{\eta}_{n,0} = \max_j e_j'Y_{n,0}$ is the maximum component of $Y_{n,0}$. In this case $\mathcal{V}^{lo}_{n,0}$ corresponds to the second-largest component of $Y_{n,0}$, i.e. $\max_{j\neq \hat j}  e_j'Y_{n,0}$, for $\hat{j}$ the location of the maximum, and $\mathcal{V}^{up}_{n,0}=\infty$. The conditional test thus rejects if $\hat{\eta}_{n,0}$ exceeds the $1-\alpha$ quantile of the standard normal distribution truncated to $[\mathcal{V}^{lo}_{n,0}, \infty]$.

\paragraph{Non-unique dual solutions.\label{paragraph:nonunique dual}} So far we have assumed the existence of a unique dual solution, $\hat\gamma = \gamma$. If $\Sigma_0$ is not full-rank, however, then there may be multiple solutions to the dual problem with positive probability.\footnote{Since the dual objective is $\hat\eta_{n,0} = \max\{\gamma_{(1)}'Y_{n,0},..., \gamma_{(J)}'Y_{n,0} \}$ and  $\gamma_{(j)}\neq \gamma_{(j')}$ for $j\neq j',$ the dual has a unique solution with probability 1 so long as $\Sigma_0$ is full rank.} In Appendix \ref{appendix: nonunique dual}, we consider a version of the conditional test that, when the dual solution is non-unique, calculates $(\mathcal{V}^{lo}_{n,0}, \mathcal{V}^{up}_{n,0})$ via (\ref{eq: V^lo, V^up})  by selecting an element of the dual solution set, $\gamma = h(\hat\gamma)$. We show that in the finite sample normal model, with probability 1 the critical values do not depend on how the optimal vertex is chosen, so the test obtained does not depend on the choice of $h(\cdot)$. Further, we show in Appendix \ref{appendix: nonunique dual} that this test controls size in the finite-sample normal model. Our sufficient conditions for uniform asymptotic size control in Section \ref{sec: Asymptotics} below imply that the dual solution will be unique with probability tending to 1, however, so we focus primarily on the case where the dual solution is unique. 

\paragraph{Insensitivity to Slack Moments} In contrast with the LF test, the conditional test has the desirable property that it is insensitive to the inclusion of slack moments. Specifically, our next result shows that the conditional test is insensitive to slack moments in the strong sense that as a moment becomes arbitrarily slack the conditional test converges to the conditional test that drops that moment ex-ante.  Intuitively, this happens because (under mild conditions) sufficiently slack moments make no contribution to $\hat{\eta}_{n,0},$ $\mathcal{V}^{lo}_{n,0},$ or $\mathcal{V}^{up}_{n,0},$ and so have no impact on the conditional test. To state this result formally, define $Y_{n,0}^{j,d}=Y_{n,0}-e_{j}\cdot d$ as a version of $Y_{n,0}$ which decreases the $j$th moment by $d$. Let $Y_{n,0}^{-j}$ collect
the rows of $Y_{n,0}$ other than the $j$th, and define $X_{n,0}^{-j}$
and $\Sigma^{-j}_0$ accordingly. Define $\hat{\eta}_{n,0}^{j,d}$
and $\hat{\eta}_{n,0}^{-j}$ as versions of $\hat{\eta}_{n,0}$ based on $\left(Y_{n,0}^{j,d},X_{n,0},\Sigma_0\right)$
and $\left(Y_{n,0}^{-j},X_{n,0}^{-j},\Sigma^{-j}_0\right)$,
respectively, and let $\phi_{C}^{j,d}$ and $\phi_{C}^{-j}$ denote
the corresponding tests.

\begin{lemma} \label{lem: slack moments}
	
	For any $Y_{n,0}$ such that
	$\gamma'Y_{n,0}\neq\tilde{\gamma}'Y_{n,0}$ for all distinct $\gamma,\tilde\gamma\in V\left(X_{n,0},\sigma_0\right)$ and
	$\hat{\eta}_{n,0}^{-j}\neq c_{\alpha,C}\left(Y_{n,0}^{-j},X_{n,0}^{-j},\Sigma_{0}^{-j}\right)$, we have $\lim_{d\to\infty}\phi_{C}^{j,d}=\phi_{C}^{-j}$.
\end{lemma}

The conditions of Lemma \ref{lem: slack moments} hold for Lebesgue almost every $Y_{n,0}$, and hold with probability 1 under (\ref{eqn: finite sample normal model}) provided that $\gamma' \Sigma_{0} \gamma > 0$ and $(\gamma-\tilde\gamma)' \Sigma_0 (\gamma-\tilde\gamma) > 0$ for all distinct $\gamma,\tilde\gamma \in V(X_{n,0}, \sigma_0)$, so that the variables $\gamma' Y_{n,0}$ have positive variance and are not perfectly correlated with one another. The only other tests we are aware of that both control size in the finite-sample normal model and are unaffected by the inclusion of arbitrarily slack moments in the sense of Lemma \ref{lem: slack moments} are those of \cite{cox_simple_2020}.

\paragraph{Power with Multiple Violated Moments.} Although the conditional test exhibits a desirable insensitivity to the inclusion of slack moments, it may exhibit poor power in cases where two (or more) moments are approximately equally violated. This is most easily seen in the example of uncorrelated moments from above, where $\mathcal{V}^{lo}_{n,0}$ corresponds with the value of the second-largest sample moment, and the critical value is the $1-\alpha$ quantile of the standard normal distribution truncated to $[\mathcal{V}^{lo}_{n,0}, \infty]$. If two moments are approximately equally violated, then the largest and second largest sample moments ($\hat{\eta}_{n,0}$ and $\mathcal{V}^{lo}_{n,0}$, respectively) may be close together, so the conditional test need not reject even if both of these are large. This phenomenon is highlighted in parts of the parameter space in our simulations in Section \ref{sec: Monte Carlo}.

\subsection{Hybrid Tests}

To mitigate the possible power losses of the conditional test when multiple moments are approximately equally violated, we next introduce a hybrid test that combines the least favorable and conditional approaches. For some $0<\kappa<\alpha$, we define the size-$\alpha$ hybrid test to reject whenever the size-$\kappa$ least favorable test does. If the least favorable test does not reject, we then consider a size-$\frac{\alpha-\kappa}{1-\kappa}$ test that conditions on both $\hat\gamma=\gamma$ and the event that the least-favorable test did not reject. Specifically, the same argument used to prove Lemma \ref{lem:conditional distribution of etahat} yields that
$$
\hat{\eta}_{n,0} \,|\, \left\{ \hat\gamma=\gamma, S_{n,0,{\gamma}} = s, \phi_{LF,\kappa} =0 \right\} \sim TN( \gamma'\mu_{n,0}, \gamma' \Sigma_0 \gamma, [\mathcal{V}^{lo}_{n,0}, \mathcal{V}^{up,H}_{n,0}]  ), 
$$
\noindent where $ \mathcal{V}^{up,H}_{n,0} = \min\{\mathcal{V}^{up}_{n,0}, c_{\alpha,LF} \}$. We then construct the second-stage critical value $\bar{c}_{\frac{\alpha-\kappa}{1-\kappa}, H}=\bar{c}_{\frac{\alpha-\kappa}{1-\kappa}, H}(Y_{n,0},X_{n,0},\Sigma_0)$ analogously to the conditional critical value $c_{\frac{\alpha-\kappa}{1-\kappa}, C}$ except using the modified truncation point $ \mathcal{V}^{up,H}_{n,0}$. Letting $c_{\frac{\alpha-\kappa}{1-\kappa}, H} = \min\{c_{\kappa,LF}, \bar{c}_{\frac{\alpha-\kappa}{1-\kappa}, H} \}$, the hybrid test is then $\phi_H = 1\{ \hat{\eta}_{n,0} > c_{\frac{\alpha-\kappa}{1-\kappa}, H} \}$. Observe that the critical value for the hybrid test approaches that of the LF test as $\kappa \rightarrow \alpha$, while it approaches that of the conditional test as $\kappa \rightarrow 0$. 

As argued above, the first-stage LF test for the hybrid rejects with probability not exceeding $\kappa$ under the null in the finite-sample normal model. Likewise, by arguments analogous to those for the conditional test, the second stage test rejects with probability no more than $\frac{\alpha - \kappa}{1-\kappa}$ conditional on the first stage not rejecting. It follows that when $\mu_{n,0} \in \mathcal{M}_{n,0}$, the hybrid test rejects with probability 
$$E[\phi_{LF,\kappa}]  + \left(1-E[\phi_{LF,\kappa}]\right)E\left[\hat{\eta}_{n,0} > \bar{c}_{\frac{\alpha-\kappa}{1-\kappa}, H} | \phi_{LF,\kappa} = 0\right]  \leq \kappa + (1-\kappa )\frac{\alpha-\kappa}{1-\kappa} = \alpha,$$ 
and so controls size in the finite sample normal model.

The hybrid test proposed above always rejects whenever a simple Bonferroni combination of a size-$\kappa$ LF test and size-($\alpha - \kappa$) conditional test would reject, and can reject in cases where the simple Bonferroni does not. The proposed method improves upon the simple Bonferroni approach in two ways, first modifying the second-stage test to condition on the event that the LF test does not reject (which truncates the distribution above and so reduces the critical value), and then using a size $\frac{\alpha-\kappa}{1-\kappa} > \alpha-\kappa$ critical value. This helps to reduce the conservativeness usually associated with Bonferroni approaches.

 \paragraph{Sensitivity to Slack Moments} The hybrid test will be sensitive to the inclusion of slack moments via its dependence on the LF critical values. However, this sensitivity will be small when $\kappa$ is close to zero, since in this case the critical values will tend to be close to those of the conditional test, which as shown above do not depend on the inclusion of slack moments. Similar to \cite{Romanoetal2014}, we consider $\kappa=\alpha/10$ in our simulations below. 

\section{Asymptotic Validity}\label{sec: Asymptotics}

We conduct our analysis conditional on a sequence of values for the
instruments, $\left\{ Z_{i}\right\} =\left\{ Z_{i}\right\} _{i=1}^{\infty},$
where the data are independent but potentially not identically distributed conditional on $\left\{ Z_{i}\right\}$,
$
D_{i}\ind D_{i'}|\left\{ Z_{j}\right\} \mbox{ for all }i\neq i'$.  
Recall that $\mathcal{P}_{D|Z}$ is the class of conditional distributions for $D_i$ given
$Z_i$, and let $B_I(P_{D|Z})$ denote the conditional identified set for $\beta$ given 
$\left\{ Z_{i}\right\},$
\[
B_I(P_{D|Z})=\left\{ \beta :\text{ there exists } \delta~s.t.~E_{P_{D|Z}}\left[Y_{i}(\beta)-X_i(\beta)\delta|Z_{i}\right]\le0\mbox{ for all }i\right\} .
\]
Note that for $B_I(P)$ as defined in \eqref{eq: ID set}, $B_I(P)\subseteq B_I(P_{D|Z})$ for almost every $\left\{ Z_{i}\right\}.$ 
 We provide conditions under which our tests uniformly control asymptotic rejection probabilities over $P_{D|Z}\in\mathcal{P}_{D|Z}$ and $\beta_0\in B_I(P_{D|Z})$.
For brevity, we will leave the conditioning on $\left\{ Z_{i}\right\}$ implicit when this is without loss of clarity.

Our first assumption is that, conditional on $Z_i$, $Y_i(\beta_0)$ can be written as a known linear transformation of a vector $U_i(\beta_0)$,
whose average conditional variance
given $Z_{i}$ converges uniformly to a bounded and full-rank limit.

\begin{assumption}\label{Assumption: Variance Convergence}
Suppose
that we can write $Y_{i}(\beta_0)=T U_{i}(\beta_0)+\zeta_{i}(\beta_0)$, where $T$ is a known $k \times l$ matrix while $\zeta_{i}(\beta_0) \in \reals^k$
is known and non-random conditional on $\left\{ Z_{i}\right\}$.  Further suppose that, (i), for
some $\Omega\left(P_{D|Z},\beta_0\right)$, 
\begin{equation}
\lim_{n\to\infty}~\sup_{P_{D|Z}\in\mathcal{P}_{D|Z}}~\sup_{\beta_0\in B_I(P_{D|Z})}\left\Vert \frac{1}{n}\sum_{i=1}^n Var_{P_{D|Z}}\left(U_{i}(\beta_0)|Z_{i}\right)-\Omega\left(P_{D|Z},\beta_0\right)\right\Vert \to0\label{eq: variance convergence}
\end{equation}
and, (ii), for $\bar{\lambda}>0$ a finite constant, $\Omega\left(P_{D|Z},\beta_0\right) \in \mathbf{\Omega}_{\bar\lambda}$ for all $P_{D|Z}\in\mathcal{P}_{D|Z}$, $\beta_0\in B_I(P_{D|Z})$, where 
\[
\mathbf{\Omega}_{\bar\lambda} = \{ \Omega \,|\,  \bar{\lambda}^{-1}\le\lambda_{\min}\left(\Omega\right)\le\lambda_{\max}\left(\Omega\right)\le\bar{\lambda} \}
\]
is the set of matrices with minimal and maximal eigenvalues bounded by $\bar\lambda^{-1}$ and $\bar\lambda$.\end{assumption}
\noindent Note that if the variance of $Y_i(\beta_0)$ is full-rank (as in Examples 2 and 3 above) then the moments can trivially be written as $Y_{i}(\beta_0)=T U_{i}(\beta_0)+\zeta_{i}(\beta_0)$ for $T=I$, $U_{i}(\beta_0)=Y_i(\beta_0)$, and $\zeta_i(\beta_0)=0$. The structure in Assumption \ref{Assumption: Variance Convergence} also commonly arises in moment inequality settings where the variance of $Y_i(\beta_0)$ is not full-rank. For example, consider the case of interval-valued regression (Example 1 above) where the upper- and lower-bounds of the interval are perfectly collinear, $Y_i^U = Y_i^L + c$ for fixed constant $c$. Then $Y_{i}(\beta_0)=T U_{i}(\beta_0)+\zeta_{i}(\beta_0)$ with $T = [I,~ -I]'$, $U_i(\beta_0) = Y_i^L - W_i\beta_0$, and $\zeta_i(\beta_0) = [0,~ -c]'$.  Settings with moment equalities represented as inequalities can similarly be expressed in this form --- if all the moments are of this form, for example, then we can take $T=[I,~-I]'$ and $\zeta_i(\beta_0)=0$.

Assumption \ref{Assumption: Variance Convergence} implies that the average conditional variance of $Y_i(\beta_0)$ given $Z_i$ converges, $\frac{1}{n}\sum Var_{P_{D|Z}}\left(Y_{i}(\beta_0)|Z_{i}\right)\to \Sigma(P_{D|Z},\beta_0)=T\Omega\left(P_{D|Z},\beta_0\right) T'$.   Although $\Omega\left(P_{D|Z},\beta_0\right)$ has full rank, $\Sigma(P_{D|Z},\beta_0)$ may have reduced rank since e.g. the dimension of $\Sigma\left(P_{D|Z},\beta_0\right)$
may exceed that of $\Omega\left(P_{D|Z},\beta_0\right)$. 
We next assume that we have a uniformly consistent estimator for $\Omega(P_{D|Z},\beta_0)$, and thus for $\Sigma(P_{D|Z},\beta_0)$.

\begin{assumption}\label{Assumption: Variance Estimator}
$\widehat{\Sigma}_{n,0}=T'\widehat{\Omega}_{n,0} T$, where $\widehat{\Omega}_{n,0}$ is uniformly consistent for 
 $\Omega\left(P_{D|Z},\beta_0\right)$,
\[
\lim_{n\to\infty}~\sup_{P_{D|Z}\in\mathcal{P}_{D|Z}}~\sup_{\beta_0\in B_I(P_{D|Z})}Pr_{P_{D|Z}}\left\{ \left\Vert \widehat{\Omega}_{n,0}-\Omega\left(P_{D|Z},\beta_0\right)\right\Vert >\varepsilon\right\} =0 \text{ for all }\varepsilon>0.
\] \end{assumption}
We discuss sufficient conditions for uniform consistency of $\widehat{\Omega}_{n,0}$
in Appendix \ref{sec: Variance Estimation}.  Note that $\widehat{\Omega}_{n,0}$ depends on the null parameter value $\beta_0$ considered,
where we again suppress this dependence for brevity of notation.

We further assume that the scaled sample average of $U_i(\beta_0)$ is uniformly
asymptotically normal once recentered around its mean. To state
this assumption we use the fact that uniform convergence in distribution
is equivalent to uniform convergence in bounded Lipschitz metric (see e.g.
Theorem 1.12.4 of van der Vaart and Wellner, 1996).

\begin{assumption}\label{Assumption: Uniform CLT}For $BL_{1}$ the
class of real-valued functions which are bounded in absolute value by one and
have Lipschitz constant bounded by one, $U_{n,0}=\frac{1}{\sqrt{n}}\sum U_i(\beta_0),$ $\pi_{i}(\beta_0)=E_{P_{D|Z}}[U_i(\beta_0)|Z_i]$, $\pi_{n,0}=\frac{1}{\sqrt{n}}\sum_i \pi_i(\beta_0),$ and $\xi_{P_{D|Z}}\sim N\left(0,\Omega\left(P_{D|Z},\beta_0\right)\right),$
\[
\lim_{n\to\infty}~\sup_{P_{D|Z}\in\mathcal{P}_{D|Z}}~\sup_{\beta_0\in B_I(P_{D|Z})}~\sup_{f\in BL_{1}}\left|E_{P_{D|Z}}\left[f\left(U_{n,0}-\pi_{n,0}\right)\right]-E\left[f\left(\xi_{P_{D|Z}}\right)\right]\right|=0.
\]
\end{assumption}Under Assumption \ref{Assumption: Variance Convergence}, the following lower-level condition is sufficient for Assumption \ref{Assumption: Uniform CLT}.

\begin{lemma}\label{lem: uniform CLT}Under Assumption \ref{Assumption: Variance Convergence},
if for all $\varepsilon>0$
\[
\limsup_{n\to\infty}~\sup_{P_{D|Z}\in\mathcal{P}_{D|Z}}~\sup_{\beta_0\in B_I(P_{D|Z})}\frac{1}{n}\sum_{i=1}^n E_{P_{D|Z}}\left[\left\Vert U_{i}(\beta_0)-\pi_i(\beta_0)\right\Vert^2 1\left\{ \left\Vert U_{i}(\beta_0)-\pi_i(\beta_0)\right\Vert >\varepsilon\sqrt{n}\right\} |Z_{i}\right]=0,
\]
then Assumption \ref{Assumption: Uniform CLT} holds.\end{lemma}

Our final assumption, which is needed for the conditional and hybrid approaches, restricts $T$ and $X_{n,0}$. Before stating this assumption, we note that the structure imposed by Assumption \ref{Assumption: Variance Convergence} allows us to consider a subset of the vertices $V(X_{n,0}, \sigma_0)$ discussed in the previous section. Intuitively, the optimal vertex $\hat\gamma$ corresponds to a vector of Lagrange multipliers for the primal problem (\ref{eq: Projection Linear Program}), and thus $\hat\gamma$ must satisfy the complementary slackness conditions. Assumption \ref{Assumption: Variance Convergence} then implies that certain vertices can never be optimal when the test rejects -- for example, if the matrix $T$ encodes moment equalities as inequalities, then the positive and negative copies of a given moment cannot bind simultaneously unless $\hat\eta_{n,0}=0$, in which case our tests do not reject. The following lemma shows that we can ignore such ``never-optimal'' vertices when establishing size control.

\begin{lemma} \label{lem: can consider subset of vertices} Suppose Assumption \ref{Assumption: Variance Convergence} holds, and let $\hat\sigma_{n,0} = \sqrt{diag(\widehat{\Sigma}_{n,0})}\in \mathbb{R}^k$. Then:
\begin{enumerate}
	\item 
	$V(X_{n,0}, \hat\sigma_{n,0}) = \{\lambda_{(1)}(X_{n,0},\hat\sigma_{n,0}) \gamma_{(1)}(X_{n,0}),...,\lambda_{(J)}(X_{n,0},\hat\sigma_{n,0}) \gamma_{(J)}(X_{n,0}) \}$, where the $\lambda_{(j)}(\cdot,\cdot)$ are scalar functions of $X$ and $\sigma$, while $\gamma_{(1)}(X_{n,0}),...,\gamma_{(J)}(X_{n,0})$ are the elements of $V(X_{n,0}, \upsilon)$ for $\upsilon = \sqrt{Diag(TT')}$. 
	
	\item
	Let $\Upsilon_{n,0} = \{ T u + \zeta_{n,0} | u \in \mathbb{R}^l \}$, where $\zeta_{n,0} = \frac{1}{\sqrt n} \sum_i \zeta_i(\beta_0)$. Let $V_{\dagger}(X_{n,0}, \hat\sigma_{n,0}) $ be the subset of $V(X_{n,0}, \hat\sigma_{n,0})$ corresponding with the indices $j$ such that there exists some $\sigma>0$ and some $y \in \Upsilon_{n,0}$ such that $\lambda_{(j)}(X_{n,0},\sigma) \gamma_{(j)}(X_{n,0}) \in \argmax_{\tilde\gamma \in V(X_{n,0}, \sigma)} \tilde\gamma' y $ and $\lambda_{(j)}(X_{n,0},\sigma) \gamma_{(j)}(X_{n,0})' y > 0$. Suppose $V_\dagger(X_{n,0},\hat{\sigma}_{n,0})$ is non-empty.\footnote{If not, then $\hat\eta_{n,0} \leq 0$ with probability 1, and thus none of our tests ever rejects for $\alpha<0.5$.} Then for any $\alpha < 0.5$, the LF, Conditional, and Hybrid tests constructed using $V(X_{n,0}, \hat{\sigma}_{n,0})$  reject only if their analogs constructed using $V_\dagger(X_{n,0}, \hat{\sigma}_{n,0})$ also reject.
\end{enumerate}
 
\end{lemma}

\noindent With the definition of $V_\dagger(X_{n,0},\hat{\sigma}_{n,0})$ in hand, we can now state our final assumption.

\begin{assumption}\label{Assumption: Nondegeneracy} For $n$ sufficiently
	large and all $\beta_0,$ $X_{n,0}\in\mathcal{X}^{*}$ for $\mathcal{X}^{*}$ a closed
	set such that
		$$
		\inf_{\Omega \in \mathbf{\Omega}_{\bar\lambda}} \hspace{.2cm} \inf_{X\in\mathcal{X}^{*}} \hspace{.2cm} \inf_{\gamma,\tilde\gamma \in V_\dagger(X,\sigma(\Omega)),\gamma\neq\tilde\gamma,c\in\mathbb{R}_{\geq0}}\left(\gamma-c\cdot \tilde\gamma\right)'T\Omega T'\left(\gamma-c\cdot \tilde\gamma\right)>0,
		$$
where $\sigma(\Omega) = \sqrt{Diag(T \Omega T')}$. 
\end{assumption}

Together with the structure for the variance matrix $\Sigma$ imposed
in Assumption \ref{Assumption: Variance Convergence}, Assumption \ref{Assumption: Nondegeneracy}
ensures that (i) $\gamma'Y_{n,0}$ has nonvanishing asymptotic variance for all dual vertices $\gamma \in V_\dagger(X_{n,0},\hat{\sigma}_{n,0})$, and (ii) for distinct dual vertices $\gamma$ and $\tilde\gamma$ in $V_\dagger(X_{n,0},\hat{\sigma}_{n,0})$, $\gamma'Y_{n,0}$ and $\tilde\gamma'Y_{n,0}$ are not perfectly positively correlated asymptotically.  The former implies that $\hat\eta_{n,0}$ is continuously distributed in large samples, while the latter ensures that the dual problem $\max_{\gamma\in V_\dagger(X_{n,0},\hat{\sigma}_{n,0})}\gamma'Y_{n,0}$ has a unique solution with probability tending to one. 

In Appendix \ref{sec: sufficient conditions for nondegeneracy}, we provide lower-level sufficient conditions for Assumption \ref{Assumption: Nondegeneracy} in settings where either $\Sigma(P_{D|Z},\beta_0)$ is full-rank or degeneracy in $\Sigma(P_{D|Z},\beta_0)$ arises from matching moments of opposite signs (e.g. moment equalities cast as inequalities). In these settings, we show that Assumption \ref{Assumption: Nondegeneracy} holds automatically when $X_{n,0}$ is constant up to scale (as occurs, e.g., in the difference-in-differences setting of \citet{rambachan_more_2022}). When $X_{n,0}$ is non-constant, a sufficient condition is that $X_{n,0}$ lies in a set $\mathcal{X}$ such that the distance between distinct vertices of $V(X,\upsilon)$ is bounded away from zero over $X \in \mathcal{X}$, where again $\upsilon = \sqrt{diag(TT')}$. Intuitively, this assumption requires that distinct vertices in $V(X_{n,0}, \upsilon)$ not ``converge to each other.'' 

We also note that we do not require any additional assumptions about how $V(X,\sigma)$ depends on $\sigma$, since the proof of Lemma \ref{lem: can consider subset of vertices} shows that $\sigma$ affects $V(X,\sigma)$ only through a continuous re-scaling of the vertices of $V(X,\upsilon)$. This enables us to establish size control when $\sigma_{n,0}$ is replaced with a consistent estimate $\hat\sigma_{n,0}$ without further assumptions.

It is worth highlighting that Assumption 4 involves the variance of $Y_{n,0}$ but not its mean $\mu_{n,0}$. This contrasts with linear independence constraint qualification (LICQ) assumptions that have been considered in other work \citep[e.g.,][]{cho_simple_2021, Garfarov2016}, which restrict the set of moments that can bind in population and thus the value of $\mu_{n,0}$ (see \citet{kaido_constraint_2021} for discussion). In the simplest case without nuisance parameters ($X_{n,0} = 0$), for example, Assumption 4 holds if all of the elements of $Y_{n,0}$ have positive variance and are not perfectly correlated, whereas a standard LICQ condition would impose that $\mu_{n,0}$ has a unique maximum element.\footnote{\citet{rambachan_more_2022} show that in a special setting where $\beta_0$ enters the moments linearly, a population version of LICQ implies that our conditional test has optimal local asymptotic power.} We explore the connections between LICQ and Assumption \ref{Assumption: Nondegeneracy} more formally in Appendix \ref{sec: LICQ}, where we show that LICQ implies that there is a unique solution to a ``population version'' of the dual for $\hat\eta_{n,0}$, whereas Assumption \ref{Assumption: Nondegeneracy} only implies uniqueness of the sample version of the problem (but not necessarily the population version). The tests proposed in \citet{cox_simple_2020}, as well as our LF test, do not require Assumption \ref{Assumption: Nondegeneracy} for uniform asymptotic validity, and thus may be attractive in settings where the researcher is not comfortable with this assumption.

Under these assumptions, feasible versions of our tests, based on the observed $(Y_{n,0},X_{n,0})$, and the estimated variance $\widehat{\Sigma}_{n,0}$, are uniformly asymptotically valid.

\begin{proposition}\label{Prop: Least Favorable Size Control-alt}Under
Assumptions \ref{Assumption: Variance Convergence}, \ref{Assumption: Variance Estimator}, and
\ref{Assumption: Uniform CLT} the least favorable test is uniformly asymptotically valid for $\alpha<0.5$,
\[
\limsup_{n\to\infty}~\sup_{P_{D|Z}\in\mathcal{P}_{D|Z}}~\sup_{\beta_0\in B_I(P_{D|Z})}Pr_{P_{D|Z}}\left\{ \hat{\eta}_{n,0}>c_{\alpha,LF}\left(X_{n,0},\widehat{\Sigma}_{n,0}\right)\right\} \le\alpha.
\]
\end{proposition}

\begin{proposition}\label{Prop: Conditional, Hybrid Size Control}Under
Assumptions \ref{Assumption: Variance Convergence}, \ref{Assumption: Variance Estimator},
\ref{Assumption: Uniform CLT},
and \ref{Assumption: Nondegeneracy}, the conditional and hybrid tests are uniformly asymptotically valid for $\alpha<0.5$,
\[
\limsup_{n\to\infty}~\sup_{P_{D|Z}\in\mathcal{P}_{D|Z}}~\sup_{\beta_0\in B_I(P_{D|Z})}Pr_{P_{D|Z}}\left\{ \hat{\eta}_{n,0}>c_{\alpha,C}\left(Y_{n,0},X_{n,0},\widehat{\Sigma}_{n,0}\right)\right\} \le\alpha,
\]
\[
\limsup_{n\to\infty}~\sup_{P_{D|Z}\in\mathcal{P}_{D|Z}}~\sup_{\beta_0\in B_I(P_{D|Z})}Pr_{P_{D|Z}}\left\{ \hat{\eta}_{n,0}>c_{\frac{\alpha-\kappa}{1-\kappa},H}\left(Y_{n,0},X_{n,0},\widehat{\Sigma}_{n,0}\right)\right\} \le\alpha.
\]
\end{proposition}

\section{Implementation}\label{sec: Practical Implementation}

We next provide practical guidance on implementing the tests described above. We also provide Matlab code to facilitate implementation.\footnote{The code is available at \href{https://github.com/jonathandroth/LinearMomentInequalities/}{https://github.com/jonathandroth/LinearMomentInequalities/}.} 

\subsection{Choice of Moments \label{sec: choice of moments}}

Researchers can use our methods whenever their model implies conditional moment inequalities of the form (\ref{eq: Linear Conditional Moments}). As discussed in Section \ref{subsec: simplifications from conditional structure}, if the model (\ref{eq: Linear Conditional Moments}) holds for a given $(Y,X)$ pair, then it also holds if $Y$ and $X$ are interacted with any non-negative function of the instruments -- i.e., if we replace $Y$ and $X$ with $\tilde Y = Y \otimes f(Z)$ and $\tilde X = X \otimes f(Z)$. An important choice in implementing our methods is thus the choice of the $k$ moments (i.e., the choice of $Y$). A formal analysis of how to optimally choose the $k$ moments is beyond the scope of this paper, but we offer some heuristic guidance. 

Intuitively, including more informative moments can tighten the identified set based on the included moments, but including too many moments relative to the sample size can harm the quality of the normal approximation. Including uninformative moments (that are not infinitely slack) can also reduce the finite-sample power of our tests. The multivariate Berry-Esseen theorem \citep[e.g.][]{bentkus_dependence_2003} suggests that the normal approximation to the distribution of the sample average should perform well when the number of moments included is sufficiently small relative to the sample size.\footnote{Specifically, as discussed in \citet{chernozhukov_central_2017}, we need the dimension of the moments ($k$) to be smaller than $o(n^{\frac{2}{7}})$ for the approximation to hold uniformly over all convex sets. If the moments are of the form $Y=TU$, as in Assumption \ref{Assumption: Variance Convergence}, then the relevant dimension is $dim(U)$ rather than $dim(Y)$.} As a heuristic, \cite{cox_simple_2020} suggest that one should ensure there are at least 15 observations per cell in cases where the instruments $f(Z)$ are binary indicators for whether $Z$ falls in a particular cell. In our Monte Carlo simulations below, where the instrument functions are continuous, we find that our proposed tests have good size control with 500 observations and up to 110 moments, although we caution that the quality of the normal approximation may depend on the specific data-generating process.  

Regarding the choice of \textit{which} $k$ moments to use, researchers should include the moments that they think will be most informative about the parameter of interest. Note that interacting an original set of moments with an instrument function $f(Z)$ will only add identifying information to the extent that $f(Z)$ is correlated with $Y$ and $X$, since if $(Y,X)$ and $f(Z)$ are uncorrelated $E_P[f(Z) (Y- X \delta) ] = E_P[f(Z)] E_P[Y-X\delta] \propto E_P[Y-X\delta]$, so adding the interaction does not shrink the set of values where the moment inequalities are satisfied on average. Heuristically, researchers should therefore include instrument functions that are likely to be strongly related to $(Y,X)$.\footnote{
As noted in Section \ref{subsec: simplifications from conditional structure} above, our approach does not deliver consistent tests in settings with continuously distributed $Z_i$. Hence, to derive general optimality results one would have to go beyond our finite-dimensional analysis.  \citet{Armstrong2014,Armstrong2018} and  \citet{Chetverikov2013} establish convergence rates for inference on the full parameter in partially identified settings, including rate-optimality results for procedures using particular kernel-based instruments and bandwidths.  Their analysis could provide a natural starting point for the study of asymptotic optimality in our setting.  We thank Tim Armstrong for bringing these connections to our attention.
} Consistent with this intuition, \citet{HoPakes2014} use instrument functions based on the distance of an individual to a hospital, since their $Y$ and $X$ relate to individuals' choices of hospitals, and distance to the hospital is known to be an important determinant of hospital choice; see Section VI.B of \citet{HoPakes2014} for an intuitive discussion of how economic knowledge can inform the choice of moments. We also emphasize that applied researchers frequently conduct inference based on a finite set of unconditional moments implied by conditional moment inequalities, so the use of our methods does not introduce a new choice relative to this common practice in empirical work.

\subsection{Forming confidence sets \label{sec: confidence sets}}

Researchers often wish to compute confidence sets for the target parameter $\beta$. This can be achieved by discretizing the parameter space for $\beta$ as $\{\beta_{(1)}, \ldots, \beta_{(L)}\}$ and testing the null hypothesis $H_0: \beta = \beta_{(l)}$ for each $l$ using the tests described above.  A confidence set can then be formed by collecting the grid points for which the test fails to reject. If the researcher is interested in a subvector of $\beta$ -- e.g. the first component of $\beta$ is of interest, whereas the remaining components are nuisance parameters that enter the moments non-linearly -- then the researcher can first form a confidence set for the full parameter vector $\beta$, and then obtain a confidence set for the parameter of interest by projection. We emphasize that test inversion is only required for $\beta$, and not for the nuisance parameters $\delta$, which can lead to substantial computational simplifications when the dimension of $\delta$ is large. For the remainder of the section, we focus on the implementation of our tests for a particular null value $\beta_0$. 

\subsection{Estimating the conditional covariance\label{subsec: implementation - Sigma}}

Our tests require an estimate of the average conditional variance, $\Omega_0 = E_P[ Var( U_i(\beta_0) | Z_i ) ]$. We briefly describe how a matching procedure proposed by \cite{AbadieImbensZheng2014} can be used to estimate $\Omega_0$ when the data are i.i.d. across $i$; see \citet{Chetverikov2013} and \citet{horowitz_adaptive_2001} for alternative estimators. Let $\widehat\Sigma_Z$ be the sample variance of $Z_i$.\footnote{The matching procedure described below assumes that $\widehat\Sigma_Z$ is non-singular. In certain applications, such as in our Monte Carlo, elements of $Z_i$ may be linearly dependent by construction, leading $\widehat\Sigma_Z$ to be singular. In this case conditioning on a maximal linearly independent subset of $Z_i$ is equivalent to conditioning on the full vector, so one can drop dependent elements from $Z_i$ until $\widehat\Sigma_Z$ is non-singular.} For each $i$, find the nearest neighbor using the Mahalanobis distance for $Z_i$:
$$
\ell_{Z}\left(i\right)=\mbox{argmin}_{j\in\left\{ 1,...,n\right\} ,j\neq i}\left(Z_{i}-Z_{j}\right)' \widehat\Sigma_Z^{-1} \left(Z_{i}-Z_{j}\right).
$$
\noindent  For ease of exposition we assume that $Z_{i}$ has at least one continuously distributed
dimension, so that $\ell_{Z}\left(i\right)$ is unique for all $i$.\footnote{If instead $Z_{i}$ is entirely discrete, one can estimate $\widehat{\Omega}_{n,0}$ using the average of the sample conditional variances across $Z_i$ cells.} The estimate of $\Omega_0$ is then:
\begin{equation}
\widehat{\Omega}_{n,0}=\frac{1}{2n}\sum_{i=1}^{n}\left(U_{i}(\beta_0)-U_{\ell_{Z}\left(i\right)}(\beta_0)\right)\left(U_{i}(\beta_0)-U_{\ell_{Z}\left(i\right)}(\beta_0)\right)' . \label{eq: conditional variance estimator}
\end{equation}
Appendix \ref{sec: Variance Estimation} provides regularity conditions under which $\widehat{\Omega}_{n,0} $ is uniformly consistent for $\Omega_0$.

\subsection{Computation of test statistic and critical values}\label{ssec: computation}

To test the null hypothesis for a particular null value $\beta_0$, one needs to compute the test statistic $\hat\eta_{n,0}$ and the critical value for the relevant test ($c_{\alpha,LF}, c_{\alpha,C},$ or $c_{\alpha,H}$). We discuss computation of each component in turn.

\subsubsection{Computing $\hat\eta_{n,0}$}

The test statistic $\hat\eta_{n,0}$ can be computed by solving the linear program (\ref{eq: Projection Linear Program}). This can be achieved using standard software, such as Matlab's \texttt{linprog} command. We recommend using the dual-simplex method in Matlab, which conveniently returns both the optimal value $\hat\eta_{n,0}$ as well as the optimal vector of Lagrange multipliers $\hat\gamma$, which is used for computing the conditional and hybrid critical values. 

\subsubsection{Computing LF critical values}

Recall that the LF critical value $c_{\alpha,LF}$ is the $1-\alpha$ quantile of $\max_{\gamma \in V(X_{n,0},\hat{\sigma}_{n,0})} \gamma' \xi$ for $\xi \sim N(0,\widehat\Sigma_{n,0})$.  By duality results for linear programming, we have that
$$\hat\eta(\xi) = \max_{\gamma \in V(X_{n,0},\hat{\sigma}_{n,0})} \gamma' \xi =  \left(\min_{\eta,\delta}~\eta 
\mbox{ subject to }\xi-X_{n,0}\delta\le\eta\cdot \hat\sigma_{n,0} \right),$$
\noindent where $\hat\sigma_{n,0} = \sqrt{Diag(\widehat{\Sigma}_{n,0})}$. To compute $c_{\alpha,LF}$, one can simulate $\xi_{(1)},...,\xi_{(S)} \sim N(0,\widehat\Sigma_{n,0})$, compute $\hat\eta(\xi_{(s)})$ using the linear program in the previous display and then take the $1-\alpha$ quantile of $\hat\eta(\xi_{(1)}),...,\hat\eta(\xi_{(S)})$.\footnote{To increase computational speed and stability across  different values of $\beta$, one can fix $Z_1,...,Z_S \sim N(0,I)$, and then set $\xi_s = \widehat\Sigma_{n,0}^{\frac{1}{2}} Z_s$.} We use $S=1000$ in our simulations below.

\subsubsection{Computing conditional and hybrid critical values}

To compute the conditional and hybrid critical values, one needs to compute $\mathcal{V}^{lo}_{n,0}$ and $\mathcal{V}^{up}_{n,0}$. Equation (\ref{eq: V^lo, V^up}) gives an analytical formula for these quantities that involves a minimum and maximum over the set of dual vertices $V(X_{n,0}, \hat\sigma_{n,0})$. Enumerating all of the vertices is, however, computationally prohibitive when there are many moments or nuisance parameters. Fortunately, we show in Appendix \ref{sec: vlo vup computation} that there are two computational shortcuts available that allow for computation of $\mathcal{V}^{lo}_{n,0}$ and $\mathcal{V}^{up}_{n,0}$ without vertex enumeration. First, when the problem for $\hat\eta_{n,0}$ has a non-degenerate solution, $\mathcal{V}^{lo}_{n,0}$ and $\mathcal{V}^{up}_{n,0}$ can each be written as the maximum/minimum of a set of at most $k$ easy-to-compute elements.\footnote{The solution to the primal problem is said to be non-degenerate if $W_{n,0,B}$ is invertible, where $W_{n,0} = (\hat\sigma_{n,0} ,~ X_{n,0})$ and $B$ indexes the set of binding moments in the primal. To use this approach, we also require that $e_1'W_{n,0,B}\geq 0$.} Second, if the problem for $\hat\eta_{n,0}$ is degenerate, $\mathcal{V}^{lo}_{n,0}$ and $\mathcal{V}^{up}_{n,0}$ can be solved using a computationally-tractable bisection approach. We thus recommend to first check whether the solution to the primal problem (\ref{eq: Projection Linear Program}) is non-degenerate, and if so, use the formula given in Lemma \ref{lem: vlo vup primal}; if not, then use the bisection approach described in Appendix \ref{sec: vlo vup computation}. We implement this approach in our publicly-available Matlab code, and find that it yields computationally tractable tests with as many as 110 moments and 11 parameters in our simulations below.

\subsubsection{Simplifications when target parameters enter the moments linearly \label{sec: computational shortcuts linear case}}

In some settings, we may have inequalities of the form
$$ E_{P_{D|Z}}[Y_i - X_{\beta,i} \beta - X_{\delta,i} \delta | Z_i ] \leq 0,$$
\noindent where $\beta$ is the parameter of interest, $\delta$ is again a nuisance parameter, $X_{\beta,i}$ and $X_{\delta,i}$ are non-random conditional on $Z_i$, and the value of $(Y_i,X_{\beta,i},X_{\delta,i})$ does not depend on $\beta$ or $\delta$. This structure arises, for example, in interval-valued regression if we are interested in the coefficient on an exogenous variable. This structure also arises in \citet{rambachan_more_2022}, who consider bounds on treatment effects in difference-in-differences settings under linear constraints on the possible violations of parallel trends. Moment inequalities of this sort can be cast into the form (\ref{eq: Linear Conditional Moments}) by setting $Y_i(\beta) = Y_i - X_{\beta,i} \beta$ and $X_{i}(\beta) = X_{\delta,i}$. The methods described above can thus be applied directly. 

The additional linear structure allows for multiple computational shortcuts, however. First, the conditional covariance matrix $E_{P}[Var_{P_{D|Z}}(Y_i(\beta) | Z_i)]$ does not depend on $\beta$, and thus the estimated variance $\widehat\Sigma_{n}$ need only be calculated once, rather than for every candidate value of $\beta$.\footnote{We write $\widehat \Sigma_n$ instead of $\widehat \Sigma_{n,0}$, since the value does not depend on the null hypothesis. We apply an analogous convention for other variables, e.g. writing $X_n$ instead of $X_{n,0}$ and $\hat\sigma_n$ instead of $\hat\sigma_{n,0}$.} Second, the LF critical value $c_{\alpha,LF}( X_{n}, \widehat\Sigma_{n} )$ likewise does not depend on the value of $\beta$. As a result, a confidence set for the LF test can be computed by solving a linear program for each of the upper and lower bounds, without any test inversion at all. For instance, the lower bound of the confidence set for the LF test can be calculated by solving
$$ \min_{\beta,\delta} \hspace{.1cm}\beta \text{ subject to } Y_{n}- X_{n,\beta}  \beta -  X_{n,\delta} \delta \leq c_{\alpha,LF} \cdot \hat\sigma_n ,$$
\noindent where $Y_{n} = \frac{1}{\sqrt{n}} \sum_i Y_i$, and $X_{n,\beta}$ and $X_{n,\delta}$ are defined analogously. Computation of confidence sets for the conditional and hybrid tests still requires test inversion over a grid for $\beta$, but will be faster because $\widehat\Sigma_n$ and the first-stage LF critical value for the hybrid need only be computed once.

\section{Simulations}
\label{sec: Monte Carlo}

\subsection{Simulation Design}
Our simulations are calibrated to  \cite{Wollmann}'s study of the bailouts
of GM and Chryslers' truck divisions.  As discussed in Example 3 above, Wollmann obtains bounds on the fixed cost of marketing a product using moment inequalities derived from revealed preference arguments. The fixed cost to firm $f$ of marketing product $j$ at time $t$ is $\beta(\delta_{c,f}+\delta_g g_j)$ if the product was marketed at time $t-1$, and $\delta_{c,f}+\delta_g g_j$ otherwise. Consistent with (\ref{eq: Linear Conditional Moments}), the parameter $\delta=(\delta_g,\{\delta_{c,f}\})$ enters the moments linearly for a fixed value of $\beta$.  

The moments we consider take the form of the example given in equation (\ref{eq: Wollmann moment example}) for the case where a product was marketed in both periods. To illustrate how performance varies with the number of parameters, we consider specifications where the intercept $\delta_{c,f}$ is constant across firms, specifications where it is allowed to vary across three groups of firms, and specifications where each of the nine firms in the data has its own intercept.  In each case, we average the moment inequalities involving $\delta_{c,f}$ across firms assumed to have the same coefficient.  We also vary the instruments used. See Appendix \ref{sec: Monte Carlo Appendix} for details on the exact construction of the moments.  Overall, the number of moments varies between 6 and 110 across our specifications.

We consider inference on three parameters of interest: the cost of marketing the truck of mean weight when it was not marketed in the prior year;\footnote{When we assume $\delta_{c,f}$ is common across firms this is $\delta_c + \delta_g \mu_g$, where $\mu_g$ is the population average weight of trucks. When we allow the estimated $\delta_c$ parameters to vary across groups, we estimate $l'\delta$, for $l = (\frac{1}{G} , \ldots, \frac{1}{G}, \mu_g)'$, where $G$ denotes the number of groups and $\delta=(\delta_{c,1},...,\delta_{c,G},\delta_g)'$. Note that since the simulation DGP holds the true value of $\delta_c$ constant across groups, the true value of the parameter is the same in all specifications.} the incremental cost of changing the weight of a product, $\delta_g$; and the non-linear parameter $\beta$, where $1-\beta$ represents the proportional cost savings from marketing a product that was previously marketed relative to a new product. For the first two target parameters, which can be written in the form $l'\delta$, we hold $\beta$ fixed at its true value and treat the component of $\delta$ orthogonal to $l'\delta$ as the nuisance parameter. This allows us to examine performance in the linear case discussed in Section \ref{ssec: computation}. In Wollman's setting the parameter $\beta$ might be calibrated based on industry knowledge about the relative cost of marketing a new versus pre-existing product.  As discussed in Section \ref{sec: confidence sets}, if we instead treated $\beta$ as unknown we could form joint confidence sets for $\beta$ along with the linear combination of interest and obtain confidence sets for the linear parameter alone by projection. For inference on $\beta$ we treat the entire vector $\delta$ as a nuisance parameter.  Overall, the number of unknown parameters varies between 2 and 11 across our specifications.

We calibrate the data-generating process in our simulations using moments reported in Wollmann -- see Appendix \ref{sec: Monte Carlo Appendix} for details.  In each simulation draw, we generate data from a cross-section of 500 independent markets.\footnote{The data in \cite{Wollmann} are a time-series but his variance estimates assume no serial correlation, so we adopt a simulation design consistent with this.}  This is substantially larger than the 27 observations used by Wollmann, but allows us to consider specifications with a widely varying number of moments.  All results are based on 500 simulations.

We consider the performance of the LF, Conditional, and Hybrid tests and compare these to several benchmarks. First, we compare to a studentized-max-statistic-based projection test which we label the least favorable projection, or LFP, test. Second, we compute the sCC and sRCC tests proposed in \citet{cox_simple_2020}. The sRCC test, which is a refinement of the sCC test, can be computationally difficult when there are many parameters. For the specifications with 10+ parameters and 100+ moments, we therefore report an upper bound for the power of the sRCC test using the fact that the refinement to the sCC test can only matter when the test statistic falls in a certain range.\footnote{Specifically, the sRCC test always rejects when the sCC test does, and can only differ from the sCC test when one moment is active ($k=1)$ and the test statistic falls between the $1-\alpha$ and $1-\alpha/2$ quantiles of the chi-squared distribution. When there are 10 or more parameters, we thus report the power of the test that rejects whenever either the sCC test rejects or the refinement could potentially lead the sRCC test to reject.} Third, we compute the projection tests of \citetalias[][AS]{AndrewsSoares2010} and \citet[][KMS]{Kaidoetal2019} using the EAM algorithm implemented in Matlab by \citet{kaido_calibrated_2017}. The AS and KMS tests can be computationally taxing when there are many parameters, and at present, the Matlab implementation of KMS by \citet{kaido_calibrated_2017} is only written for settings where the parameters enter in an additively separable way.  We therefore compute the AS and KMS tests only for the specifications when the parameters enter linearly and there are fewer than 10 parameters. See Appendix \ref{sec: Monte Carlo Appendix} for additional details on the implementation of these comparisons.

\subsection{Results}

Table \ref{tbl: size comparisons} reports the maximum null rejection probability (size) over a conservative estimate of the identified set. Since we do not have an analytical characterization of the identified set, we approximate it by the set satisfying the sample (unconditional) moment inequalities based on a simulation run with five million observations. To ensure that our estimate of the identified set is conservative, we follow \citet{chernozhukov2007estimation} and add a correction factor to the moments of $\log(n)/\sqrt{n} \approx .003$. Our estimate of the identified set is thus conservative due to both (a) the \citet{chernozhukov2007estimation} correction factor and (b) the use of unconditional rather than conditional moment inequalities.  All of the procedures nevertheless approximately control size on this set, with rejection probabilities never exceeding 0.08 for any of the procedures.

\begin{figure}
	\centering
	
	\caption{Rejection probabilities for 5\% tests of fixed cost for truck of mean weight\label{fig:mean weight power}}
	
	\subfloat[2 Parameters, 6 Moments]{\includegraphics[width=0.48\linewidth, height = .3 \textheight ]{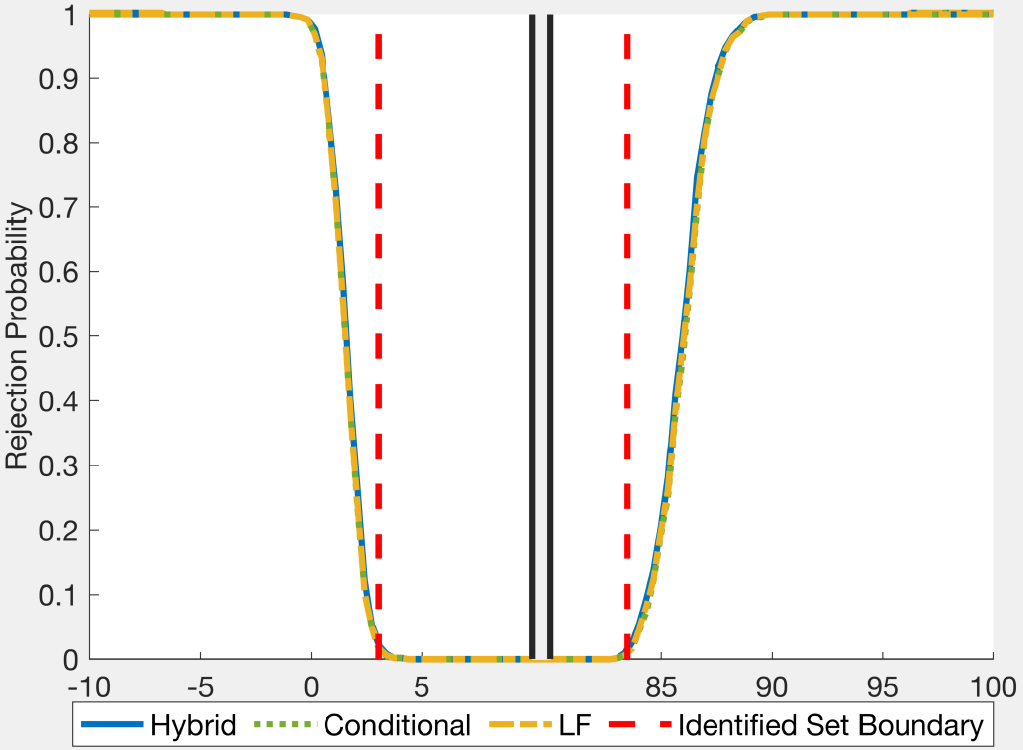} 
	}
	\hfill
	\subfloat[2 Parameters, 14 Moments]{\includegraphics[width=0.48\linewidth, height = .3 \textheight ]{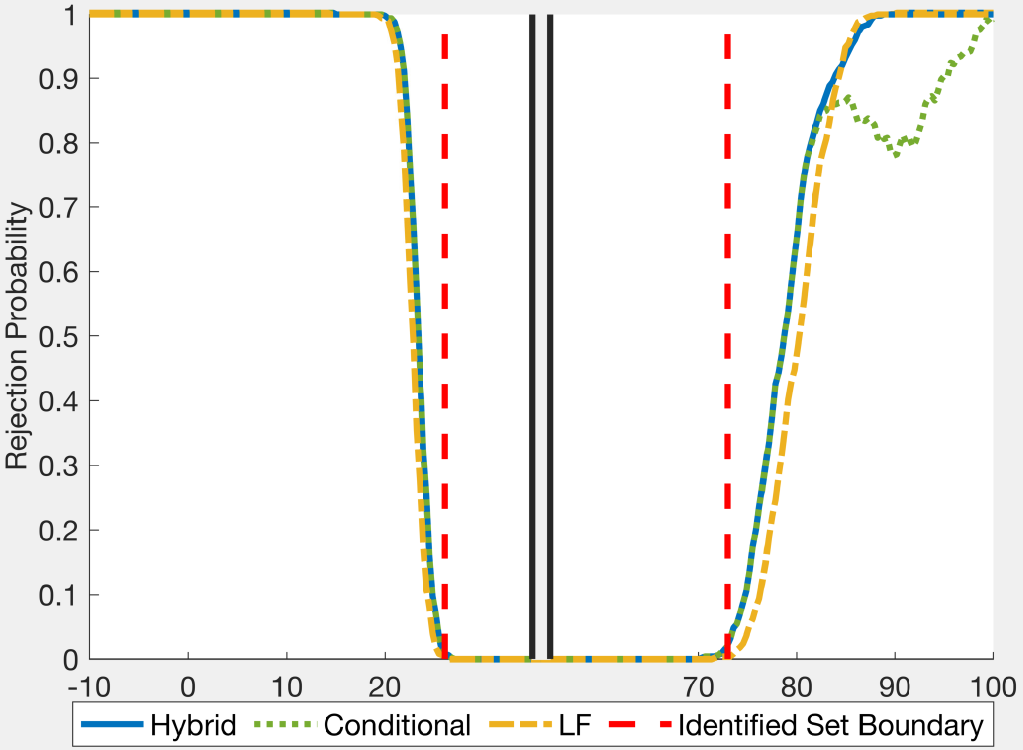} 
	}

	\subfloat[4 Parameters, 14 Moments]{\includegraphics[width=0.48\linewidth, height = .3 \textheight ]{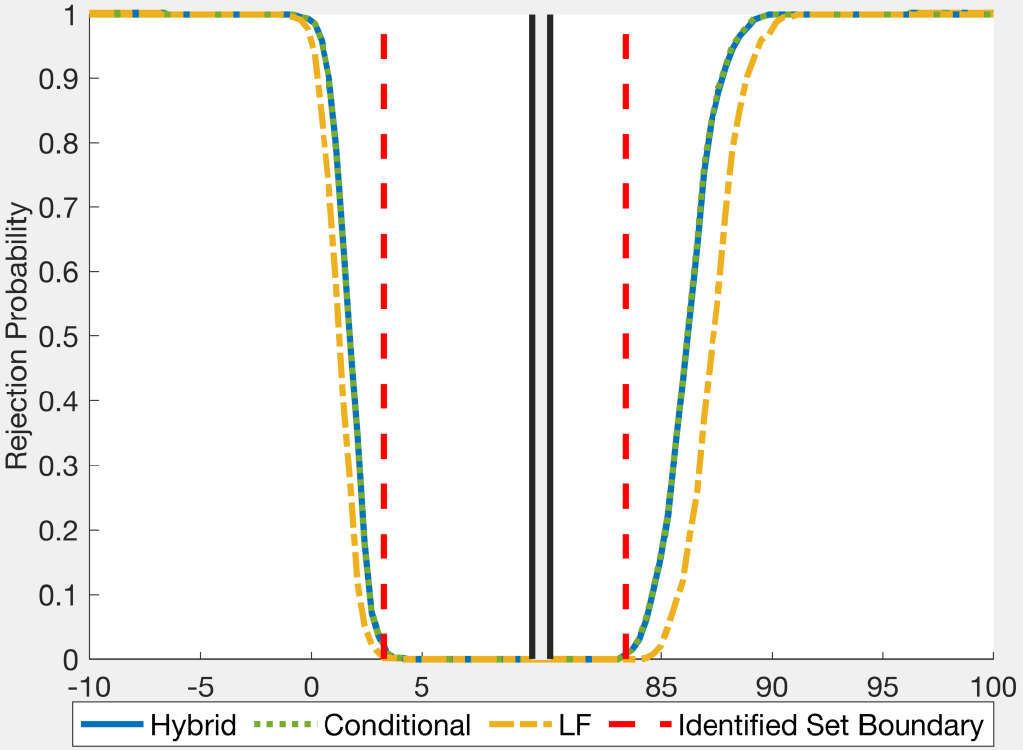} 
	}
	\hfill
	\subfloat[4 Parameters, 38 Moments]{\includegraphics[width=0.48\linewidth, height = .3 \textheight ]{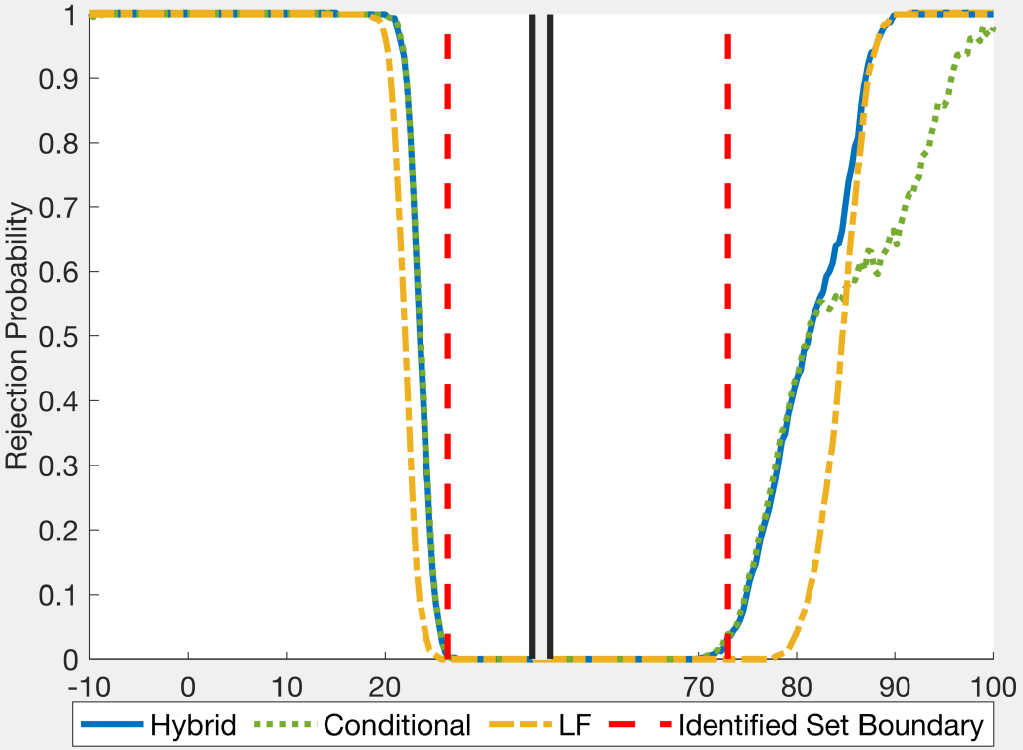} 
	}

	\subfloat[10 Parameters, 38 Moments]{\includegraphics[width=0.48\linewidth, height = .3 \textheight ]{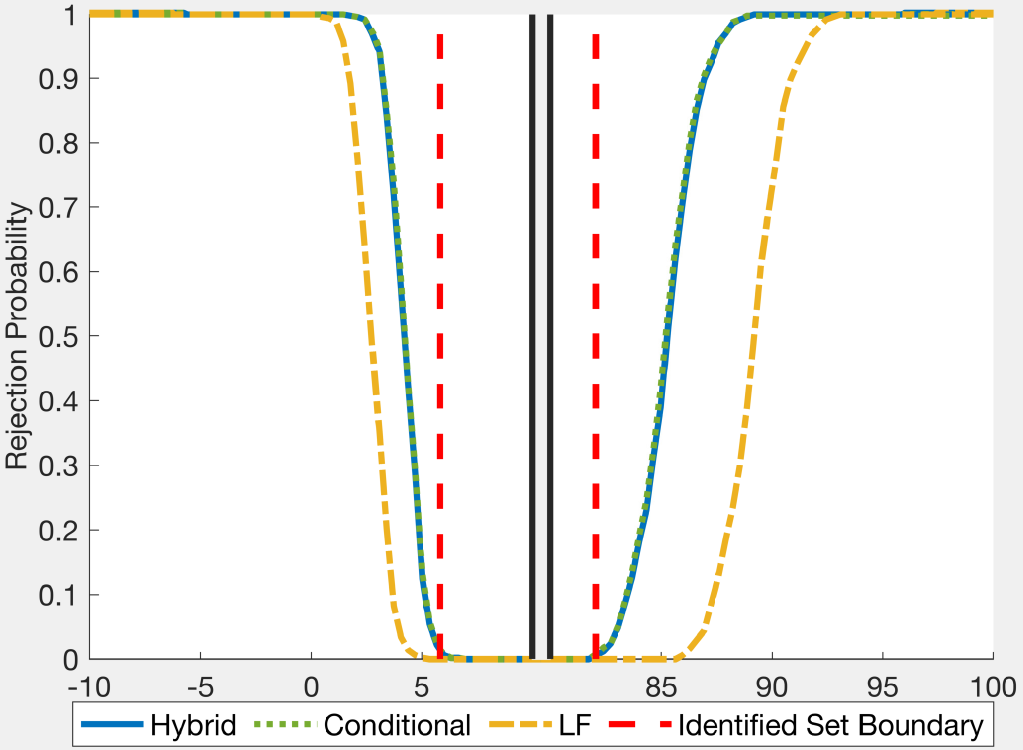} 
	}
	\hfill
	\subfloat[10 Parameters, 110 Moments]{\includegraphics[width=0.48\linewidth, height = .3 \textheight ]{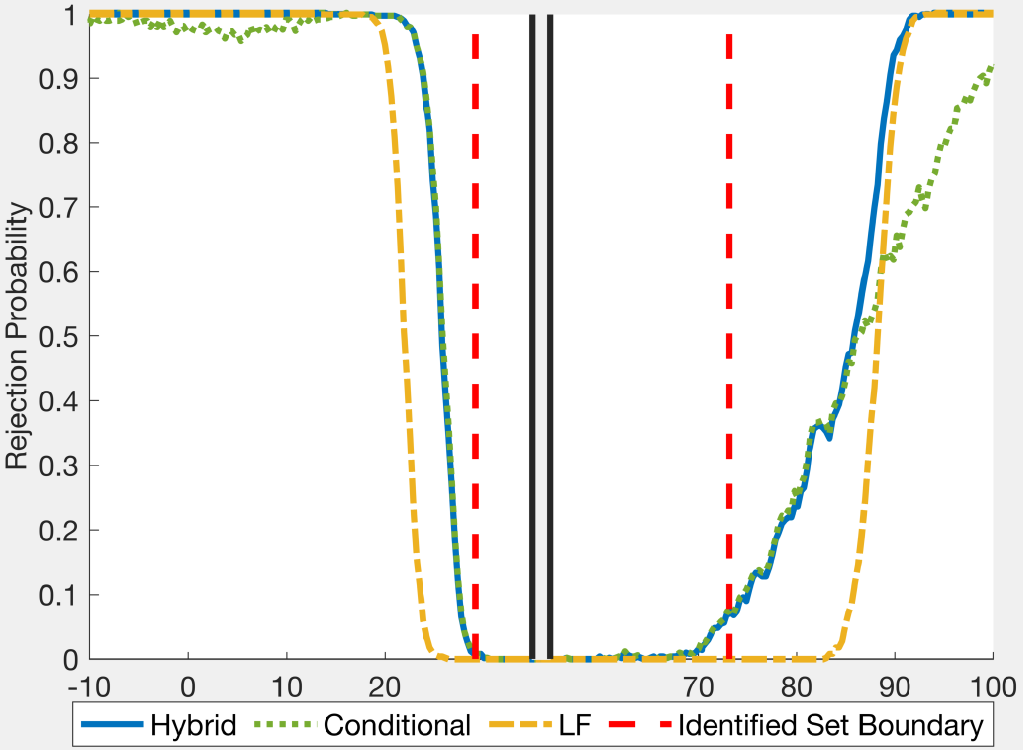} 
	}
\end{figure}

We next turn to comparisons of power. Figure \ref{fig:mean weight power} shows the rejection rates for each of our three main tests in the simulation design where the target parameter is the cost of the mean-weight truck. The vertical dashed lines denote conservative estimates of the bounds of the identified set, and the remaining curves show the probability that each of the tests rejects given a null value of the parameter of interest (holding fixed the DGP). Since the rejection probability is near-zero for all procedures in the interior of the identified set, we omit the portion of the $x$-axis well inside the identified set bounds so as to focus on the most relevant parts of the parameter space; the omitted part is grayed out in Figure \ref{fig:mean weight power} and subsequent figures. 

Overall, the figure indicates that the hybrid approach performs best among our three procedures, with rejection probabilities comparable to or above those of the LF and conditional approaches at all points in the parameter space.  To understand the superior performance of the hybrid approach, it is worth highlighting that the rejection curves for the LF and conditional approaches cross: in some specifications, the conditional approach has power substantially above that of the LF test at all parameter values (e.g. panel (e) of Figure \ref{fig:mean weight power}). In other specifications, however, the conditional approach exhibits poor power relative to the LF test in some areas of the parameter space -- e.g., in the area above the identified set in panel (d) of Figure \ref{fig:mean weight power}. We have confirmed that in this simulation design for some parameter values there are two vertices which are optimal with approximately equal probability in this part of the parameter space, which as discussed in Section \ref{sec:Inference} can lead to poor power for the conditional test. Indeed, this feature can even lead the power curves for the conditional approach to be non-monotonic, since moving farther away from the identified set can push the mean values of a pair of vertices closer together. The hybrid approach has similar power to the conditional approach in most of the parameter space, while mitigating the issues in regions of the parameter space where multiple vertices are close to binding, thus leading to better performance overall. Appendix Figures \ref{fig:theta_g power}-\ref{fig:beta power wollmann} show  results when the parameter of interest is $\delta_g$ or $\beta$: the qualitative patterns are similar, with the hybrid exhibiting power comparable to or above the other two methods throughout the parameter space. 

Table \ref{tbl: excess length comparisons} provides a comparison of our three procedures relative to the other benchmarks. We report the median excess length for confidence sets formed based on each approach, where excess length is defined as the length of the confidence set minus the length of the identified set. For reference, we also report the length of the identified set. We find that the median excess length of the hybrid confidence set is below that for the AS and KMS sets in all specifications. The median excess length for the hybrid is also better or equal to that for the sCC and sRCC sets in most specifications, although the sRCC set outperforms the hybrid for three of the specifications with  target parameter $\beta$.\footnote{Appendix Figures \ref{fig:meanweight power cox and shi}-\ref{fig:beta power cox and shi} show a comparison of the power curves of the hybrid and the sCC and sRCC tests. The figures show that for several specifications the rejection curves for the hybrid and sRCC tests cross.}  The ranking of the hybrid and sRCC approaches in these results differs from that in the simulations in \citet{cox_simple_2020}, who find better performance for sRCC. One potential factor is that the hybrid test is based on the max statistic whereas the sRCC test uses a QLR statistic, so the hybrid may be more powerful in settings where one moment is violated to a large extent, whereas the sRCC test may be more powerful when several moments are locally violated. Finally, it is worth highlighting that all of the procedures considered have better power than the LFP test in nearly all specifications. Appendix Figures \ref{fig:meanweight power cox and shi}-\ref{fig:thetag power as and kms} display comparisons of the full power curves of the hybrid relative to the LFP, sCC, sRCC, AS, and KMS tests.

In our simulations the excess length of KMS intervals sometimes exceeds that of AS intervals.  This is potentially surprising, since by construction the KMS test should reject whenever the AS test rejects, and thus should yield confidence intervals with uniformly shorter excess length.  In practice, however, the bounds of the projected confidence intervals are approximated using a finite number of objective evaluations of the Evaluation-Approximation-Maximization algorithm studied by KMS, and thus are subject to optimization error. As a consequence of these optimization errors we find the median excess length of AS to be slightly smaller than that of KMS in two of our specifications (although by less than 2\%). We have verified in an example where these issues arise that providing the EAM algorithm for AS with the optimal solution for KMS as a starting point leads to an AS interval that is a superset of the KMS interval. For simplicity, however, we report results from applying the EAM algorithm for AS directly.\footnote{We also found that reducing the objective tolerance to half the default value reduced (but did not fully eliminate) this issue, but were unable to reduce the tolerance further owing to computational constraints.}


Lastly, Table \ref{tbl:runtime} reports runtimes in minutes to calculate confidence sets for each parameter, averaging over 20 runs on a 2022 MacStudio (with M1 Ultra processor, 64GM RAM) without parallelizing the test inversion. Perhaps the most remarkable feature of the table is that our proposed tests are computationally tractable even in settings with as many as 11 parameters and 110 moments. Our preferred test, the hybrid, has runtimes under 5 minutes for all specifications in panels (a) and (b), where all of the parameters enter the moments linearly, and under 2 hours in all specifications in panel (c), where the target parameter enters the moments non-linearly. We emphasize that these runtimes could be further improved by parallelizing the test inversion. 

We highlight a few noteworthy comparisons of runtimes across both procedures and specifications. First, the runtime of the hybrid test can be either faster or slower than the runtime of the sCC and sRCC tests proposed by \citet{cox_simple_2020} depending on the specification.\footnote{The refinement for the sRCC test is needed relatively rarely, and thus the reported runtimes for the sRCC and sCC test are identical to two decimal places.} The hybrid test is faster in the majority of simulations where all parameters enter the moments linearly; this is because the LF test used in the first-stage of the hybrid is particularly fast for these specifications, as the LF confidence set can be calculated without any test inversion (see Section \ref{ssec: computation}). The \citet{cox_simple_2020} tests are faster in most of the specifications in panel (c), where the target parameter enters the moments non-linearly and thus the LF critical value must be re-calculated for each candidate value of $\beta$, with the exception of the specification with the most moments and parameters in which the hybrid is faster. Second, the runtimes for the hybrid tests are faster than for the AS and KMS projection tests in nearly all specifications, with larger differences in settings with more moments/parameters.\footnote{Runtimes between the hybrid and sCC/sRCC tests are directly comparable, since both tests use test inversion over the same grid. Comparing runtimes between the hybrid and AS/KMS projection confidence sets is somewhat more difficult, since the former depends on the grid resolution while the latter depend on the stopping criteria for the EAM algorithm. Given that the EAM algorithm relies on several stopping criteria \citep[see][p. 8]{kaido_calibrated_2017}, it is not entirely obvious how to align these parameters so that the computational accuracy of the tests is comparable. Note, however, that if the lower bound for the AS confidence set computed by the EAM algorithm is larger than that for the KMS confidence set, then the computational error in the former must be at least as large as the difference between the two computed endpoints. In the specification corresponding with the first row in Table \ref{tbl:runtime}, this difference is larger than the grid resolution used for the hybrid test in 13 percent of the cases, which provides suggestive evidence that the computational errors of the two approaches are often of a similar order of magnitude.\label{fn:runtime-comparisons}} In the specification in the fourth row of panel (b), for example, the hybrid test is over 14 times faster than both AS and KMS.\footnote{We ran a single iteration of AS for the specification with 10 parameters and 38 moments, which took 5.5 hours to complete (and the EAM algorithm for the upper bound reached the maximum of 1000 iterations without converging).} It is intuitive that the computation time is faster for the hybrid since it exploits the linear conditional structure present in our setting, whereas the EAM algorithm used to calculate the AS/KMS CIs is designed for a larger class of potentially non-linear problems and thus does not make use of this additional structure. Third, both the conditional and hybrid tests are somewhat slower when the target parameter is $\delta_g$ (panel b) relative to the cost of the mean-weight truck (panel a). The reason is that the primal solution for $\hat\eta_{n,0}$ is often degenerate, and thus we must use the slower bisection method to calculate the $\mathcal{V}^{lo}_{n,0}$ and $\mathcal{V}^{up}_{n,0}$, as described in Appendix \ref{sec: vlo vup computation}. 

\section{Conclusion}

This paper considers the problem of inference based on linear conditional moment inequalities, which arise in a wide variety of economic applications.  Using linear conditional structure, we develop inference procedures which remain both computationally tractable and powerful in the presence of nuisance parameters. We find good performance for our procedures under a variety of simulation designs based on \cite{Wollmann}, with especially good performance for our recommended hybrid procedure.

	\begin{table} 
	\caption{Size Comparisons\label{tbl: size comparisons}}
	\begin{adjustbox}{}
		\subfloat[Parameter: Cost of Mean-Weight Truck]{
			\begin{tabular}{rr} %
				\\ %
				\\%
				$\#$Params & $\#$Moments %
				\\ \hline \\[-1.8ex] %
				2&6\\ 2&14 \\ 4&14 \\ 4&38 \\ 10&38 \\ 10&110%
			\end{tabular} %
			\begin{tabular}{R{0.9cm}R{0.9cm}R{0.9cm}R{0.9cm}R{0.9cm}R{0.9cm}R{0.9cm}R{0.9cm}} %
				\multicolumn{6}{c}{Max Size}    \\ \cline{1-8}\\ %
				LF & Cond. & Hybrid & LFP & sCC & sRCC &  AS & KMS %
				\\ \hline \\[-1.8ex] %
				$0.02$ & $0.02$ & $0.02$ & $0.00$ & $0.01$ & $0.02$ & $0.02$ & $0.02$ \\ 
$0.00$ & $0.02$ & $0.02$ & $0.00$ & $0.01$ & $0.02$ & $0.02$ & $0.02$ \\ 
$0.00$ & $0.02$ & $0.02$ & $0.00$ & $0.01$ & $0.02$ & $0.03$ & $0.05$ \\ 
$0.00$ & $0.04$ & $0.04$ & $0.00$ & $0.01$ & $0.03$ & $0.00$ & $0.00$ \\ 
$0.00$ & $0.02$ & $0.01$ & $0.00$ & $0.01$ & $0.02$ & $ $ & $ $ \\ 
$0.00$ & $0.07$ & $0.07$ & $0.00$ & $0.00$ & $0.00$ & $ $ & $ $\\ %
			\end{tabular} %
	}\end{adjustbox}
	\\
	\begin{adjustbox}{}
		\subfloat[Parameter: $\delta_g$]{
			\begin{tabular}{rr} %
				\\ %
				\\%
				$\#$Params & $\#$Moments %
				\\ \hline \\[-1.8ex] %
				2&6\\ 2&14 \\ 4&14 \\ 4&38 \\ 10&38 \\ 10&110%
			\end{tabular} %
			\begin{tabular}{R{0.9cm}R{0.9cm}R{0.9cm}R{0.9cm}R{0.9cm}R{0.9cm}R{0.9cm}R{0.9cm}} %
				\multicolumn{6}{c}{Max Size}    \\ \cline{1-8}\\ %
				LF & Cond. & Hybrid & LFP & sCC & sRCC &  AS & KMS %
				\\ \hline \\[-1.8ex] %
				$0.04$ & $0.04$ & $0.06$ & $0.01$ & $0.02$ & $0.04$ & $0.03$ & $0.03$ \\ 
$0.02$ & $0.05$ & $0.05$ & $0.00$ & $0.03$ & $0.05$ & $0.02$ & $0.02$ \\ 
$0.03$ & $0.04$ & $0.05$ & $0.00$ & $0.03$ & $0.04$ & $0.04$ & $0.05$ \\ 
$0.00$ & $0.05$ & $0.05$ & $0.00$ & $0.03$ & $0.05$ & $0.07$ & $0.08$ \\ 
$0.00$ & $0.05$ & $0.05$ & $0.00$ & $0.03$ & $0.05$ & $ $ & $ $ \\ 
$0.00$ & $0.03$ & $0.03$ & $0.00$ & $0.02$ & $0.02$ & $ $ & $ $\\ %
			\end{tabular} %
		}
	\end{adjustbox}\\
	\begin{adjustbox}{}
		\subfloat[Parameter: $\beta$]{
			\begin{tabular}{rr} %
				\\ %
				\\%
				$\#$Params & $\#$Moments %
				\\ \hline \\[-1.8ex] %
				3&6\\ 3&14 \\ 5&14 \\ 5&38 \\ 11&38 \\ 11&110%
			\end{tabular} %
			\begin{tabular}{R{0.9cm}R{0.9cm}R{0.9cm}R{0.9cm}} %
				\multicolumn{4}{r}{Max Size}    \\ \cline{1-4}\\ %
				LF & Cond. & Hybrid & LFP %
				\\ \hline \\[-1.8ex] %
				$0.00$ & $0.00$ & $0.00$ & $0.00$ \\ 
$0.00$ & $0.01$ & $0.01$ & $0.00$ \\ 
$0.00$ & $0.01$ & $0.01$ & $0.00$ \\ 
$0.00$ & $0.03$ & $0.02$ & $0.00$ \\ 
$0.00$ & $0.01$ & $0.01$ & $0.00$ \\ 
$0.00$ & $0.05$ & $0.04$ & $0.00$\\ %
			\end{tabular}%
			\begin{tabular}{R{0.9cm}R{0.9cm}} %
				\\ \cline{1-2}\\ %
				sCC & sRCC %
				\\ \hline \\[-1.8ex] %
				$0.00$ & $0.00$ \\ 
$0.00$ & $0.01$ \\ 
$0.01$ & $0.01$ \\ 
$0.02$ & $0.02$ \\ 
$0.00$ & $0.01$ \\ 
$0.01$ & $0.01$\\ %
			\end{tabular}
		}
	\end{adjustbox}
\end{table}

	\begin{table} 
	\caption{Excess Length Comparisons\label{tbl: excess length comparisons}}
	\begin{adjustbox}{}
		\subfloat[Parameter: Cost of Mean-Weight Truck]{
			\begin{tabular}{rrR{1.2cm}R{0.9cm}R{0.9cm}R{0.9cm}R{0.9cm}R{0.9cm}R{0.9cm}R{0.9cm}R{0.9cm}} %
				&& & \multicolumn{8}{c}{Median Excess Length}    \\ \cline{4-11}\\ %
				$\#$Params & $\#$Moments & ID Set & %
				LF & Cond. & Hybrid & LFP & sCC & sRCC &  AS & KMS %
				\\ \hline \\[-1.8ex] %
				$2$ & $6$ & $80.42$ & $3.99$ & $4.08$ & $3.76$ & $5.33$ & $4.73$ & $4.08$ & $4.12$ & $4.14$ \\ 
$2$ & $14$ & $46.89$ & $10.30$ & $10.31$ & $8.36$ & $12.57$ & $9.66$ & $8.36$ & $9.67$ & $9.80$ \\ 
$4$ & $14$ & $80.13$ & $5.92$ & $4.37$ & $4.37$ & $7.57$ & $5.02$ & $4.37$ & $5.82$ & $5.38$ \\ 
$4$ & $38$ & $46.61$ & $16.14$ & $14.49$ & $11.56$ & $18.88$ & $12.86$ & $12.54$ & $15.90$ & $15.41$ \\ 
$10$ & $38$ & $76.21$ & $10.21$ & $4.72$ & $4.72$ & $12.71$ & $5.37$ & $4.72$ & $ $ & $ $ \\ 
$10$ & $110$ & $43.95$ & $22.24$ & $17.80$ & $14.25$ & $25.50$ & $18.45$ & $18.45$ & $ $ & $ $\\ %
			\end{tabular} %
	}\end{adjustbox}
	\\
	\begin{adjustbox}{}
		\subfloat[Parameter: $\delta_g$]{
			\begin{tabular}{rrR{1.2cm}R{0.9cm}R{0.9cm}R{0.9cm}R{0.9cm}R{0.9cm}R{0.9cm}R{0.9cm}R{0.9cm}} %
				&& \multicolumn{6}{c}{Median Excess Length}    \\ \cline{4-11}\\ %
				$\#$Params & $\#$Moments & ID Set &%
				LF & Cond. & Hybrid & LFP & sCC & sRCC & AS & KMS %
				\\ \hline \\[-1.8ex] %
				$2$ & $6$ & $120.05$ & $4.29$ & $4.20$ & $3.95$ & $6.04$ & $4.95$ & $4.20$ & $4.94$ & $4.74$ \\ 
$2$ & $14$ & $120.05$ & $5.41$ & $4.45$ & $4.20$ & $6.93$ & $5.20$ & $4.45$ & $5.31$ & $5.26$ \\ 
$4$ & $14$ & $120.07$ & $5.19$ & $4.43$ & $4.18$ & $6.99$ & $5.18$ & $4.43$ & $5.48$ & $5.13$ \\ 
$4$ & $38$ & $120.07$ & $6.68$ & $4.43$ & $4.43$ & $7.97$ & $5.43$ & $4.43$ & $6.23$ & $6.08$ \\ 
$10$ & $38$ & $120.07$ & $6.58$ & $4.43$ & $4.43$ & $8.09$ & $5.43$ & $4.43$ & $ $ & $ $ \\ 
$10$ & $110$ & $120.07$ & $7.69$ & $5.18$ & $5.18$ & $9.11$ & $7.43$ & $7.18$ & $ $ & $ $\\ %
			\end{tabular} %
		}
	\end{adjustbox}\\
	\begin{adjustbox}{}
		\subfloat[Parameter: $\beta$]{
			\begin{tabular}{rrR{1.2cm}R{0.9cm}R{0.9cm}R{0.9cm}R{0.9cm}} %
				&& \multicolumn{4}{r}{Median Excess Length}    \\ \cline{4-7}\\ %
				$\#$Params & $\#$Moments & ID Set &%
				LF & Cond. & Hybrid & LFP %
				\\ \hline \\[-1.8ex] %
				3 & 6 & 16.89 & 61.87 & 42.93 & 36.62 & 118.69 \\ 
3 & 14 & 1.41 & 0.55 & 0.45 & 0.35 & 0.76 \\ 
5 & 14 & 8.71 & 7.78 & 6.01 & 5.30 & 10.25 \\ 
5 & 38 & 1.31 & 0.66 & 0.96 & 0.45 & 0.86 \\ 
11 & 38 & 2.99 & 1.01 & 1.01 & 0.81 & 1.41 \\ 
11 & 110 & 1.01 & 0.66 & 2.57 & 0.55 & 0.86\\ %
			\end{tabular} %
			\begin{tabular}{R{0.9cm}R{0.9cm}} %
				\\ \cline{1-2}\\ %
				sCC & sRCC %
				\\ \hline \\[-1.8ex] %
				60.61 & 42.93 \\ 
0.45 & 0.35 \\ 
6.36 & 5.66 \\ 
0.40 & 0.35 \\ 
0.71 & 0.71 \\ 
0.45 & 0.45\\ %
			\end{tabular} %
		}
	\end{adjustbox}
\end{table}

	\begin{table}
	\caption{Computational Time Comparison\label{tbl:runtime}}
	\begin{adjustbox}{}
		\subfloat[Parameter: Cost of Mean-Weight Truck]{
			\begin{tabular}{rrR{0.9cm}R{0.9cm}R{0.9cm}R{0.9cm}R{0.9cm}R{0.9cm}R{0.9cm}R{0.9cm}} %
				&& \multicolumn{8}{c}{Average Runtime in Minutes}    \\ \cline{3-10}\\ %
				$\#$Params & $\#$Moments & %
				LF & Cond. & Hybrid & LFP & sCC & sRCC &  AS & KMS %
				\\ \hline \\[-1.8ex] %
				$2$ & $6$ & $0.12$ & $0.24$ & $0.23$ & $0.03$ & $1.06$ & $1.06$ & $3.23$ & $3.63$ \\ 
$2$ & $14$ & $0.05$ & $0.22$ & $0.22$ & $0.00$ & $3.05$ & $3.05$ & $0.52$ & $0.95$ \\ 
$4$ & $14$ & $0.12$ & $0.42$ & $0.42$ & $0.03$ & $2.48$ & $2.48$ & $22.46$ & $27.95$ \\ 
$4$ & $38$ & $0.06$ & $0.38$ & $0.38$ & $0.00$ & $19.39$ & $19.39$ & $18.59$ & $22.44$ \\ 
$10$ & $38$ & $0.05$ & $1.40$ & $1.39$ & $0.00$ & $19.16$ & $19.16$ & $ $ & $ $ \\ 
$10$ & $110$ & $0.10$ & $0.75$ & $0.79$ & $0.01$ & $208.65$ & $208.65$ & $ $ & $ $\\ %
			\end{tabular} %
	}\end{adjustbox}
	\\
	\begin{adjustbox}{}
		\subfloat[Parameter: $\delta_g$]{
			\begin{tabular}{rrR{0.9cm}R{0.9cm}R{0.9cm}R{0.9cm}R{0.9cm}R{0.9cm}R{0.9cm}R{0.9cm}} %
				&& \multicolumn{6}{c}{Average Runtime in Minutes}    \\ \cline{3-10}\\ %
				$\#$Params & $\#$Moments & %
				LF & Cond. & Hybrid & LFP & sCC & sRCC & AS & KMS %
				\\ \hline \\[-1.8ex] %
				$2$ & $6$ & $0.05$ & $5.11$ & $2.14$ & $0.00$ & $0.74$ & $0.74$ & $0.14$ & $0.36$ \\ 
$2$ & $14$ & $0.05$ & $2.58$ & $0.67$ & $0.00$ & $2.49$ & $2.49$ & $3.72$ & $2.78$ \\ 
$4$ & $14$ & $0.05$ & $4.23$ & $2.35$ & $0.01$ & $2.22$ & $2.22$ & $11.97$ & $19.38$ \\ 
$4$ & $38$ & $0.05$ & $6.04$ & $3.83$ & $0.00$ & $13.77$ & $13.77$ & $59.27$ & $55.17$ \\ 
$10$ & $38$ & $0.06$ & $6.03$ & $4.04$ & $0.00$ & $13.10$ & $13.10$ & $ $ & $ $ \\ 
$10$ & $110$ & $0.10$ & $5.80$ & $3.93$ & $0.01$ & $127.46$ & $127.46$ & $ $ & $ $\\ %
			\end{tabular} %
		}
	\end{adjustbox}\\
	\begin{adjustbox}{}
		\subfloat[Parameter: $\beta$]{
			\begin{tabular}{rrR{0.9cm}R{0.9cm}R{0.9cm}R{0.9cm}R{0.9cm}R{0.9cm}} %
				&& \multicolumn{6}{c}{Average Runtime in Minutes}    \\ \cline{3-8}\\ %
				$\#$Params & $\#$Moments & %
				LF & Cond. & Hybrid & LFP & sCC & sRCC %
				\\ \hline \\[-1.8ex] %
				$3$ & $6$ & $47.66$ & $0.47$ & $47.86$ & $0.24$ & $1.32$ & $1.32$ \\ 
$3$ & $14$ & $48.23$ & $0.61$ & $48.39$ & $0.41$ & $3.31$ & $3.31$ \\ 
$5$ & $14$ & $47.71$ & $6.45$ & $49.59$ & $0.36$ & $2.84$ & $2.84$ \\ 
$5$ & $38$ & $52.23$ & $7.80$ & $53.57$ & $1.24$ & $22.47$ & $22.47$ \\ 
$11$ & $38$ & $52.52$ & $7.24$ & $55.01$ & $1.13$ & $18.24$ & $18.24$ \\ 
$11$ & $110$ & $98.13$ & $14.69$ & $99.59$ & $7.75$ & $251.21$ & $251.21$\\ %
			\end{tabular} %
		}
	\end{adjustbox}
\end{table}	

\clearpage

\bibliographystyle{agsm}
\bibliography{Conditional_MI_Tests}

\clearpage
\appendix

\counterwithin{assumption}{section}
\counterwithin{lemma}{section}
\counterwithin{proposition}{section}
\counterwithin{corollary}{section}
\counterwithin{definition}{section}
\counterwithin{figure}{section}
\counterwithin{table}{section}
\counterwithin{algorithm}{section}

\noindent \begin{center}
{\large{}Supplement to the paper}
\par\end{center}{\large \par}

\noindent \begin{center}
{\LARGE{}Inference for Linear Conditional\\ Moment Inequalities}
\par\end{center}{\LARGE \par}

\medskip{}

\begin{center}
{\large{} Isaiah Andrews ~~~ Jonathan Roth  ~~~ Ariel Pakes}
\par\end{center}{\large \par}

\noindent \begin{center}
\today
\par\end{center}

\vspace*{10pt}

This supplement provides proofs and additional results for the paper ``Inference for Linear Conditional Moment Inequalities.''  Appendix \ref{sec: proofs} proves the results stated in the main text.  
Appendix \ref{appendix: nonunique dual} proves validity of our tests in the finite-sample normal model when the dual problem has a non-unique solution.
Appendix \ref{sec: Variance Estimation} discusses an estimator for the variance $\Omega(P_{D|Z},\beta_0),$ and provides sufficient conditions for it to be uniformly consistent. Appendix \ref{sec: sufficient conditions for nondegeneracy} provides sufficient conditions for Assumption \ref{Assumption: Nondegeneracy} in the main text.  Appendix \ref{sec: vlo vup computation} discusses how to quickly compute the bounds $\mathcal{V}^{lo}_{n,0}$ and $\mathcal{V}^{up}_{n,0}$ used by the conditional and hybrid tests. Finally, Appendix \ref{sec: LICQ} discusses connections to LICQ conditions considered in the previous literature, while Appendix \ref{sec: Monte Carlo Appendix} provides further details on our simulations.

\section{Proofs for Results in Main Text}\label{sec: proofs}

\paragraph{Proof of Lemma \ref{lem:conditional distribution of etahat}}

Observe that $\hat\gamma=\gamma$ only if $Y_{n,0}$ lies in the polyhedron $\{y : (\gamma - \tilde\gamma)' y \geq 0, \, \forall \tilde\gamma \in V(X_{n,0}, \sigma_0)\}$. The result is then immediate from Lemma 5.1 in \citet{Leeetal2016}.

\paragraph{Proof of Lemma \ref{lem: slack moments}}

Let 
\[
V^{*}\left(X_{n,0}^{-j},\sigma_0^{-j}\right)=\left\{ \gamma\in\mathbb{R}^{k}:e_{j}'\gamma=0,\gamma^{-j}\in V\left(X_{n,0}^{-j},\sigma_0^{-j}\right)\right\} 
\]
be the $k$-dimensional version of $V\left(X_{n,0}^{-j},\sigma_0^{-j}\right)$,
and note that $V^{*}\left(X_{n,0}^{-j},\sigma_0^{-j}\right)\subseteq V\left(X_{n,0},\sigma_0\right)$
by construction. Let $F(X_{n,0}, \sigma_0) =  \{\gamma \,|\, \gamma \geq 0, \gamma' X_{n,0} = 0, \gamma'\sigma_0 = 1 \} $ denote the dual feasible set using $(X_{n,0}, \sigma_0)$, and define $F(X_{n,0}^{-j}, \sigma_0^{-j})$ analogously.  Observe that for any $\gamma\in V\left(X_{n,0},\sigma_0\right)\setminus V^{*}\left(X_{n,0}^{-j},\sigma_0^{-j}\right),$
either $e_{j}'\gamma>0$ or $\gamma^{-j}\in F\left(X_{n,0}^{-j},\sigma_0^{-j}\right)$. 

We first show that $\hat{\eta}_{n,0}^{j,d}\to\hat{\eta}_{n,0}^{-j}.$ To this
end, consider $\gamma\in V\left(X_{n,0},\sigma_0\right)\setminus V^{*}\left(X_{n,0}^{-j},\sigma_0^{-j}\right)$.
If $e_{j}'\gamma>0,$ then $\gamma'Y_{n,0}^{j,d}\to-\infty$ as $d\to\infty$.
Hence, if $V\left(X_{n,0}^{-j},\sigma_0^{-j}\right)\neq\emptyset$
(i.e. if the dual problem for $(X_{n,0}^{-j},\sigma_0^{-j})$
is feasible) then for $d$ sufficiently large we must have $\gamma\not\in\arg\max_{\gamma\in V\left(X_{n,0},\sigma_0\right)}\gamma'Y_{n,0}^{j,d}.$
If instead $e_{j}'\gamma=0$ then $\gamma^{-j}\in F\left(X_{n,0}^{-j},\sigma_0^{-j}\right)$,
so $\gamma'Y_{n,0}^{j,d}\le\max_{\gamma\in V^{*}\left(X_{n,0}^{-j},\sigma_0^{-j}\right)}\gamma'Y_{n,0}^{j,d}=\hat{\eta}_{n,0}^{-j}$
for all $d$, and either $\hat{\gamma}^{j,d}\in V^{*}\left(X_{n,0}^{-j},\sigma_0^{-j}\right)$
for $d$ sufficiently large or there exists $\tilde{\gamma}\in V^{*}\left(X_{n,0}^{-j},\sigma_0^{-j}\right)$
such that $\gamma'Y_{n,0}=\tilde{\gamma}'Y_{n,0}$, which we rule out by assumption. Hence, either
$\hat{\eta}_{n,0}^{j,d}=\hat{\eta}_{n,0}^{-j}$ and $\hat\gamma^{j,d}\in V^{*}\left(X_{n,0}^{-j},\sigma_0^{-j}\right)$ for $d$ sufficiently large or
the dual is infeasible and $\hat{\eta}_{n,0}^{j,d}\to-\infty$. Infeasibility
of the dual corresponds to unboundedness of the primal, so in this
case $\hat{\eta}_{n,0}^{-j}=-\infty$ and we again have $\hat{\eta}_{n,0}^{j,d}\to\hat{\eta}_{n,0}^{-j}.$

By the definition of the conditional test, if $\hat{\eta}_{n,0}^{j,d}\to\hat{\eta}_{n,0}^{-j}=-\infty$
then $\phi_{C}^{j,d}\to\phi_{C}^{-j}=0$. Hence, for the remainder
of the proof we consider the case with $\hat{\eta}_{n,0}^{-j}>-\infty$.
In this case, the argument above implies that $e_{j}'\hat{\gamma}^{j,d}=0$
for $d$ sufficiently large. It is straightforward to verify that if $\hat{\gamma}^{j,d} \in V^*(X_{n,0}^{-j}, \sigma_0^{-j})$, then $S_{n,0,\hat\gamma^{-j}}^{-j} = M_{-j} S_{n,0,\hat\gamma^{j,d}}^{j,d}$, where $M_{-j}$ is the matrix that selects all of the rows except row $j$. It follows that
\begin{align*}
\mathcal{V}^{lo,-j}_{n,0}  &= \text{max}_{\gamma^{-j}\in V(X_{n,0}^{-j},\sigma_0^{-j}):(\hat{\gamma}^{-j})'\Sigma_{0}^{-j}(\hat{\gamma}^{-j})>(\hat{\gamma}^{-j})'\Sigma_{0}^{-j} (\gamma^{-j})}\frac{(\hat{\gamma}^{-j})'\Sigma_{0}(\hat{\gamma}^{-j})\cdot (\gamma^{-j})'S_{n,0,\hat{\gamma}^{-j}}^{-j}}{(\hat{\gamma}^{-j})'\Sigma_{0}(\hat{\gamma}^{-j})-(\hat{\gamma}^{-j})'\Sigma_{0}(\gamma^{-j})} \\
&= 	\text{max}_{\gamma\in V^*(X_{n,0},\sigma_0):\hat{\gamma}_{jd}'\Sigma_{0}\hat{\gamma}_{jd}>\hat{\gamma}_{jd}'\Sigma_{0}\gamma}\frac{\hat{\gamma}_{jd}'\Sigma_{0}\hat{\gamma}_{jd}\cdot\gamma'S_{n,0,\hat{\gamma}_{jd}}^{j,d}}{\hat{\gamma}_{jd}'\Sigma_{0}\hat{\gamma}_{jd}-\hat{\gamma}_{jd}'\Sigma_{0}\gamma}
\end{align*}
\noindent for $d$ sufficiently large, where for brevity of notation we write $\hat\gamma_{jd}$ instead of $\hat\gamma^{j,d}$. Considering $\gamma\in V\left(X_{n,0},\sigma_0\right)\setminus V^{*}\left(X_{n,0}^{-j},\sigma_0^{-j}\right)$,
note that if $e_{j}'\gamma>0$ then $\gamma'S_{n,0,\hat{\gamma}_{jd}}^{j,d}\to-\infty$
as $d\to\infty,$ which implies that either 
\[
\gamma\not\in\text{argmax}_{\tilde\gamma\in V(X_{n,0},\sigma_0):\hat{\gamma}_{jd}'\Sigma_{0}\hat{\gamma}_{jd}>\hat{\gamma}_{jd}'\Sigma_{0}\tilde\gamma}\frac{\hat{\gamma}_{jd}'\Sigma_{0}\hat{\gamma}_{jd}\cdot\tilde\gamma'S_{n,0,\hat{\gamma}_{jd}}^{j,d}}{\hat{\gamma}_{jd}'\Sigma_{0}\hat{\gamma}_{jd}-\hat{\gamma}_{jd}'\Sigma_{0}\tilde\gamma}
\]
for $d$ sufficiently large or $\mathcal{V}^{lo,j,d}_{n,0}\to\mathcal{V}^{lo,-j}_{n,0}=-\infty$,
and similarly for $\mathcal{V}^{up,j,d}_{n,0}$. 

If instead $e_{j}'\gamma=0$, then as noted above $\gamma^{-j}\in F\left(X_{n,0}^{-j},\sigma_0^{-j}\right)$,
so for any $y\in\mathbb{R}^{k}$
\[
\gamma'y\le\max_{\tilde{\gamma}\in V^{*}\left(X_{n,0}^{-j},\sigma_0^{-j}\right)}\tilde{\gamma}'y=\max_{\tilde\gamma\in V\left(X_{n,0}^{-j},\sigma_0^{-j}\right)}\tilde{\gamma}'y^{-j}.
\]
Lemma 5.1 of \cite{Leeetal2016} implies, however, that

$$\mathcal{V}^{lo,j,d}_{n,0}= \min_{y}	 (\hat\gamma^{j,d})' y, \text { s.t. } (\hat\gamma^{j,d})'y \geq \max_{\tilde\gamma \in V(X_{n,0}, \sigma_0 )} \tilde\gamma'y \text{ and } S(y,\hat\gamma^{j,d}) = S_{n,0,\hat\gamma^{j,d}}^{j,d} ,$$

\noindent where $S(y,\hat\gamma) =  \left(I-\frac{\Sigma_{0}\hat{\gamma}\hat{\gamma}'}{\hat{\gamma}'\Sigma_{0}\hat{\gamma}}\right)y$. The previous two displays together imply that
$$\mathcal{V}^{lo,j,d}_{n,0}= \min_{y}	(\hat\gamma^{j,d})' y, \text { s.t. } (\hat\gamma^{j,d})'y \geq \max_{\tilde\gamma \in V(X_{n,0}, \sigma_0 )\setminus \{ \gamma\}} \tilde\gamma'y \text{ and } S(y,\hat\gamma^{j,d}) = S_{n,0,\hat\gamma^{j,d}}^{j,d} .$$
\noindent Applying Lemma 5.1 of \cite{Leeetal2016} in the opposite direction, 
\[
\max_{\tilde\gamma\in V(X_{n,0},\sigma_0):\hat{\gamma}_{jd}'\Sigma_{0}\hat{\gamma}_{jd}>\hat{\gamma}_{jd}'\Sigma_{0}\tilde\gamma}\frac{\hat{\gamma}_{jd}'\Sigma_{0}\hat{\gamma}_{jd}\cdot\tilde\gamma'S_{n,0,\hat{\gamma}_{jd}}^{j,d}}{\hat{\gamma}_{jd}'\Sigma_{0}\hat{\gamma}_{jd}-\hat{\gamma}_{jd}'\Sigma_{0}\tilde\gamma} = \max_{\tilde\gamma\in V(X_{n,0},\sigma_0 )\setminus \{\gamma\}:\hat{\gamma}_{jd}'\Sigma_{0}\hat{\gamma}_{jd}>\hat{\gamma}_{jd}'\Sigma_{0}\tilde\gamma}\frac{\hat{\gamma}_{jd}'\Sigma_{0}\hat{\gamma}_{jd}\cdot\tilde\gamma'S_{n,0,\hat{\gamma}_{jd}}^{j,d}}{\hat{\gamma}_{jd}'\Sigma_{0}\hat{\gamma}_{jd}-\hat{\gamma}_{jd}'\Sigma_{0}\tilde\gamma}.
\]
\noindent Iterating this argument, we obtain that
\[
\max_{\tilde\gamma\in V(X_{n,0},\sigma_0):\hat{\gamma}_{jd}'\Sigma_{0}\hat{\gamma}_{jd}>\hat{\gamma}_{jd}'\Sigma_{0}\tilde\gamma}\frac{\hat{\gamma}_{jd}'\Sigma_{0}\hat{\gamma}_{jd}\cdot\tilde\gamma'S_{n,0,\hat{\gamma}_{jd}}^{j,d}}{\hat{\gamma}_{jd}'\Sigma_{0}\hat{\gamma}_{jd}-\hat{\gamma}_{jd}'\Sigma_{0}\tilde\gamma} = \max_{\tilde\gamma\in V^*(X_{n,0},\sigma_0 ):\hat{\gamma}_{jd}'\Sigma_{0}\hat{\gamma}_{jd}>\hat{\gamma}_{jd}'\Sigma_{0}\tilde\gamma}\frac{\hat{\gamma}_{jd}'\Sigma_{0}\hat{\gamma}_{jd}\cdot\tilde\gamma'S_{n,0,\hat{\gamma}_{jd}}^{j,d}}{\hat{\gamma}_{jd}'\Sigma_{0}\hat{\gamma}_{jd}-\hat{\gamma}_{jd}'\Sigma_{0}\tilde\gamma} ,
\]
\noindent where we showed above that the expression on the right-hand side is equal to $\mathcal{V}^{lo,-j}_{n,0} $ for $d$ sufficiently large. A similar argument applies for $\mathcal{V}^{up,j,d}_{n,0}$. We have thus shown that $\left(\mathcal{V}^{lo,j,d}_{n,0},\mathcal{V}^{up,j,d}_{n,0}\right)\to\left(\mathcal{V}^{lo,-j}_{n,0},\mathcal{V}^{up,-j}_{n,0}\right)$
as $d\to\infty$.

This convergence, combined with the fact that $\hat{\gamma}^{j,d}\in V^{*}\left(X_{n,0}^{-j},\sigma_0^{-j}\right)$
for $d$ sufficiently large and the fact that for $\gamma\in V^{*}\left(X_{n,0}^{-j},\sigma_0^{-j}\right),$
$\gamma'\Sigma_{0}\gamma=\gamma^{-j}\Sigma_0^{-j}\gamma^{-j}$,
implies that $c_{\alpha,C}\left(Y_{n,0}^{j,d},X_{n,0},\Sigma_{0}\right)\to c_{\alpha,C}\left(Y_{n,0}^{-j},X_{n,0}^{-j},\Sigma_{0}^{-j}\right).$
Hence, so long as $\hat{\eta}_{n,0}^{-j}\neq c_{\alpha,C}\left(Y_{n,0}^{-j},X_{n,0}^{-j},\Sigma_{0}^{-j}\right),$
$\phi_{C}^{j,d}\to\phi_{C}^{-j}$, as desired. $\Box$

\paragraph{Proof of Lemma \ref{lem: uniform CLT}}

Towards contradiction, suppose the conclusion of the lemma fails.
Then there exists a sequence of distributions, null parameter values, and sample sizes $\left\{ P_{D|Z,n_m},\beta_{0,n_m},n_{m}\right\} $ with $\beta_{0,n_m}\in B_I(P_{D|Z,n_m})$ for all $m,$ and a constant $\varepsilon>0$ such that 
\begin{equation}
\liminf_{m\to\infty}\sup_{f\in BL_{1}}\left|E_{P_{D|Z,n_m}}\left[f\left(U_{n_{m},0}-\pi_{n_{m},0}\right)\right]-E\left[f\left(\xi_{P_{D|Z,n_m}}\right)\right]\right|>\varepsilon.\label{eq: non-vanishing BL}
\end{equation}
Since the set of possible variances $\Omega$ consistent with Assumption
\ref{Assumption: Variance Convergence} is compact, there exists a
subsequence $\left\{ P_{D|Z,n_l},\beta_{0,n_l},n_{l}\right\} \subseteq\left\{ P_{D|Z,n_m},\beta_{0,n_m},n_{m}\right\} $
along which $\Omega\left(P_{D|Z,n_l},\beta_{0,n_l}\right)\to\Omega^*$
for some $\Omega^*.$ Under this subsequence, however, the Lindeberg-Feller Central Limit Theorem (see e.g. Proposition 2.27 in \citet{VanderVaart2000}), along with the assumptions of the lemma, implies that
\[
U_{n_{l},0}-\pi_{n_{l},0}\to_{d}N\left(0,\Omega^*\right),
\]
and thus that 
\[
\lim_{l\to\infty}\sup_{f\in BL_{1}}\left|E_{P_{D|Z,n_l}}\left[f\left(U_{n_{l},0}-\pi_{n_{l},0}\right)\right]-E\left[f\left(\xi_{P_{D|Z,n_l}}\right)\right]\right|=0.
\]
This contradicts (\ref{eq: non-vanishing BL}), completing the proof.
$\Box$

The following result characterizes the vertices of the dual vertex set.

\begin{lemma}\label{lem: vertex characterization}
Suppose $\gamma \in F(X,\sigma)$. Then $\gamma \in V(X,\sigma)$ if and only if $\gamma = A_B(X,\sigma)^{-1} e_1$, for $e_1$ the first standard basis vector in $\mathbb{R}^k$,
	$$A(X,\sigma) =  \left( \begin{array}{l}
		\sigma' \\
		X'\\
		-I
	\end{array} \right)  ,$$
	and $B \subset \{1,...,p+k+1\}$ with $|B| = k$ and $1 \in B$, where $M_B$ denotes the rows of the matrix $M$ contained in $B$. 
\end{lemma}

\paragraph{Proof of Lemma \ref{lem: vertex characterization}}
From Theorem 8.4 and statement (23) in Section 8.5 in \citet{Schrijver1986},  $v \in \{x \in \mathbb{R}^k : Wx \leq b\}$ is a vertex of $\{x \in \mathbb{R}^k : Wx \leq b\}$ if and only if there exists $B \subset \{1,...k\}$ such that $W_B$ is invertible and $W_B x = b_B$, where $W_B$ denotes the rows of $W$ corresponding with the indices in $B$, and $b_B$ is defined analogously. Observe that $F(X,\sigma)$ takes the form $\{ \gamma \in \mathbb{R}^k : W \gamma \leq b \}$, where 
$$W = \left( \begin{array}{r}
	\sigma' \\
	-\sigma'\\
	X'\\
	-X'\\
	-I
\end{array} \right) \text{ and } b = \left( \begin{array}{r}
	1 \\
	-1\\
	0\\
	0\\
	0
\end{array} \right),$$

\noindent where $W$ is $(2(p+1)+k)\times k$ and $b$ is $(2(p+1)+k)\times 1$. Thus, $\gamma \in F(X,\sigma)$ is a vertex if and only if $\gamma = W_B^{-1} b_B$ for some index set $B \subset \{1,...,2(p+1) +k \}$ with $|B| =k$ such that $W_B$ is invertible.

Next, observe that $\gamma \in F(X,\sigma)$ satisfies $\gamma'\sigma = 1$ and thus must be non-zero. Since $b_B = 0$ unless $B$ contains an index corresponding with a row of $W$ containing either $\sigma'$ or $-\sigma'$, it follows that if there is a vertex corresponding with $B$ then $B$ must always contain one such index. Moreover, it's clear that $B$ can select at most one of each pair of inequalities of the opposite sign, since $W_B$ is full-rank. Further, we claim that every vertex corresponds with an index $B$ that only selects from the rows of the matrix $Q := (\sigma,~X)'$ and not from the matrix $-(\sigma,~X)'$. To show this, let $B \subset \{1,...,2(p+1) +k \}$ with $|B| =k$ such that $W_B$ is invertible, and suppose there is a vertex corresponding to $B$. Let $\tilde{B}$ be the analogous index that replaces all the indices of $B$ corresponding to rows of $-Q$ with the analogous rows of $Q$. By the preceeding argument, $B$ selects exactly one of the rows of $Q$ corresponding to $\sigma'$ or $-\sigma'$. Suppose first that $B$ selects the row corresponding to $-\sigma$. Without loss of generality, order the remaining rows of $W$ so that $B$ and $\tilde{B}$ differ in the first $w$ positions and agree otherwise. Then we can write

$$W_B = \left( \begin{array}{ll}
	-I_w & 0 \\
	0 & I_{k-w}
\end{array}\right)  W_{\tilde{B}}.$$

\noindent It follows that 

$$W_B^{-1} =  W_{\tilde{B}}^{-1} \left( \begin{array}{ll}
	-I_w & 0 \\
	0 & I_w
\end{array}\right)^{-1}  =  W_{\tilde{B}}^{-1} \left( \begin{array}{ll}
	-I_w & 0 \\
	0 & I_w
\end{array}\right).$$

\noindent However, $b_{\tilde{B}} = e_1$ while $b_B = -e_1$, which combined with the previous display implies that $W_B^{-1}b_B = W_{\tilde{B}}^{-1} b_{\tilde{B}}$. Similarly, suppose that $B$ selects the row corresponding with $\sigma'$. Order the remaining elements of $W$ so that $B$ differs from $\tilde{B}$ in positions $2,...,w+1$. Then we can write 

$$W_B = \left( \begin{array}{lll}
	1& 0 & 0 \\
	0 & -I_w & 0 \\
	0 & 0 & I_{k-w-1}
\end{array}\right)  W_{\tilde{B}}$$

\noindent and hence

$$W_B^{-1} =  W_{\tilde{B}}^{-1} \left( \begin{array}{lll}
	1& 0 & 0 \\
	0 & -I_w & 0 \\
	0 & 0 & I_{k-w-1}
\end{array}\right) $$

\noindent But $b_B = e_1 = b_{\tilde{B}}$, which together with the previous display implies that $W_B^{-1}b_B = W_{\tilde{B}}^{-1} W_{\tilde{B}}$, as we wished to show. We have thus established that $\gamma \in F(X,\sigma)$ is a vertex if and only if it takes the form $A_B^{-1} e_1$, where 
$$A = \left( \begin{array}{l}
	\sigma' \\
	X'\\
	-I
\end{array} \right)  ,$$
\noindent and $B \subset \{1,...,p+k+1\}$ with $|B|=p+1$ and $1 \in B$. $\Box$

To prove our remaining results it is helpful to introduce some additional notation.  Let $\Gamma(X,\sigma)$ be a matrix whose rows collect the elements of $V(X,\sigma)$,
$$V(X,\sigma)=\left\{\gamma\in\mathbb{R}^k:\gamma'=e_j'\Gamma(X,\sigma) \text{ for some }j\in\{1,...,\dim(\Gamma(X,\sigma)\sigma)\}\right\}.$$ We first prove a lemma describing how $\Gamma(X,\sigma)$ varies with $\sigma.$

\begin{lemma} \label{lem: Gamma structure} Suppose Assumption \ref{Assumption: Variance Convergence} holds. For $\upsilon=\sqrt{Diag\left(TT'\right)}$ and $\sigma = \sqrt{Diag(T\Omega T')}$ for some positive-definite $\Omega$,
$\Gamma\left(X,\sigma\right)=\Lambda\left(X,\sigma\right)\Gamma\left(X,\upsilon\right)$
where $\Lambda\left(X,\sigma\right)$ is a diagonal matrix with $\Lambda_{jj}\left(X,\sigma\right)=\frac{1}{e_{j}'\Gamma\left(X,\upsilon\right)\sigma}.$
\end{lemma}

\paragraph{Proof of Lemma \ref{lem: Gamma structure}}

This follows by an argument as in Lemma A.1 of \citet{rambachan_more_2022}, but is included for completeness. Recall that the elements of $\Gamma(X,\sigma)$ take the form $A_B(X,\sigma)^{-1} e_1$ for $B$ such that $A_B(X,\sigma)$ is invertible and $A_B(X,\sigma)^{-1} e_1 \geq 0$. Fix a $B$ corresponding to a vertex in $V(X,\sigma)$. Write $$A_B(X,\sigma) =  \left( \begin{array}{l} \sigma' \\ (X')_{B_1} \\ -I_{B_2} \end{array} \right) $$
\noindent where $B_1$ and $B_2$ are the subsets of $B$ corresponding to the rows of $X'$ and $-I$ respectively. Since $A_B(X,\sigma)$ has rank $k$, it follows that $L := \left[ \begin{array}{l} (X')_{B_1} \\ -I_{B_2}  \end{array}\right]$ has rank $k-1$. Thus, the space of vectors $v$ such that $L v = 0$ is a 1-dimensional linear subspace. Note, however, that by construction if $\vartheta = A_B(X,\tilde\sigma)^{-1}e_1$ for some $\tilde\sigma$ such that $A_B(X,\tilde\sigma)$ is full-rank, then $A_B(X,\tilde\sigma) \vartheta = e_1$ and hence $L \vartheta = 0$. It follows that if $A_B(X,\upsilon)$ is also full rank then $A_B(X,\sigma) \propto A_B(X,\upsilon)$. Note further that from the definition of the vertex set, we must have that $(A_{B}(X,\sigma)^{-1} e_1)' \sigma = 1$. Thus, if $A_B(X,\sigma)$ and $A_B(X,\upsilon)$ both have full rank then 
$$A_B(X,\sigma)^{-1} e_1 = \frac{(A_{B}(X,\sigma)^{-1} e_1)' \sigma  }{(A_{B}(X,\upsilon)^{-1} e_1)' \sigma } A_B(X,\upsilon)^{-1} e_1 =  \frac{1}{(A_{B}(X,\upsilon)^{-1} e_1)' \sigma } A_B(X,\upsilon)^{-1} e_1.$$
\noindent Note that Lemma \ref{lem: vertex characterization} implies that $A_B(X,\upsilon)^{-1} e_1 \in V(X,\upsilon)$, since $A_B(X,\upsilon) \propto A_B(X,\sigma) \geq 0$ and $A_B(X,\upsilon) \upsilon = 1$ by construction. By an analogous argument reversing the roles of $\sigma$ and $\upsilon$, we can show that if $B$ corresponds to a vertex of $V(X,\upsilon)$, then a re-scaling of $A_B(X,\upsilon)^{-1} e_1$ is also a vertex of $V(X,\sigma)$ provided that $A_B(X,\upsilon)$ is full-rank. 

It thus remains to show that $A_B(X,\sigma)$ has full rank and satisfies $A_B(X,\sigma)^{-1} e_1 \geq 0$ if and only if $A_B(X,\upsilon)$ does. To this end, suppose that $A_B(X,\upsilon)$ has full rank and $A_B(X,\upsilon)^{-1} e_1 \geq 0$. Let $\vartheta = A_B(X,\upsilon)^{-1} e_1$ and note that by construction $\vartheta \geq 0$, $\upsilon'\vartheta= 1$, and $L \vartheta = 0$. Note, however, that the structure of $\sigma$ implies that $\upsilon_{j}=0$ if and only if $\sigma_j =0$, so $\upsilon'\vartheta = 1$ and $\vartheta \geq 0$ implies that $\sigma' \vartheta> 0$. Hence, since $L \vartheta = 0$ while $\sigma' \vartheta >0$, we see that $\sigma'$ is linearly independent of $L$, and thus $A_B(X,\sigma)$ has full rank. Moreover, by the argument above, we have that $A_B(X,\sigma)^{-1} e_1$ is a positive rescaling of $A_B(X,\upsilon)e_1$, and thus $A_B(X,\sigma)^{-1} e_1 \geq 0$, as needed. Since we can repeat the same argument reversing the roles of $\sigma$ and $\upsilon$, we have established the desired result. $\square$

\paragraph{Proof of Lemma \ref{lem: can consider subset of vertices}}
The first part of the Lemma follows immediately from Lemma \ref{lem: Gamma structure} above. To show the second part, let $\hat\eta_{\dagger} = \max_{\gamma \in V_\dagger(X_{n,0},\hat{\sigma}_{n,0})} \gamma'Y_{n,0}$ denote the analog to $\hat\eta_{n,0}$ using $V_\dagger$ instead of $V$, and define other variables subscripted with $\dagger$ analogously. Observe that by construction, $\hat\eta_{\dagger}= \hat\eta_{n,0}$ unless $\hat\eta_{n,0} \leq 0$. Next, consider the modified least favorable critical value, $c_{\alpha,LF,\dagger}$, which is the $1-\alpha$ quantile of $\max_{\gamma \in V_\dagger(X_{n,0},\hat{\sigma}_{n,0})} \gamma' \xi$, for $\xi \sim N(0,\widehat{\Sigma}_{n,0})$. By construction, $\max_{\gamma \in V_\dagger(X_{n,0},\hat{\sigma}_{n,0})} \gamma' \xi = \max_{\gamma \in V(X_{n,0}, \hat{\sigma}_{n,0})} \gamma' \xi $ unless $\max_{\gamma \in V(X_{n,0}, \hat{\sigma}_{n,0})} \gamma' \xi \leq 0$. Now, for any $\gamma_{1,\dagger}\in V_\dagger(X_{n,0},\hat{\sigma}_{n,0})$, we have that $\gamma_{1,\dagger}'\xi \leq \max_{\gamma \in V_\dagger(X_{n,0},\hat{\sigma}_{n,0})} \gamma' \xi $, and $\gamma_{1,\dagger}' \xi \sim N(0, \gamma_{1,\dagger}' \widehat{\Sigma}_{n,0} \gamma_{1,\dagger})$, which has median of zero. It follows that for $\alpha < 0.5$, the $1-\alpha$ quantile of $\max_{\gamma \in V_\dagger(X_{n,0},\hat{\sigma}_{n,0})} \gamma' \xi $ is weakly positive, and hence that $c_{\alpha,LF} = c_{\alpha,LF,\dagger}$. We have thus established the result for the LF test. 

Next consider the conditional test. By construction the conditional test never rejects when $\hat\eta_{n,0} \leq 0$, so we will consider the case where $\hat\eta_{n,0} >0$. As argued above, in this case $\hat\eta_{n,0} = \hat\eta_{\dagger}$, and moreover, $\hat\gamma = \hat\gamma_{\dagger}$ from the definition of $V_\dagger(X_{n,0},\hat{\sigma}_{n,0})$. Finally, recall that Lemma 5.1 in \cite{Leeetal2016} implies that $\mathcal{V}^{lo}_{n,0}$ and $\mathcal{V}^{up}_{n,0}$ are the minimum and maximum of the set $$\left\{\hat\gamma' y \,|\, y \text{ s.t. } \hat\gamma'y \geq \max_{\tilde\gamma \in V(X_{n,0}, \hat{\sigma}_{n,0} )} \tilde\gamma'y \text{ and } S(y,\hat\gamma) = S_{n,0,\hat\gamma} \right\} .$$

\noindent Since $\max_{\tilde\gamma \in V(X_{n,0}, \hat{\sigma}_{n,0} )} \tilde\gamma'y$ is equal to $\max_{\tilde\gamma \in V_\dagger(X_{n,0}, \hat{\sigma}_{n,0} )} \tilde\gamma'y $ whenever the former is positive, we see that $\mathcal{V}^{up}_{n,0} = \mathcal{V}^{up}_\dagger$, since $\mathcal{V}^{up}_{n,0} \geq \hat\eta_{n,0} > 0$. Further, since $V_\dagger(X_{n,0},\hat{\sigma}_{n,0}) \subseteq V(X_{n,0}, \hat{\sigma}_{n,0})$, we have that $\hat\gamma'y \geq \max_{\tilde\gamma \in V_\dagger(X_{n,0}, \hat{\sigma}_{n,0} )} \tilde\gamma'y$ whenever $\hat\gamma'y \geq \max_{\tilde\gamma \in V(X_{n,0}, \hat{\sigma}_{n,0} )} \tilde\gamma'y$. It follows that $\mathcal{V}^{lo}_{\dagger} \leq \mathcal{V}^{lo}_{n,0}$. Note, however, that the critical value for the conditional test is increasing in the value of $\mathcal{V}^{lo}_{n,0}$, and thus $c_{\alpha,C} \geq c_{\alpha,C,\dagger}$. It follows that $\hat\eta_{n,0} > c_{\alpha,C}$ only if $\hat\eta_{\dagger} > c_{\alpha,C,\dagger}$, as we wished to show. The desired result for the hybrid test follows immediately from the arguments for the LF and conditional tests. $\Box$

Following \citetalias{Andrewsetal2017}, we establish size control using a subsequencing argument.

\begin{lemma} \label{lem: Size limit problem} Under Assumptions
\ref{Assumption: Variance Convergence}, \ref{Assumption: Variance Estimator},
and \ref{Assumption: Uniform CLT}, to show that a test $\phi$ which (i)
depends on the data through $\left(Y_{n,0},X_{n,0},\widehat{\Sigma}_{n,0}\right)$ and (ii) does not reject when $\hat\eta_{n,0}=-\infty$
has uniformly correct asymptotic size, 
\[
\limsup_{n\to\infty}~\sup_{P_{D|Z}\in\mathcal{P}_{D|Z}}~\sup_{\beta_0\in B_I(P_{D|Z})}E_{P_{D|Z}}\left[\phi\right]\le\alpha,
\]
it suffices to show that $\limsup_{l\to\infty}E_{P_{D|Z},n_{l}}\left[\phi\right]\le\alpha$
for all subsequences $\left\{ n_{l}\right\} \subseteq\left\{ n\right\} $,
$\left\{ P_{D|Z,n_{l}}\right\} \in\mathcal{P}_{D|Z}^{\infty}=\times_{l=1}^{\infty}\mathcal{P}_{D|Z},$ $\{\beta_{0,n_l}\}\in\times_{l=1}^{\infty} B_I(P_{D|Z,n_l})$
with
\begin{enumerate}
\item $\min_\delta \max_j e_j'X_{n_l,0}\delta>-\infty$ and $\Omega\left(P_{D|Z,n_{l}},\beta_{0,n_l}\right)\to\Omega^{*}$ for some $\Omega^{*}\in \mathbf{\Omega}_{\bar\lambda}$
\item For each $j$ and 
$
\psi_{j,n_{l}}=\sqrt{e_{j}'\Gamma\left(X_{n_{l},0},\upsilon\right)TT'\Gamma\left(X_{n_{l},0},\upsilon\right)e_{j}},
$
either $\psi_{j,n_{l}}=0$ for all $l$ or $\psi_{j,n_{l}}\neq0$
for all $l$
\item If $\psi_{j,n_{l}}>0$ for some $j$ then for $\psi_{n_{l}}=\max_{j}\psi_{j,n_{l}}$,
 $\psi_{n_{l}}^{-1}\Gamma\left(X_{n_{l},0},\upsilon\right)T\to\Pi^{*}$
for  $\Pi^{*} \neq 0$
\item If $\psi_{n_{l}}>0$, then $\psi_{n_{l}}^{-1}\Gamma\left(X_{n_{l},0},\upsilon\right)\mu_{n_{l},0}\to\nu^{*}\in\left[-\infty,0\right]^{\dim\left(Y_{n,0}\right)}$
\item For $\sigma(\Omega)=\sqrt{Diag(T'\Omega T)}$ and $\Lambda(X,\sigma)$ as defined in Lemma \ref{lem: Gamma structure}, $\Lambda(X_{n_l,0},\sigma(\Omega(P_{D|Z,n_l},\beta_{0,n_l})))\to \Lambda^*$ for $\Lambda^*$ a diagonal, positive-definite matrix. Likewise, $\Lambda(X_{n_l,0},\hat\sigma_{n_l,0} )\to_{p} \Lambda^*$ for $\hat\sigma_{n_l,0} = \sigma( \hat\Omega_{n_l,0})$.
\end{enumerate}
\end{lemma}

\paragraph{Proof of Lemma \ref{lem: Size limit problem}}

We establish that if size control fails, then there always exists a sequence satisfying the conditions of the lemma under which size control also fails. 

If size control fails, then 
\[
\limsup_{n\to\infty}~\sup_{P_{D|Z}\in\mathcal{P}_{D|Z}}~\sup_{\beta_0\in B_I(P_{D|Z})}E_{P_{D|Z}}\left[\phi\right]\ge\alpha+2\varepsilon
\]
for some $\varepsilon>0$. This implies that there exists a subsequence
$\left\{ n_{t}^{1}\right\} \subseteq\left\{ n\right\} $, $\left\{ P_{D|Z,n_{t}^{1}}\right\} \in\mathcal{P}_{D|Z}^{\infty},$ $\{\beta_{0,n_t^1}\}\in\times_{t=1}^\infty B_I(P_{D|Z,n_{t}^{1}})$
such that $\liminf_{t\to\infty}E_{P_{D|Z,n_{t}^{1}}}\left[\phi\right]\ge\alpha+\varepsilon$.
Since $\phi$ is assumed not to reject when $\hat\eta_{n,0}=-\infty$, it must be that $\min_\delta \max_j e_j'X_{n_t,0}\delta$ is finite for all $t$, since otherwise $\hat\eta_{n_t,0}=-\infty$ with probability 1 and the test never rejects. 
Since $\Omega\left(P_{D|Z,n_{t}^{1}},\beta_{0,n_{t}^{1}}\right)\in  \mathbf{\Omega}_{\bar\lambda}$
for all $t$ by assumption, and $ \mathbf{\Omega}_{\bar\lambda}$ is compact, there exists a further subsequence $\left\{ n_{t}^{2}\right\} \subseteq\left\{ n_{t}^{1}\right\} $ with
$\Omega\left(P_{D|Z,n_{t}^{2}},\beta_{0,n_{t}^{2}}\right)\to\Omega^{*}\in  \mathbf{\Omega}_{\bar\lambda}$.

For each $t$, $\Gamma\left(X_{n_{t}^{2},0},\upsilon\right)$ is a matrix
with $\dim\left(Y_{n,0}\right)$ columns, and a uniformly bounded number of rows. Hence there exists a subsequence $\left\{ n_{t}^{3}\right\} \subseteq\left\{ n_{t}^{2}\right\} $
along which the dimension of $\Gamma\left(X_{n_{t}^{3},0},\upsilon\right)$
is constant. For each $j$ and any subsequence $\left\{ n_{r}\right\} \subseteq\left\{ n\right\} $,
either $\psi_{j,n_{r}}=0$ infinitely often or not. We can thus extract
a further subsequence $\left\{ n_{t}^{4}\right\} \subseteq\left\{ n_{t}^{3}\right\} $
along which part (2) of the lemma holds. If $\psi_{j,n_{t}^{4}}=0$
for all $j$ then part (3) of the lemma is vacuous, while if $\psi_{j,n_{t}^{4}}>0$
for some $j$, $\psi_{j,n_{t}^{4}}^{-1}\left\Vert e_{j}'\Gamma\left(X_{n_{t}^{4},0},\upsilon\right)T\right\Vert =1$
by construction, so $\psi_{n_{t}^{4}}^{-1}\left\Vert e_{j}'\Gamma\left(X_{n_{t}^{4},0},\upsilon\right)T\right\Vert \le1$
for all $j$, and there exists a subsequence $\left\{ n_{t}^{5}\right\} \subseteq\left\{ n_{t}^{4}\right\} $
along which $\psi_{n_{t}^{5}}^{-1}\Gamma\left(X_{n_{t}^{5},0},\upsilon\right)T\to\Pi^{*},$ where $\Pi^{*}\neq0$ since $\psi_{n_{t}^{5}}^{-1}\left\Vert e_{j}'\Gamma\left(X_{n_{t}^{5},0},\upsilon\right)T\right\Vert=1$ for at least one $j$, thus
establishing part (3) of the lemma.

Part (4) of the lemma is again vacuous if $\psi_{n_{l}}=0$. Otherwise, note that since
\[
\max_{j}\hspace{.1cm}e_{j}'\Gamma\left(X_{n,0},\upsilon\right)\mu_{n,0}=\min_{\delta}\hspace{.1cm}\max_{j} \hspace{.1cm}e_{j}'\left(\mu_{n,0}-X_{n,0}\delta\right)
\]
whenever the solution is finite, $\Gamma\left(X_{n_t^5,0},\upsilon\right)\mu_{n_t^5,0}\le0$ for all $t$. For
any subsequence $\left\{ n_{r}\right\} \subseteq\left\{ n_{t}^{5}\right\} $
and any $j$, $\psi_{n_{r}}^{-1}e_{j}'\Gamma\left(X_{n_{r},0},\upsilon\right)\mu_{n_{r},0}$
is either bounded or unbounded as $r\to\infty$, allowing us to extract
a further subsequence $\left\{ n_{t}^{6}\right\} \subseteq\left\{ n_{t}^{5}\right\} $
along which $\psi_{n_{t}^{6}}^{-1}e_{j}'\Gamma\left(X_{n_{t}^{6},0},\upsilon\right)\mu_{n_{t}^{6},0}\to\nu_{j}^{*}\in\left[-\infty,0\right]$.
Starting from $\left\{ n_{t}^{5}\right\} $ and iterating this argument
over the rows of $\psi_{n_{t}^{5}}^{-1}\Gamma\left(X_{n_{t}^{5},0},\upsilon\right)\mu_{n_{t}^{5},0}$
delivers a subsequence $\left\{ n_{s}\right\} $ satisfying properties
(1)-(4) of the lemma.


Next, let $M$ be the matrix that selects the non-zero rows of $T$, and observe that $M$ also selects the non-zero elements of $\upsilon$ and of $\sigma(\Omega)$ for any positive definite $\Omega$.  Let $\gamma_{n,j}' = e_j'(\Gamma(X_{n,0}, \upsilon))$. By construction, $\gamma_{n,j}' \upsilon = (M \gamma_{n,j})'(M \upsilon) = 1$. Since $M\upsilon >0 $ and $M \gamma_{n,j} \geq 0$ by construction, it follows that $||M \gamma_{n,j}||$ is bounded. However, for $\sigma_{n,0} = \sigma(\Omega(P_{D|Z,n},\beta_{0,n}))$, we have $|\gamma_{n,j}'\sigma_{n,0}| = | (M \gamma_{n,j})' (M\sigma_{n,0})| \leq ||M \gamma_{n,j}|| \cdot ||M \sigma_{n,0}||$, where part (ii) of Assumption \ref{Assumption: Variance Convergence} implies that $||M \sigma_{n,0}||$ is also bounded. It follows that there exists a subsequence $\left\{ n_{l}^j\right\} \subseteq\left\{ n_{s}\right\}$ such that $\gamma_{n_l^j,j}'\sigma_{n_l^j,0}$ converges. Moreover, the limit must be strictly positive, since by construction $\gamma_{n_l^j,j}'\upsilon = 1$ and $\gamma_{n_l^j,j}\geq0$, whereas the fact that the eigenvalues of $\Omega_{n_l^j,0}$ are bounded from below implies $\sigma_{n_l^j,0} \geq c \upsilon$ for some $c>0$. Iterating this argument for each $j$, we obtain a subsequence $\left\{ n_{l}\right\} \subseteq\left\{ n_{s}\right\}$ such that $\gamma_{n_l,j}'\sigma_{n_l,0}$ converges to a positive limit for all $j$. The $j$th diagonal element of $\Lambda\left(X_{n_{l},0},\sigma(\Omega(P_{D|Z,n_l},\beta_{0,n_l}))\right)$ is $1/(\gamma_{n_l,j}'\sigma_{n_l,0}) $, and hence  $\Lambda\left(X_{n_{l},0},\sigma(\Omega(P_{D|Z,n_l},\beta_{0,n_l}))\right)\to\Lambda^{*}$ for $\Lambda^{*}$ a positive-definite and diagonal matrix, which establishes that the sequence also meets the first part of condition (5). To establish the second part of condition (5), observe that $$| \gamma_{n,j}' \hat{\sigma}_{n_l,0} - \gamma_{n,j}' \sigma_{n_l,0}|  =   | (M\gamma_{n,j})' M(\hat{\sigma}_{n_l,0} - \sigma_{n_l,0})|   \leq  || M\gamma_{n,j} || \cdot || M(\hat{\sigma}_{n_l,0} - \sigma_{n_l,0})|| \to_{p} 0. $$ However, the $j$th diagonal element of $\Lambda(X_{n_l,0}, \sigma_{n_l,0})$ is equal to $1/ (\gamma_{n,j}' \sigma_{n_l,0})$, which we showed above converges to a positive constant $e_j' \Lambda^* e_j$. The continuous mapping theorem thus implies that $e_j' \Lambda( X_{n_l,0} , \hat\sigma_{n_l,0}) e_j = 1/(\gamma_{n,j}' \hat\sigma_{n_l,0}) \to_{p} e_j' \Lambda^* e_j$.


We have thus established that there exists a sequence satisfying the conditions of the lemma under which size control fails, as we wished to show. $\square$

\paragraph{Proof of Proposition \ref{Prop: Least Favorable Size Control-alt}}

By construction, the least favorable test never rejects when $\hat\eta_{n,0}=-\infty$.
Hence, by Lemma \ref{lem: Size limit problem}, it suffices to show size
control for sequences $\left\{ n_{l},P_{D|Z,n_{l}},\beta_{0,n_l}\right\} $ satisfying
the conditions of the lemma.

Note that by Lemma \ref{lem: Gamma structure}
we can write 
\[
\hat{\eta}_{n_l,0}=\max_{j} \left\{ e_{j}'\Gamma\left(X_{n_{l},0},\hat{\sigma}_{n_l,0}\right)Y_{n_{l},0}\right\}=\max_{j} \{ e_{j}'\Lambda\left(X_{n_{l},0},\hat{\sigma}_{n_l,0}\right)\Gamma\left(X_{n_{l},0},\upsilon\right)Y_{n_{l},0} \}
\]
\[
=\max_{j} \left\{ e_{j}'\Lambda\left(X_{n_{l},0},\hat{\sigma}_{n_l,0}\right)\left(\Gamma\left(X_{n_{l},0},\upsilon\right)\left(Y_{n_{l},0}-\mu_{n_{l},0}\right)+\Gamma\left(X_{n_{l},0},\upsilon\right)\mu_{n_{l},0}\right) \right\}.
\]
 Assumption \ref{Assumption: Variance Convergence} implies that
we can re-write $Y_{n_{l},0}-\mu_{n_{l},0}$ as $T(U_{n_l,0}-\pi_{n_l,0})$. Hence, 
\[
\hat{\eta}_{n_l,0}= \max_{j} \left\{e_{j}'\Lambda\left(X_{n_{l},0},\hat{\sigma}_{n_l,0}\right)\left(\Gamma\left(X_{n_{l},0},\upsilon\right)T(U_{n_l,0}-\pi_{n_l,0})+\Gamma\left(X_{n_{l},0},\upsilon\right)\mu_{n_{l},0}\right) \right\}.
\]

First consider the case where $\psi_{n_{l}}=0$. This implies that
$\Gamma\left(X_{n_{l},0},\upsilon\right)T=0$ for all $l$, which in
turn implies that $\Gamma\left(X_{n_{l},0},\upsilon\right)Y_{n_{l},0}\le0$
with probability one since $\beta_{0,n_l} \in B_I(P_{D|Z,n_l})$ by construction and thus $\Gamma\left(X_{n_{l},0},\upsilon\right)\mu_{n_{l},0} \leq 0$. The least favorable
test never rejects in this case, since $\alpha<\frac{1}{2}$ implies
that $c_{\alpha,LF}\left(X_{n,0},\widehat{\Sigma}_{n,0}\right)\ge0$.

Next consider the case where $\psi_{n_{l}}>0$. Assumption \ref{Assumption: Uniform CLT}
implies that $Y_{n_{l},0}-\mu_{n_{l},0}\to_{d}N\left(0,T\Omega^{*}T'\right)$.
Parts (3) and (4) of Lemma \ref{lem: Size limit problem} thus imply that 
\[
\psi_{n_{l}}^{-1}\left(\Gamma\left(X_{n_{l},0},\upsilon\right)T(U_{n_l,0}-\pi_{n_l,0})+\Gamma\left(X_{n_{l},0},\upsilon\right)\mu_{n_{l},0}\right)\to N\left(\nu^{*},\Pi^{*}\Omega^{*}\Pi^{*'}\right)
\]
By part (5) of Lemma \ref{lem: Size limit problem}, $\Lambda\left(X_{n_{l},0},\hat{\sigma}_{n_l,0}\right)\to_{p}\Lambda^{*}$, for $\Lambda^*$ diagonal and positive definite, so by the continuous mapping theorem,
\[
\psi_{n_{l}}^{-1}\Lambda\left(X_{n_{l},0},\hat{\sigma}_{n_l,0}\right)\left(\Gamma\left(X_{n_{l},0},\upsilon\right)T(U_{n_l,0}-\pi_{n_l,0})+\Gamma\left(X_{n_{l},0},\upsilon\right)\mu_{n_{l},0}\right)
\]
\[
\to_{d}G^{*}\sim N\left(\Lambda^{*}\nu^{*},\Lambda^{*}\Pi^{*}\Omega^{*}\Pi^{*'}\Lambda^{*}\right).
\]
Hence, by another application of the continuous mapping theorem, $\psi_{n_{l}}^{-1}\hat{\eta}_{n_l,0}\to_{d}\max_{j}e_{j}'G^{*},$
where since $\Lambda^{*}\nu^{*}\le0$, the limiting distribution is
continuous at all strictly positive values.

To show size control for the least favorable test, we must further
show convergence of the critical value. To this end, note that Assumptions
\ref{Assumption: Variance Convergence} and \ref{Assumption: Variance Estimator},
together with convergence of $\Lambda\left(X_{n_{l},0},\hat{\sigma}_{n_l,0}\right)$,
imply that 
\[
\psi_{n_{l}}^{-2}\Gamma\left(X_{n_{l},0},\hat{\sigma}_{n_l,0}\right)\widehat{\Sigma}_{n,0}\Gamma\left(X_{n_{l},0},\hat{\sigma}_{n_l,0}\right)'\to_{p}\Lambda^{*}\Pi^{*}\Omega^{*}\Pi^{*'}\Lambda^{*},
\]
where the limit is nonzero. Note, moreover, that 
\[
c_{\alpha,LF}\left(X_{n_l,0},\widehat{\Sigma}_{n,0}\right)=\psi_{n_{l}}\cdot c_{\alpha,LF}\left(X_{n_l,0},\psi_{n_{l}}^{-2}\cdot\widehat{\Sigma}_{n,0}\right).
\]
Hence, $c_{\alpha,LF}\left(X_{n_l,0},\psi_{n_{l}}^{-2}\cdot\widehat{\Sigma}_{n,0}\right)$
converges in probability to $c_{\alpha,LF}^{*},$ the $1-\alpha$ quantile
of $\max_{j}e_{j}'\tilde{G}$ for $\tilde{G}\sim N\left(0,\Lambda^{*}\Pi^{*}\Omega^{*}\Pi^{*'}\Lambda^{*}\right)$,
where $c_{\alpha,LF}^{*}>0$ for $\alpha<\frac{1}{2}$. Note further
that
\[
\phi_{LF}=1\left\{ \hat{\eta}_{n_l,0}>c_{\alpha,LF}\left(X_{n_l,0},\widehat{\Sigma}_{n,0}\right)\right\} =1\left\{ \psi_{n_{l}}^{-1}\hat{\eta}_{n_l,0}>c_{\alpha,LF}\left(X_{n_l,0},\psi_{n_{l}}^{-2}\cdot\widehat{\Sigma}_{n,0}\right)\right\} ,
\]
so by another application of the continuous mapping theorem,
\[
\phi_{LF}\to_{d}1\left\{ \left( \max_{j} \hspace{.1cm }e_{j}'G^{*} \right)>c_{\alpha,LF}^{*}\right\} ,
\]
which implies that $\limsup_{s\to\infty}E_{P_{D|Z},n_{l}}\left[\phi_{LF}\right]\le\alpha,$
as we wanted to show. $\Box$

\paragraph{Proof of Proposition \ref{Prop: Conditional, Hybrid Size Control}}

We first prove the result for the conditional test. As in Lemma \ref{lem: Size limit problem}, we use a subsequencing argument.
Specifically, begin with sequences of sample sizes, data generating processes, and null parameter values $\{n_s\}\subseteq\{n\}$, $\{P_{D|Z,n_s}\}\in\mathcal{P}_{D|Z}^{\infty}$, and $\{\beta_{0,n_s}\}\in\times_{s=1}^\infty B_I(P_{D|Z,n_s})$. Observe that whether $V_{\dagger}(X_{n_s,0}, \hat\sigma_{n_s,0})$ is empty depends only on $X_{n_s,0}$. If $X_{n_s,0}$ is such that $V_{\dagger}(X_{n,0}, \hat\sigma_{n_s,0})$ is empty, then $\hat\eta_{n,0}\leq 0$ with probability 1, and thus the conditional and hybrid tests never reject. For the remainder of the proof, we therefore consider sequences where $X_{n_s,0}$ is such that $V_{\dagger}(X_{n_s,0}, \hat\sigma_{n_s,0})$ is non-empty, which implies that $\min_\delta \max_j e_j' X_{n,0} \delta > -\infty$, and thus $\hat\eta_{n_s,0}$ is finite with probability 1. It then suffices to establish size control for the test $\phi_{C,\dagger}$,  since $\phi_{C}\leq \phi_{C,\dagger}$ with probablity 1 by Lemma \ref{lem: can consider subset of vertices}.  

Let $M$ be the selection matrix such that $M'T$ picks out the nonzero rows of $T$, and note that by construction $\Gamma_{\dagger}\left(X_{n,0},\upsilon\right)MM'\upsilon=\iota,$ where $\Gamma_{\dagger}$ denotes the subset of rows of $\Gamma$ corresponding with vertices in $V_{\dagger}(X_{n,0},\upsilon)$ and $\iota$ is the vector of ones.
Since $M'\upsilon$ is strictly positive, $\Gamma_{\dagger}\left(X_{n,0},\upsilon\right)M$
is a non-negative matrix with a uniformly bounded number of rows and
uniformly bounded row-sums. There thus exists a subsequence of sample
sizes $\left\{ n_{r}\right\} \subseteq\left\{ n_{s}\right\} $ such
that $\Gamma_{\dagger}\left(X_{n_{r},0},\upsilon\right)M$ has fixed dimensions and $\Gamma_{\dagger}\left(X_{n_{r},0},\upsilon\right)M\to\Gamma^{*}_{\dagger} M$ for $\Gamma^{*}_{\dagger}$
a non-negative matrix with $\Gamma^{*}_{\dagger} \upsilon=\iota$. 
 Since $\Omega\left(P_{D|Z,n_{r}},\beta_{0,n_r}\right)\in \mathbf{\Omega}_{\bar\lambda}$
for all $r$ by assumption, and $\mathbf{\Omega}_{\bar\lambda}$ is compact, there exists a further subsequence $\left\{ n_{t}\right\} \subseteq\left\{ n_{r}\right\} $ with
$\Omega\left(P_{D|Z,n_{t}},\beta_{0,n_t}\right)\to\Omega^{*}\in \mathbf{\Omega}_{\bar\lambda}.$

Note, next, that \begin{align}
\Gamma_{\dagger}\left(X_{n_{t},0},\upsilon\right)Y_{n_{t},0}&=\Gamma_{\dagger}\left(X_{n_{t},0},\upsilon\right)\left(Y_{n_{t},0}-\mu_{n_{t},0}\right)+\Gamma_{\dagger}\left(X_{n_{t},0},\upsilon\right)\mu_{n_{t},0}  \nonumber \\
&=\Gamma_{\dagger} \left(X_{n_{t},0},\upsilon\right)MM'T\left(U_{n_{t},0}-\pi_{n_t,0}\right)+\Gamma_{\dagger}\left(X_{n_{t},0},\upsilon\right)\mu_{n_{t},0}, \label{eqn: gammadagger Y decom}
\end{align}
where $\Gamma_{\dagger}\left(X_{n_{t},0},\upsilon\right)\mu_{n_{t},0}\le0$
for all $t$ since $\beta_{0,n_t}\in B_I(P_{D|Z,n_t})$. Assumptions \ref{Assumption: Variance Convergence} and \ref{Assumption: Uniform CLT} imply
that 
\[
U_{n_{t},0}-\pi_{n_t,0}\to_{d}N\left(0,\Omega^{*}\right),
\]
so for $\Sigma^*=T\Omega^*T'$,
\begin{equation}
\Gamma_{\dagger} \left(X_{n_{t},0},\upsilon\right)MM'T\left(U_{n_{t},0}-\pi_{n_t,0}\right)\to_{d}N\left(0,\Gamma^{*}_{\dagger}MM'\Sigma^* MM'\Gamma^{*'}_{\dagger}\right)=N\left(0,\Gamma^{*}_{\dagger}\Sigma^*\Gamma^{*'}_{\dagger}\right) \label{eqn: convergence gammadagger U}
\end{equation}
by the continuous mapping theorem, where Assumption \ref{Assumption: Nondegeneracy}
implies that the diagonal elements of $\Gamma^{*}_{\dagger}T \Omega^* T'\Gamma^{*'}_{\dagger} =  \Gamma^{*}_{\dagger} \Sigma^* \Gamma^{*'}_{\dagger}$ are bounded away from zero. As argued in the proof of Lemma \ref{lem: Size limit problem}, we can extract a further subsequence $\{n_l\}$ where
\[
\Gamma_{\dagger}\left(X_{n_{l},0},\upsilon\right)\mu_{n_{l},0}\to\nu^{*}\in\left[-\infty,0\right]^{\dim\left(\Gamma_{\dagger}^{*}\upsilon\right)}.
\]
\noindent By an argument analogous to that for part (5) of Lemma \ref{lem: Size limit problem}, we can also choose $\{n_l\}$ such that,  for $\sigma_{n_l,0}= \sigma( \Omega(P_{D|Z,n_l},\beta_{0,n_l}) )$ and $\hat\sigma_{n_l,0} =\sigma( \hat\Omega_{n_l,0})$, $\Lambda_{\dagger}\left(X_{n_{l},0},\sigma_{n_l,0} \right)\to\Lambda^{*}_{\dagger}$ and $\Lambda_{\dagger}\left(X_{n_{l},0},\hat\sigma_{n_l,0}\right)\to_p \Lambda^{*}_{\dagger}$ for $\Lambda^{*}_{\dagger}$ diagonal and positive definite.


Note next that if $\hat\eta_{\dagger}\to_p -\infty$ (because $\nu_j^*=-\infty$ for all $j$) then the rejection probability of the test $\phi_{C,\dagger}$  converges to zero.  If instead $\hat\eta_{\dagger} \not\to_p-\infty,$ then it must be that $\nu_j^*>-\infty$ for some $j$.  Let $M_+$ be a selection matrix such that $M_+\nu^*$ picks out the finite elements of $\nu^*$.  Note that for any $\gamma$ corresponding to a row of $\Gamma_{\dagger}(X_{n_l,0},\hat{\sigma}_{n,0})$ not selected by $M_+$, $Pr_{P_{D|Z,n_l}}\left\{\hat\gamma_{\dagger}=\gamma\right\}\to 0$, and thus asymptotically neither $\hat\gamma_{\dagger}$ nor $\hat\eta_\dagger$ is affected by $\gamma' Y_{n_l,0}$. By an argument analogous to that in the proof to Lemma \ref{lem: slack moments}, one can also show that asymptotically $\gamma' Y_{n_l,0}$ does not affect the values of $\mathcal{V}^{lo}_{n,0,\dagger}$ or $\mathcal{V}^{lo}_{n,0,\dagger}$. The asymptotic behavior of the $\phi_{C,\dagger}$ test is thus determined by $(M_+ \Gamma_{\dagger}(X_{n_l,0},\hat{\sigma}_{n,0}) Y_{n_l,0},M_+\Gamma_{\dagger}(X_{n_l,0},\hat{\sigma}_{n,0})\widehat{\Sigma}_{n,0}\Gamma_{\dagger}(X_{n_l,0},\hat{\sigma}_{n,0})' M'_+).$ 

Next, observe from equations (\ref{eqn: gammadagger Y decom}) and (\ref{eqn: convergence gammadagger U}), combined with the fact that $\Gamma_{\dagger}(X_{n,0},\hat\sigma_{n,0}) = \Lambda_{\dagger}(X_{n,0}, \hat\sigma_{n,0}) \Gamma_{\dagger}(X_{n,0}, \upsilon )$, that $$M_+\Gamma_{\dagger}(X_n,\hat{\sigma}_{n,0})(Y_{n,0}-\mu_{n,0})\to_d N(0,M_+\Lambda^*_{\dagger}\Gamma^*_{\dagger}\Sigma^* \Gamma^{*'}_{\dagger} \Lambda_{\dagger}^*M_+').$$  Further, since $M_+\Gamma_{\dagger}(X_{n_l,0},\upsilon)\mu_{n_l,0}$ converges to a finite vector by construction, we have that
$$
M_+\left(\Gamma_{\dagger}(X_{n_l,0},\hat{\sigma}_{n,0})-\Gamma_{\dagger}(X_{n_l,0},\sigma_{n_l,0})\right)\mu_{n_l,0}=
M_+(\Lambda_{\dagger}(X_{n_l,0}, \hat\sigma_{n_l,0})-\Lambda_{\dagger}(X_{n_l,0}, \sigma_{n_l,0}))\Gamma_{\dagger}(X_{n_l,0},\upsilon)\mu_{n_l,0}\to_p 0,
$$
where we use the fact that $\Lambda_{\dagger}\left(X_{n_{l},0},\sigma_{n_l,0}\right)\to\Lambda^{*}_{\dagger}$ and $\Lambda_{\dagger}\left(X_{n_{l},0},\hat\sigma_{n_l,0} \right)\to_p \Lambda^{*}_{\dagger}$. Hence,  
$$
M_+\Gamma_{\dagger}(X_{n_l,0},\hat{\sigma}_{n_l,0})Y_{n_l,0}-M_+\Gamma_{\dagger}(X_{n_l,0},\sigma_{n_l,0})\mu_{n_l,0}\to_d G^* \sim N(0,M_+\Lambda^*_{\dagger}\Gamma^*_{\dagger}\Sigma^* \Gamma^{*'}_{\dagger}\Lambda^*_{\dagger}M_+'),
$$
where Assumption \ref{Assumption: Nondegeneracy} implies (i) that
the diagonal elements of the limiting variance are nonzero and (ii)
that no two rows of $G^{*}$ are perfectly positively correlated.
Further, by the continuous mapping theorem
\[
M_+\Gamma_{\dagger}\left(X_{n_{l},0},\hat{\sigma}_{n_l,0}\right)\widehat{\Sigma}_{n,0}\Gamma_{\dagger}\left(X_{n_{l},0},\hat{\sigma}_{n_l,0}\right)'M_+'\to_{p}M_+\Lambda^{*}_{\dagger}\Gamma^{*}_{\dagger}\Sigma^*\Gamma^{*'}_{\dagger}\Lambda^{*}_{\dagger}M_+'.
\]
These are precisely
the conditions assumed in \cite{Andrewsetal2018}, which we shorthand as AKM, to establish uniform asymptotic size control, so we can use their results to establish size control in our setting.

Specifically, to connect our setting to that in AKM, let $X_n$ and $Y_n$ in the notation of AKM both be equal to $M_+\Gamma_{\dagger}(X_{n_l,0},\hat{\sigma}_{n,0})Y_{n_l,0},$ and let $\mu_{X,n}$ and $\mu_{Y,n}$ both be equal to $M_+\Gamma_{\dagger}(X_{n_l,0},\sigma_{n_l,0})\mu_{n_l,0}$. Let $\hat{j}$ be the row of $M_+\Gamma_{\dagger}(X_{n_l,0},\hat{\sigma}_{n,0})$ corresponding to $\hat{\gamma}_{\dagger}$, and let $\hat{\gamma}_{\dagger,*}$ be the $\hat{j}$th row of $M_+\Gamma_{\dagger}(X_{n_l,0},\sigma_{n_l,0})$. We have established that Assumptions 2-4 of AKM hold under the sequence $\{n_l,P_{D|Z,n_l},\beta_{0,n_l}\},$ so Proposition 10 in AKM establishes that for $\hat{\mu}_{\alpha,n_l}$ the $\alpha$-quantile unbiased estimator for $\hat\gamma_{\dagger,*}'\mu_{n_l,0}$ (see AKM for details), 
$$
\limsup_{l\to\infty}\left|Pr_{P_{D|Z,n_l}}\left\{\hat\mu_{\alpha,n}\geq \hat\gamma_{\dagger,*}'\mu_{n_l,0}\right\}-\alpha\right|=0.
$$
The quantile unbiased estimator is closely related to our conditional test, however: the $\phi_{C,\dagger}$ test rejects if and only if $\hat\mu_{\alpha,n_l}>0$ and $\hat\eta_{\dagger} > 0$, provided that the test statistic and critical value for the $\phi_{C,\dagger}$ test are determined only by the vertices in $M_+ \Gamma_{\dagger}(X_{n_l,0}, \hat\sigma_{n_l,0})$, which we have established occurs w.p.a. 1.  Since $\hat\gamma_{\dagger,*}'\mu_{n_l,0}\le0$ under the null hypothesis, this suffices to establish that $\limsup_{l\to\infty}Pr_{P_{D|Z,n_l}}\left\{\phi_{C,\dagger}=1\right\}\le\alpha,$ as we wanted to show.  As in the proof of Lemma \ref{lem: Size limit problem}, this implies size control for the conditional test.

Next consider the hybrid test.  For $\hat{\mu}_{\alpha,n_l}^H$ the $\alpha$-quantile hybrid estimator of AKM with conditioning event $\left\{\hat\eta \le c_{\kappa,LF,\dagger}(X_{n_l,0},\widehat{\Sigma}_{n_l,0}), \hat\gamma_{\dagger} = \gamma \right\},$ Proposition 12 of AKM implies that 
$$
\limsup_{l\to\infty}\left|Pr_{P_{D|Z,n_l}}\left\{\hat\mu_{\alpha,n_l}^H\ge \hat\gamma_{\dagger,*}'\mu_{n_l,0} |\hat\eta_{\dagger} \le c_{\kappa,LF,\dagger}(X_{n_l,0},\widehat{\Sigma}_{n_l,0}), \hat\gamma_{\dagger} = \gamma\right\}-\alpha\right|Pr_{P_{D|Z,n_l}}\left\{\hat\eta _{\dagger}\le c_{\kappa,LF,\dagger}(X_{n_l,0}\widehat{\Sigma}_{n_l,0}), \hat\gamma_{\dagger} = \gamma \right\}
$$
\noindent is equal to 0. Since the vertex set is finite, it follows that 
$$
\limsup_{l\to\infty}\left|Pr_{P_{D|Z,n_l}}\left\{\hat\mu_{\alpha,n_l}^H\ge\hat\gamma_{\dagger,*}'\mu_{n_l,0} |\hat\eta_{\dagger} \le c_{\kappa,LF,\dagger}(X_{n_l,0},\widehat{\Sigma}_{n_l,0}) \right\}-\alpha\right|Pr_{P_{D|Z,n_l}}\left\{\hat\eta _{\dagger}\le c_{\kappa,LF,\dagger}(X_{n_l,0}\widehat{\Sigma}_{n_l,0}) \right\}=0.
$$
Note, however, that the $\phi_{H,\dagger}$ test rejects only if $\hat\eta_{\dagger} > c_{\kappa,LF,\dagger}$ or $\hat\mu_{\frac{\alpha-\kappa}{1-\kappa},n_l}^H>0$ (again, assuming the test is determined only by the vertices of $M_+ \Gamma_{\dagger}(X_{n_l,0}, \hat\sigma_{n_l,0})$), and $0 \geq \hat\gamma_{\dagger,*}'\mu_{n_l,0} $, so
$$
Pr_{P_{D|Z,n_l}}\left\{\phi_{H,\dagger}=1\right\}\le Pr_{P_{D|Z,n_l}}\left\{\hat\eta_{\dagger}>c_{\alpha,LF,\dagger}(X_{n_l,0},\widehat{\Sigma}_{n_l,0})\right\}+
$$
$$
Pr_{P_{D|Z,n_l}}\left\{\hat\mu_{\frac{\alpha-\kappa}{1-\kappa},n}^H \geq \hat\gamma_{\dagger,*}'\mu_{n_l,0}  |\hat\eta_{\dagger} \le c_{\alpha,LF,\dagger}(X_{n_l,0},\widehat{\Sigma}_{n_l,0})\right\}Pr_{P_{D|Z,n_l}}\left\{\hat\eta_{\dagger}\le c_{\alpha,LF,\dagger}(X_{n_l,0},\widehat{\Sigma}_{n_l,0})\right\}.
$$
 Proposition $\ref{Prop: Least Favorable Size Control-alt}$ establishes that $\liminf_{l\to\infty}Pr_{P_{D|Z,n_l}}\left\{\hat\eta_{\dagger} \le c_{\kappa,LF,\dagger}\right\}\ge 1-\kappa$, so 
$$
\limsup_{l\to\infty}Pr_{P_{D|Z,n_l}}\left\{\phi_{H,\dagger}=1\right\}\le \kappa+\frac{\alpha-\kappa}{1-\kappa}(1-\kappa)=\alpha,
$$ 
implying size control for the hybrid test.  $\Box$

\section{Non-Unique Dual Solutions \label{appendix: nonunique dual}}

We now consider the behavior of the conditional test in the finite sample normal model without assuming that the dual solution is unique.  Recall that we define $\hat\gamma$ as the argmax in the dual problem, so $\hat\gamma$ is set-valued when the dual solution is non-unique. We show that a version of the conditional test which chooses an arbitrary dual solution when there is multiplicity is well-defined with probability 1 in the finite-sample normal model and also controls size.

We first show that we can partition the set of vertices into disjoint subsets $V_1,...,V_m$ such that the set of optimal vertices is one of the $V_j$ with probability 1.

\begin{lemma} \label{lem: Finite Solution Set}
	For every $(\mu_{n,0}, X_{n,0},\Sigma_0)$, there exists a finite collection of disjoint sets $\mathbf{V}= \{V_1,...,V_m\}$ such that $V(X_{n,0},\sigma_0) = V_1 \cup ...\cup V_m$ and $Pr\{\hat\gamma \in \mathbf{V} \} = 1$ under the finite-sample normal model (\ref{eqn: finite sample normal model}).   
\end{lemma}

\paragraph{Proof of Lemma \ref{lem: Finite Solution Set}}

Let $\gamma,\tilde\gamma,\check{\gamma} \in V(X_{n,0},\sigma_0)$. Observe that $\gamma, \tilde\gamma \in \hat\gamma$ only if $\gamma'Y_{n,0} = \tilde\gamma' Y_{n,0}$. However, for $Y_{n,0} \sim N(\mu_{n,0}, \Sigma_0)$, $$Pr\{\gamma'Y_{n,0} = \tilde\gamma'Y_{n,0}\} \in \{ 0,1\} .$$ 
\noindent Moreover, $Pr\{\gamma'Y_{n,0} = \tilde\gamma'Y_{n,0}\} =1 $ and $Pr\{\gamma'Y_{n,0} = \check\gamma'Y_{n,0}\} =1$ if and only if $Pr\{\gamma'Y_{n,0} = \tilde\gamma'Y_{n,0} = \check\gamma'Y_{n,0}\} =1$. It follows that we can partition $V(X_{n,0}, \sigma_0)$ into distinct equivalence classes $V_1,...,V_m$ where $\gamma,\tilde\gamma \in V(X_{n,0},\sigma)$ are contained in the same $V_j$ if and only if $Pr\{\gamma'Y_{n,0} = \tilde\gamma'Y_{n,0}\}=1$. Towards contradiction, suppose that $Pr\{\hat\gamma \in \mathbf{V}\} < 1$. Then it must be that either (i) there exists $\gamma, \tilde\gamma \in V_j$ such that $Pr\{ \gamma \in \hat\gamma, \tilde\gamma \not\in \hat\gamma\}>0$, or (ii) there exists $\gamma \in V_j$, $\tilde\gamma \in V_{j'}$ for $j \neq j'$ such that $Pr\{ \gamma \in \hat\gamma, \tilde\gamma \in \hat\gamma\}>0$. Note, however, that $\gamma \in \hat\gamma, \tilde\gamma \not\in \hat\gamma$ only if $\gamma'Y_{n,0} \neq \tilde\gamma'Y_{n,0}$, and by construction if $\gamma,\tilde\gamma \in V_j$ then $Pr\{\gamma'Y_{n,0} \neq \tilde\gamma'Y_{n,0}\} =0$ so (i) cannot be satisfied. Likewise, $\gamma \in \hat\gamma, \tilde\gamma \in \hat\gamma$ only if $\gamma'Y_{n,0} = \tilde\gamma'Y_{n,0}$, and by construction if $\gamma \in V_j,\tilde\gamma \in V_{j'}$ then $Pr\{\gamma'Y_{n,0} = \tilde\gamma'Y_{n,0}\} =0$ so (ii) cannot be satisfied. We have thus reached a contradiction. $\Box$\\

Our next result establishes that if one computes the conditional test using the formulas for $\mathcal{V}^{lo}_{n,0},\mathcal{V}^{up}_{n,0}$ in (\ref{eq: V^lo, V^up}), then one obtains the same values regardless of which element of $V_j$ one chooses. Together with the previous lemma, this result implies that a modified version of the conditional test which chooses arbitrarily among the optimal vertices is well-defined with probability 1 in the finite sample normal model.

\begin{lemma} \label{lem: same vlo vlup}
	Let $V_1,...,V_m$ be as defined in Lemma \ref{lem: Finite Solution Set}. Suppose $Y_{n,0}$ follows the finite sample normal model (\ref{eqn: finite sample normal model}). If $\gamma_{(1)}, \gamma_{(2)} \in V_j$ for some $j$, then with probability 1 the values for $\mathcal{V}^{lo}_{n,0}$ and $\mathcal{V}^{up}_{n,0}$ given in (\ref{eq: V^lo, V^up}) are the same if one sets $\gamma = \gamma_{(1)}$ or $\gamma = \gamma_{(2)}$. 
\end{lemma}

\paragraph{Proof of Lemma \ref{lem: same vlo vlup}}
By construction, if $\gamma_{(1)}, \gamma_{(2)} \in V_j$ then $Pr\{\gamma_{(1)}'Y_{n,0} = \gamma_{(2)}'Y_{n,0}\} =1$ for $Y_{n,0} \sim N(\mu_{n,0}, \Sigma_0)$. It follows that $(\gamma_{(1)} - \gamma_{(2)})'\Sigma_0 = 0$ and $\gamma_{(1)}'\Sigma \gamma_{(1)} = \gamma_{(2)}'\Sigma \gamma_{(2)}$. It is then immediate that for any $\tilde\gamma \in V(X_{n,0}, \sigma_0)$, $\gamma_{(1)}'\Sigma_{0} \tilde\gamma = \gamma_{(2)}'\Sigma_{0} \tilde\gamma$. Note, however, that the formulas for $\mathcal{V}^{lo}_{n,0}$ and $\mathcal{V}^{up}_{n,0}$ in (\ref{eq: V^lo, V^up}) depend on $\gamma$ only through the expressions $\gamma' \Sigma_0 \gamma, \gamma'\Sigma_0 \tilde\gamma, \Sigma_0 \gamma$, and $\gamma'Y_{n,0}$. Since we have shown that with probability 1 all of these expressions obtain the same value if we set $\gamma = \gamma_{(1)}$ as if we set $\gamma = \gamma_{(2)}$, the result follows. $\Box$\\

Finally, we establish that the conditional test which chooses arbitrarily among the optimal dual vertices controls size in the finite-sample normal model. 

\begin{proposition} \label{prop: size control dual multiple} 
	Consider a version of the conditional test where the critical values are determined by the formulas for $\mathcal{V}^{lo}_{n,0},\mathcal{V}^{up}_{n,0}$ in (\ref{eq: V^lo, V^up}) setting $\gamma = h(\hat\gamma)$ for any arbitrary (possibly randomized) function $h(\cdot)$ that selects among the elements of $\hat\gamma$. Let $\phi_C^h$ denote the indicator for whether the test rejects. Then under the finite sample normal model (\ref{eqn: finite sample normal model}), $E[\phi_C^h]\leq \alpha$ whenever $\mu_{n,0} \in \mathcal{M}_{n,0}$. 
\end{proposition}

\paragraph{Proof of Proposition \ref{prop: size control dual multiple}}
Observe that the proof to Lemma \ref{lem:conditional distribution of etahat} does not rely on uniqueness of the dual, and thus the statement of Lemma \ref{lem:conditional distribution of etahat} holds replacing the conditioning event $\hat\gamma = \gamma$ with $\gamma \in \hat\gamma$. Moreover, by Lemma \ref{lem: Finite Solution Set}, there is some $j$ such that $Pr\{ 1\{\gamma \in \hat\gamma\} = 1\{\hat\gamma = V_j\} \}=1$. It follows that the statement of Lemma \ref{lem:conditional distribution of etahat} also holds if we replace the conditioning event $\hat\gamma = \gamma$ with $\hat\gamma = V_j$. Additionally, by Lemma \ref{lem: same vlo vlup}, the values of $\mathcal{V}^{lo}_{n,0},\mathcal{V}^{up}_{n,0}$ are the same for all $\gamma \in V_j$. Thus, the conclusion of Lemma \ref{lem:conditional distribution of etahat} holds if we condition on $\hat\gamma = V_j$ and replace all instances of $\gamma$ with $h(\hat\gamma)$. By the same argument as in Section \ref{subsec: conditional test} for the unique-solution case, it then follows that $E[\phi_C^h | \hat\gamma = V_j ] \leq \alpha$ for $\mu_{n,0} \in \mathcal{M}_{n,0}$. But Lemma \ref{lem: Finite Solution Set} implies that $E[\phi_C^h] = \sum_j E[\phi_C^h | \hat\gamma = V_j] P\{\hat\gamma = V_j\}$, from which unconditional size control is immediate. $\Box$\\

By analogous arguments, one can also establish that the hybrid test is well-defined with probability 1 and controls size in the finite sample normal model when there is multiplicity in the dual.

\section{Asymptotic Variance Estimation \label{sec: Variance Estimation}}

Assumption \ref{Assumption: Variance Estimator} requires the existence
of a uniformly consistent estimator $\widehat{\Omega}_{n,0}$ for the conditional
variance $\Omega\left(P_{D|Z},\beta_0\right).$ Here, we establish the uniform
consistency of the matching estimator discussed in Section \ref{subsec: implementation - Sigma} under mild conditions. For brevity, we shorthand $U_i(\beta_0)$
as $U_{i,0}$.

Following \citet{AbadieImbensZheng2014}, we consider the nearest-neighbor
variance estimator given in (\ref{eq: conditional variance estimator}). The intuition for the estimator $\widehat{\Omega}_{n,0}$ is straightforward:
provided the conditional mean and variance of $U_{i,0}$ given $Z_{i}=z$
are smooth in $z$, if $Z_{\ell_{Z}\left(i\right)}$ is close
to $Z_{i}$, then the mean and variance of $U_{i,0} | Z_i$ will be nearly the same as the mean and variance of $U_{\ell_Z(i),0}|Z_{\ell_Z(i)} $.  Hence,
the variance of $U_{i,0}-U_{\ell_{Z}\left(i\right),0}$ will be approximately
twice the variance of $U_{i,0}|Z_i$, and the approximation error will vanish
as $Z_{\ell_{Z}\left(i\right)}$ approaches $Z_{i}$. If the support
of $Z_{i}$ is compact, however, then with a large enough sample we
are guaranteed to have observations quite ``close'' to almost all
of our observations, and $\widehat{\Omega}_{n,0}$ will converge to the average
conditional variance $\Omega\left(P_{D|Z},\beta_0\right).$ The next assumption
formalizes the conditions needed for this argument.

\begin{assumption}\label{Assumption: Variance Estimation}For $\lambda_{\max}\left(A\right)$
	the maximal eigenvalue of a matrix $A$, the following conditions
	hold
	\begin{enumerate}
		\item $\left\{ Z_{i}\right\} _{i=1}^{\infty}\subseteq\mathcal{Z}$ for $\mathcal{Z}$
		a compact set
		\item $\limsup_{n\to\infty}\sup_{P_{D|Z}\in\mathcal{P}_{D|Z}}\sup_{\beta_0\in B_I(P_{D|Z})}\frac{1}{n}\sum E_{P_{D|Z}}\left[\left\|U_{i,0}\right\|^4|Z_{i}\right]$ is finite
		\item $\mu_{P_{D|Z}}\left(z,\beta_0\right)=E_{P_{D|Z}}\left[U_{i,0}|Z_{i}=z\right]$
		is Lipschitz in $z$ with Lipschitz constant uniformly bounded over
		$P_{D|Z}\in\mathcal{P}_{D|Z}$, $\beta_0\in B_I(P_{D|Z})$, and is uniformly bounded over
		$P_{D|Z}\in\mathcal{P}_{D|Z}$, $\beta_0\in B_I(P_{D|Z})$
		\item $V_{P_{D|Z}}\left(z,\beta_0\right)=E_{P_{D|Z}}\left[U_{i,0}U_{i,0}'|Z_{i}=z\right]$
		is Lipschitz in $z$ with Lipschitz constant uniformly bounded over
		$P_{D|Z}\in\mathcal{P}_{D|Z},$ $\beta_0\in B_I(P_{D|Z})$
		\item $\sup_{P_{D|Z}\in\mathcal{P}_{D|Z}}\sup_{\beta_0\in B_I(P_{D|Z})}\sup_{z\in\mathcal{Z}}\lambda_{\max}\left(Var_{P_{D|Z}}\left(U_{i,0}|Z_{i}=z\right)\right)$
		is finite
		\item For $\widehat{\Sigma}_{Z} = \widehat{{Var}}(Z_i)$ the sample variance of $Z_i$,  $\widehat{\Sigma}_{Z}\to \Sigma_Z$ for a positive-definite limit $\Sigma_Z$
	\end{enumerate}
\end{assumption}

Assumption \ref{Assumption: Variance Estimation}(1) is used only
to establish that the average distance between $Z_{i}$ and $Z_{\ell_{Z}\left(i\right)}$
converges to zero, $\frac{1}{n}\sum\left\Vert Z_{i}-Z_{\ell_{Z}\left(i\right)}\right\Vert \to0$.
Hence, one may instead assume this condition directly. Assumption
\ref{Assumption: Variance Estimation}(2) and (5) restrict the variance
and fourth moment of $U_{i,0}$, and are satisfied under a wide
range of data generating processes. Assumption \ref{Assumption: Variance Estimation}(3)
and (4) impose Lipschitz continuity on the mean and second moment
of $U_{i,0}$, consistent with the heuristic argument given above. Finally,
Assumption \ref{Assumption: Variance Estimation}(6) requires only that $\widehat{\Sigma}_{Z}$
converge to a positive-definite limit.

\begin{proposition}\label{Prop: Variance Consistency}Under Assumptions
	\ref{Assumption: Variance Convergence} and \ref{Assumption: Variance Estimation},
	for $\widehat{\Omega}_{n,0}$ as defined in (\ref{eq: conditional variance estimator}) and
	all $\varepsilon>0$ 
	\[
	\lim_{n\to\infty}~\sup_{P_{D|Z}\in\mathcal{P}_{D|Z}}~\sup_{\beta_0\in B_I(P_{D|Z})}Pr_{P_{D|Z}}\left\{ \left\Vert \widehat{\Omega}_{n,0}-\Omega\left(P_{D|Z},\beta_0\right)\right\Vert >\varepsilon\right\} =0,
	\]
	so Assumption \ref{Assumption: Variance Estimator} holds.\end{proposition}

\subsection{Proof of Variance Consistency\label{subsec: Variance Consistency}}

We first prove two auxiliary lemmas, which we then use to prove  Proposition
\ref{Prop: Variance Consistency}.

\begin{lemma}\label{lem: Second Moment Convergence} Under Assumption
	\ref{Assumption: Variance Estimation}, 
	\[
	\frac{1}{n}\sum_{i=1}^{n}\left(U_{\ell_{Z}\left(i\right),0}U_{\ell_{Z}\left(i\right),0}'-V_{P_{D|Z}}\left(Z_{i},\beta_0\right)\right)\to_{p}0
	\]
	uniformly over $P_{D|Z}\in\mathcal{P}_{D|Z},$ $\beta_0\in B_I(P_{D|Z})$.
	
\end{lemma}

\paragraph{Proof of Lemma \ref{lem: Second Moment Convergence}}

Note that we can write 
\[
\frac{1}{n}\sum_{i=1}^{n}\left(U_{\ell_{Z}\left(i\right),0}U_{\ell_{Z}\left(i\right),0}'-V_{P_{D|Z}}\left(Z_{i},\beta_0\right)\right)=
\]
\[
\frac{1}{n}\sum_{i=1}^{n}\left(U_{\ell_{Z}\left(i\right),0}U_{\ell_{Z}\left(i\right),0}'-V_{P_{D|Z}}\left(Z_{\ell_{Z}\left(i\right)},\beta_0\right)\right)+\frac{1}{n}\sum_{i=1}^{n}\left(V_{P_{D|Z}}\left(Z_{\ell_{Z}\left(i\right)},\beta_0\right)-V_{P_{D|Z}}\left(Z_{i},\beta_0\right)\right),
\]
so to prove the result it suffices to show that both terms tend to
zero. To show that the second term tends to zero, note that by the
triangle inequality and Assumption \ref{Assumption: Variance Estimation}(4),
\[
\left\Vert \frac{1}{n}\sum_{i=1}^{n}\left(V_{P_{D|Z}}\left(Z_{\ell_{Z}\left(i\right)},\beta_0\right)-V_{P_{D|Z}}\left(Z_{i},\beta_0\right)\right)\right\Vert \le\frac{1}{n}\sum_{i=1}^{n}\left\Vert V_{P_{D|Z}}\left(Z_{\ell_{Z}\left(i\right)},\beta_0\right)-V_{P_{D|Z}}\left(Z_{i},\beta_0\right)\right\Vert 
\]
\[
\le\frac{K}{n}\sum_{i=1}^{n}\left\Vert Z_{i}-Z_{\ell_{Z}\left(i\right)}\right\Vert 
\]
for $K$ the upper bound on the Lipschitz constant. Note, next, that
since $\mathcal{Z}$ is compact by Assumption \ref{Assumption: Variance Estimation}(1),
the proof of Lemma 1 of \citet{AbadieImbens2008} implies that 
\[
\frac{1}{n}\sum_{i=1}^{n}\left\Vert Z_{i}-Z_{\ell_{Z}\left(i\right)}\right\Vert \to0.
\]
Thus, we immediately see that $\frac{1}{n}\sum_{i=1}^{n}\left(V_{P_{D|Z}}\left(Z_{\ell_{Z}\left(i\right)},\beta_0\right)-V_{P_{D|Z}}\left(Z_{i},\beta_0\right)\right)\to0$
uniformly over $P_{D|Z}\in\mathcal{P}_{D|Z}$ and $\beta_0\in B_I(P_{D|Z}).$ 

We next show that 
\[
\frac{1}{n}\sum_{i=1}^{n}\left(U_{\ell_{Z}\left(i\right),0}U_{\ell_{Z}\left(i\right),0}'-V_{P_{D|Z}}\left(Z_{\ell_{Z}\left(i\right)},\beta_0\right)\right)\to_{p}0.
\]
To do so, note first that the number of observations that can be
matched to a given $Z_{i}$, $\left| \left\{ j:\ell_{Z}\left(j\right)=i\right\} \right|,$
is bounded above by the so-called ``kissing number'' which is a
finite function $\mathcal{K}\left(\dim\left(Z_{i}\right)\right)$ of the dimension
of $Z$ (see \citet{AbadieImbensZheng2014}). Since $U_{i,0}$ is independent
across $i$, this implies that for $(A)_{jk}$ the $(j,k)$ element of a matrix $A,$
\[
Var\left(\frac{1}{n}\sum_{i=1}^{n}\left(U_{\ell_{Z}\left(i\right),0}U_{\ell_{Z}\left(i\right),0}'-V_{P_{D|Z}}\left(Z_{\ell_{Z}\left(i\right)},\beta_0\right)\right)_{jk}|\left\{ Z_{i}\right\} _{i=1}^{\infty}\right)
\]
\[
\le \mathcal{K}\left(\dim\left(Z_{i}\right)\right)^2Var\left(\frac{1}{n}\sum_{i=1}^{n}\left(U_{i,0}U_{i,0}'\right)_{jk}|\left\{ Z_{i}\right\} _{i=1}^{\infty}\right)
\]
\[
=\frac{\mathcal{K}\left(\dim\left(Z_{i}\right)\right)^2}{n^{2}}\sum_{i=1}^{n}Var\left(\left(U_{i,0}U_{i,0}'\right)_{jk}|Z_{i}\right).
\]
By Assumption \ref{Assumption: Variance Estimation}(2) and Chebyshev's inequality, however, this implies that
\[
\frac{1}{n}\sum_{i=1}^{n}\left(U_{\ell_{Z}\left(i\right),0}U_{\ell_{Z}\left(i\right),0}'-V_{P_{D|Z}}\left(Z_{\ell_{Z}\left(i\right)},\beta_0\right)\right)\to_{p}0,
\]
uniformly over $P_{D|Z}\in\mathcal{P}_{D|Z}$ and $\beta_0\in B_I(P_{D|Z}),$ which completes the
proof. $\Box$

\begin{lemma}\label{lem: Outer Product Convergence} Under Assumption
	\ref{Assumption: Variance Estimation}, 
	\[
	\frac{1}{n}\sum_{i=1}^{n}\left(U_{i,0}U_{\ell_{Z}\left(i\right),0}'-\mu_{P_{D|Z}}\left(Z_{i},\beta_0\right)\mu_{P_{D|Z}}\left(Z_{i},\beta_0\right)'\right)\to_{p}0,
	\]
	uniformly over $P_{D|Z}\in\mathcal{P}_{D|Z}$ and $\beta_0\in B_I(P_{D|Z}).$
	
\end{lemma}

\paragraph{Proof of Lemma \ref{lem: Outer Product Convergence}}

Note that we can write
\[
\frac{1}{n}\sum_{i=1}^{n}\left(U_{i,0}U_{\ell_{Z}\left(i\right),0}'-\mu_{P_{D|Z}}\left(Z_{i},\beta_0\right)\mu_{P_{D|Z}}\left(Z_{i},\beta_0\right)'\right)
\]
\[
=\frac{1}{n}\sum_{i=1}^{n}\left(U_{i,0}U_{\ell_{Z}\left(i\right),0}'-\mu_{P_{D|Z}}\left(Z_{i},\beta_0\right)\mu_{P_{D|Z}}\left(Z_{\ell_{Z}\left(i\right)},\beta_0\right)'\right)
\]
\[
+\frac{1}{n}\sum_{i=1}^{n}\left(\mu_{P_{D|Z}}\left(Z_{i},\beta_0\right)\mu_{P_{D|Z}}\left(Z_{\ell_{Z}\left(i\right)},\beta_0\right)'-\mu_{P_{D|Z}}\left(Z_{i},\beta_0\right)\mu_{P_{D|Z}}\left(Z_{i},\beta_0\right)'\right).
\]
We first show the initial term converges in probability to zero, and then do the same for the second term.

By independence, 
\[
E\left[U_{i,0}U_{\ell_{Z}\left(i\right),0}'-\mu_{P_{D|Z}}\left(Z_{i},\beta_0\right)\mu_{P_{D|Z}}\left(Z_{\ell_{Z}\left(i\right)},\beta_0\right)'|Z_{i},Z_{\ell_{Z}\left(i\right)}\right]=0,
\]
while the variance of the $jk$th element is 
\[
Var_{P_{D|Z}}\left(\left(U_{i,0}U_{\ell_{Z}\left(i\right),0}'-\mu_{P_{D|Z}}\left(Z_{i},\beta_0\right)\mu_{P_{D|Z}}\left(Z_{\ell_{Z}\left(i\right)},\beta_0\right)'\right)_{jk}|Z_{i},Z_{\ell_{Z}\left(i\right)}\right)
\]
\[
=E_{P_{D|Z}}\left[\left(U_{i,0,j}U_{\ell_{Z}\left(i\right),0,k}-\mu_{P_{D|Z},j}\left(Z_{i},\beta_0\right)\mu_{P_{D|Z},k}\left(Z_{\ell_{Z}\left(i\right)},\beta_0\right)\right)^{2}|Z_{i},Z_{\ell_{Z}\left(i\right)}\right]
\]
\[
=\begin{array}{c}
	\mu_{P_{D|Z},j}^{2}\left(Z_{i},\beta_0\right)Var_{P_{D|Z}}\left(U_{\ell_{Z}\left(i\right),0,k}|Z_{\ell_{Z}\left(i\right)}\right)+Var_{P_{D|Z}}\left(U_{i,0,j}|Z_{i}\right)\mu_{P_{D|Z},k}^{2}\left(Z_{\ell_{Z}\left(i\right)},\beta_0\right)\\
	+Var_{P_{D|Z}}\left(U_{i,0,j}|Z_{i}\right)Var_{P_{D|Z}}\left(U_{\ell_{Z}\left(i\right),0,k}|Z_{\ell_{Z}\left(i\right)}\right).
\end{array}
\]
Assumption \ref{Assumption: Variance Estimation}(5) thus implies
that for some constant $C$, 
\[
\begin{array}{c}
	Var_{P_{D|Z}}\left(\left(U_{i,0}U_{\ell_{Z}\left(i\right),0}'-\mu_{P_{D|Z}}\left(Z_{i},\beta_0\right)\mu_{P_{D|Z}}\left(Z_{\ell_{Z}\left(i\right)},\beta_0\right)'\right)_{jk}|Z_{i},Z_{\ell_{Z}\left(i\right)}\right)\\
	\le\left(\mu_{P_{D|Z},j}^{2}\left(Z_{i},\beta_0\right)+\mu_{P_{D|Z},k}^{2}\left(Z_{\ell_{Z}\left(i\right)},\beta_0\right)+C\right)C
\end{array},
\]
which, together with Assumption \ref{Assumption: Variance Estimation}(3)
and the finiteness of the ``kissing number'' $\mathcal{K}\left(\dim\left(Z_{i}\right)\right)$
(see the proof of Lemma \ref{lem: Second Moment Convergence} above)
implies that 
\[
\limsup_{n\to\infty}\sup_{P_{D|Z}\in\mathcal{P}_{D|Z}}\sup_{\beta_0\in B_I(P_{D|Z})}Var\left(\frac{1}{n}\sum_{i=1}^{n}\left(U_{i,0}U_{\ell_{Z}\left(i\right),0}'-\mu_{P_{D|Z}}\left(Z_{i},\beta_0\right)\mu_{P_{D|Z}}\left(Z_{\ell_{Z}\left(i\right)},\beta_0\right)'\right)|\left\{ Z_{i}\right\} _{i=1}^{\infty}\right)=0,
\]
and thus by Chebyshev's inequality that 
\[
\frac{1}{n}\sum_{i=1}^{n}\left(U_{i,0}U_{\ell_{Z}\left(i\right),0}'-\mu_{P_{D|Z}}\left(Z_{i},\beta_0\right)\mu_{P_{D|Z}}\left(Z_{\ell_{Z}\left(i\right)},\beta_0\right)'\right)\to_{p}0,
\]
uniformly over $P_{D|Z}\in\mathcal{P}_{D|Z},$ $\beta_0\in B_I(P_{D|Z}),$ as we wanted to show.

To complete the proof, we need only show that 
\[
\frac{1}{n}\sum_{i=1}^{n}\left(\mu_{P_{D|Z}}\left(Z_{i},\beta_0\right)\mu_{P_{D|Z}}\left(Z_{\ell_{Z}\left(i\right)},\beta_0\right)'-\mu_{P_{D|Z}}\left(Z_{i},\beta_0\right)\mu_{P_{D|Z}}\left(Z_{i},\beta_0\right)'\right).
\]
converges to zero uniformly over $P_{D|Z}\in\mathcal{P}_{D|Z},$ $\beta_0\in B_I(P_{D|Z}).$  Note, however, that 
by the triangle inequality and Assumption \ref{Assumption: Variance Estimation}(3),
\[
\left\|\frac{1}{n}\sum_{i=1}^{n}\left(\mu_{P_{D|Z}}\left(Z_{i},\beta_0\right)\mu_{P_{D|Z}}\left(Z_{\ell_{Z}\left(i\right)},\beta_0\right)'-\mu_{P_{D|Z}}\left(Z_{i},\beta_0\right)\mu_{P_{D|Z}}\left(Z_{i},\beta_0\right)'\right)\right\|
\]
\[
\le
\frac{1}{n}\sum_{i=1}^{n}\left\|\mu_{P_{D|Z}}\left(Z_{i},\beta_0\right)\mu_{P_{D|Z}}\left(Z_{\ell_{Z}\left(i\right)},\beta_0\right)'-\mu_{P_{D|Z}}\left(Z_{i},\beta_0\right)\mu_{P_{D|Z}}\left(Z_{i},\beta_0\right)'\right\|
\]
\[
\le
\frac{1}{n}\sum_{i=1}^{n}\left\|\mu_{P_{D|Z}}\left(Z_{i},\beta_0\right)\right\|\cdot\left\|\mu_{P_{D|Z}}\left(Z_{\ell_{Z}\left(i\right)},\beta_0\right)-\mu_{P_{D|Z}}\left(Z_{i},\beta_0\right)\right\|
\]

\begin{equation}\label{eq: last covariance term}
	\le
	\frac{K}{n}\sum_{i=1}^{n}\left\|\mu_{P_{D|Z}}\left(Z_{i},\beta_0\right)\right\|\cdot\left\|Z_{\ell_{Z}\left(i\right)}-Z_i\right\|\le\frac{KC}{n}\sum_{i=1}^{n}\left\|Z_{\ell_{Z}\left(i\right)}-Z_i\right\|
\end{equation}
for $K$ a Lipschitz constant and $C$ a constant.
As above, since $\mathcal{Z}$ is compact by Assumption \ref{Assumption: Variance Estimation}(1),
the proof of Lemma 1 of \citet{AbadieImbens2008} implies that 
\[
\frac{1}{n}\sum_{i=1}^{n}\left\Vert Z_{i}-Z_{\ell_{Z}\left(i\right)}\right\Vert \to0,
\]
and thus that (\ref{eq: last covariance term}) converges to zero uniformly over $P_{D|Z}\in\mathcal{P}_{D|Z},$ $\beta_0\in B_I(P_{D|Z}).$
$\Box$

\paragraph{Proof of Proposition \ref{Prop: Variance Consistency}}

Following proof of Lemma A.3 in \citet{AbadieImbensZheng2014}, note that
\[
\widehat{\Omega}_{n,0}=\frac{1}{2n}\sum_{i=1}^{n}\left(U_{i,0}-U_{\ell_{Z}\left(i\right),0}\right)\left(U_{i,0}-U_{\ell_{Z}\left(i\right),0}\right)'
\]
\[
=\frac{1}{2n}\sum_{i=1}^{n}U_{i,0}U_{i,0}'+\frac{1}{2n}\sum_{i=1}^{n}U_{\ell_{Z}\left(i\right),0}U_{\ell_{Z}\left(i\right),0}'-\frac{1}{2n}\sum_{i=1}^{n}\left(U_{i,0}U_{\ell_{Z}\left(i\right),0}'+U_{\ell_{Z}\left(i\right),0}U_{i,0}'\right).
\]
Assumption \ref{Assumption: Variance Estimation}(2) together with
Chebyshev's inequality implies that 
\[
\frac{1}{2n}\sum_{i=1}^{n}\left(U_{i,0}U_{i,0}'-V_{P_{D|Z}}\left(Z_{i},\beta_0\right)\right)\to_{p}0
\]
uniformly over $P_{D|Z}\in\mathcal{P}_{D|Z},$ $\beta_0\in B_I(P_{D|Z}).$ 
Since 
\[
Var\left(U_{i,0}|Z_{i}\right)=V_{P_{D|Z}}\left(Z_{i},\beta_0\right)-\mu_{P_{D|Z}}\left(Z_{i},\beta_0\right)\mu_{P_{D|Z}}\left(Z_{i},\beta_0\right)',
\]
however, we see that 
\[
\frac{1}{n}\sum_{i=1}^nVar_{P_{D|Z}}\left(U_{i,0}|Z_{i}\right)=\frac{1}{n}\sum_{i=1}^nV_{P_{D|Z}}\left(Z_{i},\beta_0\right)-\frac{1}{n}\sum_{i=1}^n\mu_{P_{D|Z}}\left(Z_{i},\beta_0\right)\mu_{P_{D|Z}}\left(Z_{i},\beta_0\right)'.
\]
Thus, to prove that 
\[
\widehat{\Omega}_{n,0}-\frac{1}{n}\sum_{i=1}^n Var_{P_{D|Z}}\left(U_{i,0}|Z_{i}\right)\to_{p}0,
\]
it suffices to prove that 
\[
\frac{1}{n}\sum_{i=1}^n \left(U_{\ell_{Z}\left(i\right),0}U_{\ell_{Z}\left(i\right),0}'-V_{P_{D|Z}}\left(Z_{i},\beta_0\right)\right)\to_{p}0
\]
and 
\[
\frac{1}{n}\sum_{i=1}^n \left(U_{i,0}U_{\ell_{Z}\left(i\right),0}'-\mu_{P_{D|Z}}\left(Z_{i},\beta_0\right)\mu_{P_{D|Z}}\left(Z_{i},\beta_0\right)'\right)\to_{p}0,
\]
where the first statement follows from Lemma \ref{lem: Second Moment Convergence}
and the second from Lemma \ref{lem: Outer Product Convergence}. Since
\[
\frac{1}{n}\sum_{i=1}^n Var_{P_{D|Z}}\left(U_{i,0}|Z_{i}\right)-\Omega\left(P_{D|Z},\beta_0\right)\to0
\]
uniformly over $P_{D|Z}\in \mathcal{P}_{D|Z}$ and $\beta_0\in B_I(P_{D|Z})$ by Assumption \ref{Assumption: Variance Convergence},
however, the result follows by the triangle inequality. $\Box$

\section{Sufficient Conditions for Assumption \ref{Assumption: Nondegeneracy}}\label{sec: sufficient conditions for nondegeneracy}

We now provide lower-level sufficient conditions for Assumption \ref{Assumption: Nondegeneracy} for the case where the degeneracy in $\Sigma_0$ arises from moment equalities represented as inequalities, or other moment pairs which cannot bind simultaneously. This setting is similar to that in Assumption E.3.2 in Kaido et al. (2018).

\begin{assumption}\label{Assumption: Correlation Restriction} We can write $Y_{i}(\beta_0)=T U_{i}(\beta_0)+\zeta_{i}(\beta_0)$, where $\zeta_i(\beta_0)$ is non-stochastic conditional on $Z_i$, and $U_{i}(\beta_0)$ satisfies the conditions of Assumption \ref{Assumption: Variance Convergence}. Further, we can decompose $U_{n,0}=\frac{1}{\sqrt{n}}\sum U_i(\beta_0)$ as $U_{n,0}=(U_{n,0,1}',U_{n,0,2}')'$, where the matrix $T$ takes the form
\[
T=\left[\begin{array}{cc}
I_{dim(U_{n,0,1})} & 0\\
-I_{dim(U_{n,0,1})} & 0\\
0 & I_{dim(U_{n,0,2})}
\end{array}\right],
\]
while $\zeta_i(\beta_0)=[\zeta_{i1}(\beta_0)'~\zeta_{i2}(\beta_0)'~\zeta_{i3}(\beta_0)']'$ with $\zeta_{i1}(\beta_0) + \zeta_{i2}(\beta_0) \leq 0$ (elementwise).\footnote{Observe that $e_j' E[U_i(\beta_0) + \zeta_{i1} - Q\delta | Z_i] + e_j' E[-U_i(\beta_0) + \zeta_{2i} + Q \delta | Z_i] = \zeta_{1i} + \zeta_{2i} $, regardless of $E[U_i(\beta_0) |Z_i]$, and thus the null hypothesis can only possibly be satisfied if $\zeta_{i1}+\zeta_{i2} \leq 0$.} We can likewise decompose $X_{n,0} = T Q_{n,0}$ for a comformable matrix $Q_{n,0}$.
 \end{assumption}
\noindent We note that Assumption \ref{Assumption: Correlation Restriction} is trivially satisfied with $T=I$ when $E[ Var(Y_i(\beta_0)|Z_i ) ]$ is guaranteed to be full rank.

Our second primitive condition ensures that for $n$ sufficiently large, $X_{n,0}$ lies in a set on which the distance between distinct vertices of $V(X,\upsilon)$ is bounded away from zero (where $\upsilon = \sqrt{diag(TT')}$). Let $\mathcal{B}$ denote the set of $B\subset \{1,...,k+p+1\}$ with $|B| = k$ and $1 \in B$. 

\begin{assumption} \label{Assumption: Vertices Separated}
For $n$ sufficiently large and all $\beta_0$, $X_{n,0}$ is contained in a set $\mathcal{X}$ such that for some constant $\omega>0$ and any distinct $B,B' \in \mathcal{B}$, either

\begin{enumerate}
	\item $A_B(X,\upsilon)^{-1} e_1 = A_{B'}(X,\upsilon)^{-1} e_1$ for all $X \in \mathcal{X}$ such that $A_B(X,\upsilon)$ and $A_{B'}(X,\upsilon)$ are full-rank, OR
	
	\item $||A_B(X,\upsilon)^{-1} e_1 - A_{B'}(X,\upsilon)^{-1}e_1|| \geq \omega$ for all $X \in \mathcal{X}$ such that $A_B(X,\upsilon)$ and $A_{B'}(X,\upsilon)$ are full-rank
\end{enumerate}
\noindent where the matrix $A_B(X,\upsilon)$ is as defined as in Lemma \ref{lem: vertex characterization}.
\end{assumption}

\noindent Recall from Lemma \ref{lem: vertex characterization} that each vertex in $V(X,\upsilon)$ corresponds to $A_B(X,\upsilon)^{-1} e_1$ for some $B$, so Assumption \ref{Assumption: Correlation Restriction} guarantees that the distance between distinct vertices of $V(X,\upsilon)$ is bounded from below over $X \in \mathcal{X}$. We note that Assumption \ref{Assumption: Vertices Separated} is satisfied trivially if $X_{n,0}/||X_{n,0}||$ is constant, since in that case $V(X_{n,0},\upsilon)$ is constant.

\begin{proposition} \label{Prop: Low Level Conditions Sufficient}
	Assumptions \ref{Assumption: Correlation Restriction} and \ref{Assumption: Vertices Separated} imply Assumption \ref{Assumption: Nondegeneracy}.
\end{proposition}

To prove Proposition \ref{Prop: Low Level Conditions Sufficient}, we first establish some auxilliary lemmas. In the following results, we partition a vertex $\gamma \in V(X,\upsilon)$ as $(\gamma_1', \gamma_2', \gamma_3')'$ comformably with the blocks of $T$ in Assumption \ref{Assumption: Correlation Restriction}. We also define $V_{\mathcal{B}^*}(X, \upsilon) \subset V(X,\upsilon)$ to be the subset of $V(X,\upsilon)$ such that $\max\{e_j'\gamma_1, e_j'\gamma_2\}=0$ for each $j=1,...,\dim(\gamma_1)$. Intuitively, $V_{\mathcal{B}^*}(X,\upsilon)$ is the set of vertices that have at most one positive entry corresponding with each pair of matching moments of opposite signs.

\begin{lemma} \label{lem: bound on norm of gammas}
	If Assumption \ref{Assumption: Correlation Restriction} holds, then for any $\gamma, \tilde\gamma \in V_{\mathcal{B}^*}(X,\sigma)$ and $c \geq 0$, 
	$$ || (\gamma - c \cdot \tilde\gamma)'T  || \geq k^{-\frac{1}{2}} || \gamma - c \cdot \tilde\gamma || .$$

\end{lemma}
\paragraph{Proof of Lemma \ref{lem: bound on norm of gammas}}
 To establish the result, it suffices to show that 
 \begin{equation} 
 || (\gamma - c \cdot \tilde\gamma)'T  ||_{\infty} \geq || \gamma - c \cdot \tilde\gamma ||_{\infty}, \label{eqn: bound on l0 norm}
 \end{equation} where $||x||_{\infty} = \max\{|x_1|,...,|x_k| \}$ is the $\ell_\infty$ norm. The desired result then follows from the fact that for any $x \in \mathbb{R}^k$,  $||x|| \geq ||x||_\infty \geq k^{-\frac{1}{2}} ||x||$.

	Clearly, the inequality (\ref{eqn: bound on l0 norm}) holds trivially when $\gamma - c \cdot \tilde\gamma=0$, so for the remainder of the proof we consider the case where $ || \gamma - c \cdot \tilde\gamma ||_{\infty} =m > 0$. Write 
	
	\[
	(\gamma - c \cdot \tilde\gamma)'T =
	\left(\begin{array}{c}
		\gamma_{1}-\gamma_{2}\\
		\gamma_{3}
	\end{array}\right)'-c\cdot\left(\begin{array}{c}
		\tilde\gamma_{1}-\tilde\gamma_{2}\\
		\tilde\gamma_{3}
	\end{array}\right)'.
	\]
	
	\noindent It is clear from the previous display that if $| \gamma_{3,j} - c \cdot \tilde\gamma_{3,j}| =  m$ for some $j$, then $|| (\gamma - c \cdot \tilde\gamma)'T  ||_{\infty} \geq m$. Consider next the case where $| \gamma_{1,j} - c \cdot \tilde\gamma_{1,j}| = m$ for some $j$. Suppose first that $\gamma_{1,j} > c \cdot \tilde\gamma_{1,j} \geq 0$. By the definition of $V_{\mathcal{B}^*}(X,\sigma)$, this implies that $\gamma_{2,j}=0$. Hence the $j$th element of $(\gamma - c \cdot \tilde\gamma)'T$ is equal to 
	$$ \underbrace{ \gamma_{1,j} - \tilde\gamma_{1,j} }_{=m}  + \underbrace{c \cdot \tilde\gamma_{2,j} }_{\geq 0 } \geq m, $$
	
	\noindent which implies that $||(\gamma - c \cdot \tilde\gamma)'T||_\infty \geq m$. Likewise, if $ c \cdot \tilde\gamma_{1,j} > \gamma_{1,j} \geq 0$, then we know that $\tilde\gamma_{2,j}= 0$, and thus the $j$th element of $(\gamma - c\cdot \tilde\gamma)'T$ is equal to 
	$$ \underbrace{ \gamma_{1,j} - c \cdot \tilde\gamma_{1,j} }_{= -m}  - \underbrace{\gamma_{1,j} }_{\geq 0 } \leq - m, $$
	
	\noindent which implies that $||(\gamma - c \cdot \tilde\gamma)'T||_\infty \geq m$. We have thus established that $||(\gamma - c \cdot \tilde\gamma)'T||_\infty \geq m$ when $| \gamma_{1,j} - c \tilde\gamma_{1,j}| = m$ for some $j$. The case where $| \gamma_{2,j} - c \tilde\gamma_{2,j}| = m$ for some $j$ can be handled analogously. $\Box$

\begin{lemma} \label{lem: lambda bounded}
If Assumption \ref{Assumption: Correlation Restriction} holds, then there exists a constant $c_\lambda > 0$ such that $c_\lambda^{-1} \leq \lambda_j(X,\sigma(\Omega)) \leq c_{\lambda}$ for all $\Omega \in \mathbf{\Omega}_{\bar\lambda}$ and for all $j$ and $X$, where the function $\lambda_j(X,\sigma)$ is as given in Lemma \ref{lem: can consider subset of vertices}. 
\end{lemma}

\paragraph{Proof of Lemma \ref{lem: lambda bounded}}
Recall from the proof of Lemma \ref{lem: Gamma structure} that $ \lambda_j(X,\sigma) = 1/((A_B(X,\upsilon)^{-1} e_1 )'\sigma(\Omega) ) $ for some index set $B$. Since by construction $(A_{B}(X,\upsilon)^{-1} e_1)'\upsilon =1$, we have that $$  \lambda_j(X,\sigma) = \frac{ (A_B(X,\upsilon)^{-1} e_1 )' \upsilon   }{(A_B(X,\upsilon)^{-1} e_1 )'\sigma(\Omega)  }. $$
\noindent Since $A_B(X,\upsilon)^{-1} e_1$, $\upsilon$, and $\sigma(\Omega)$ are all non-negative vectors by construction, it thus suffices to establish that $c_\lambda^{-1} \upsilon \leq  \sigma(\Omega) \leq  c_\lambda \upsilon$ (where the inequalities hold elementwise). Observe, however, that $\upsilon_{j} = ||T_j||$, whereas $\sigma(\Omega)_j = \sqrt{T_j \Omega T_j'}$. However, since the eigenvalues of $\Omega$ are bounded above and below by $\bar{\lambda}$ and $\bar{\lambda}^{-1}$ respectively, we have that for every $j$, $||T_j||^2  \bar{\lambda}^{-1} \leq T_j \Omega T_j' \leq \bar{\lambda} ||T_j||^2$, and hence $c_\lambda^{-1} v_j \leq \sigma(\Omega)_j \leq c_\lambda v_j$ for $c_\lambda =\bar{\lambda}^{\frac{1}{2}}$. 
$\Box$

\paragraph{Proof of Proposition \ref{Prop: Low Level Conditions Sufficient}}

First, we show that $V^{\dagger}(X,\sigma) \subseteq V_{\mathcal{B}_*}(X, \sigma)$ for all $\sigma$. Suppose that  $\gamma \in V^{\dagger}(X,\sigma)$. By part 1 of Lemma \ref{lem: can consider subset of vertices}, $\gamma = \lambda(\sigma) \bar\gamma$ for a scalar function $\lambda(\sigma)$ and vector $\bar\gamma$ (both depending on $X$). Under the structure imposed by Assumption \ref{Assumption: Correlation Restriction}, the fact that $\gamma \in V^{\dagger}(X,\sigma)$ implies that for some $\tilde\sigma$,  $\tilde\gamma = \lambda(\tilde\sigma) \bar\gamma$ is a Lagrange multiplier for the primal linear program 
$$
\hat\eta = \min_{\eta,\delta}\hspace{.1cm}\eta \text{ subject to } \left(Tu+\left(\begin{array}{ccc}
	\zeta_1' & \zeta_2' & \zeta_3'\end{array}\right)'-T Q \delta \le \eta\cdot \tilde\sigma\right)
$$

\noindent for some $u$ such that $\hat\eta>0$. Observe, however, that the constraints in the linear program corresponding with $\tilde\gamma_{1,j}$ and $\tilde\gamma_{2,j}$ can bind simultaneously only if 
$$e_j'(u - Q \delta^*) + e_j' \zeta_1 = \hat\eta e_j'\tilde\sigma = -e_j'(u - Q\delta^*) + e_j' \zeta_2,$$
\noindent for $\delta^*$ an optimizer to the linear program for $\hat\eta$. This implies that $\hat\eta = \frac{1}{2e_j'\tilde\sigma} e_j'(\zeta_1 + \zeta_2)\leq 0$. Since $\hat\eta > 0$, it must be that at most one of the moments corresponding with $\tilde\gamma_{1,j}$ and $\tilde\gamma_{2,j}$  is binding. Hence, complementary slackness implies that $\min\{e_j'\tilde\gamma_1, e_j'\tilde\gamma_{2} \} = 0$, and thus that $\min\{e_j'\gamma_1, e_j'\gamma_{2} \} = 0$ since $\gamma \propto \tilde\gamma$. It follows that $\gamma \in V_{\mathcal{B}_*}(X, \sigma)$, as we wished to show. 

Next, note that since every $\Omega \in \mathbf{\Omega}_{\bar\lambda}$ has eigenvalues bounded below by assumption, Assumption \ref{Assumption: Nondegeneracy} can fail only if there exists a sequence of $\Omega_m \in \mathbf{\Omega}_{\bar\lambda}$, $X_m \in \mathcal{X}$, distinct vertices $\gamma_m, \tilde\gamma_m \in V_{\dagger}(X_m, \sigma(\Omega_m))$, and values $c_m \geq 0$ such that $|| (\gamma_m - c_m \cdot \tilde \gamma_m)'T|| \rightarrow 0$ as $m\rightarrow \infty$. From Lemma \ref{lem: bound on norm of gammas} combined with the argument in the previous paragraph, it follows that Assumption  \ref{Assumption: Nondegeneracy} can fail only if there exist a sequence of distinct vertices $\gamma_m, \tilde\gamma_m \in V_{\mathcal{B}^*}(X_m, \sigma(\Omega_m))$ and values $c_m \geq 0$ such that $|| \gamma_m - c_m \cdot \tilde \gamma_m|| \rightarrow 0$ as $m\rightarrow \infty$. Towards contradiction, suppose that such a sequence exists. Since by construction $\gamma_m' \sigma_m = \tilde\gamma_m'\sigma_m = 1$, where $\sigma_m = \sigma(\Omega_m)$, we have that $| \sigma_m' (\gamma_m - c_m \cdot \tilde\gamma_m) | = |1-c_m| $. By the Cauchy-Schwarz inequality, it follows that $|| \gamma_m - c_m \cdot \tilde\gamma_m || \geq |1-c_m| / ||\sigma_m||$. However, since $\Omega_m$ has eigenvalues bounded above, $||\sigma_m||$ is bounded above, and thus it must be that $c_m \rightarrow 1$. Note further that $\sigma_{m,j}^2 = T_j \Omega_m T_j'$, where by Assumption \ref{Assumption: Correlation Restriction}, $||T_j||=1$, and thus $\sigma_{m,j}^2 \geq \bar{\lambda}^{-1}$. Since the elements of $\sigma_m > 0$ are bounded away from zero while $\gamma_m,\tilde\gamma_m\geq0$ and $\gamma_m'\sigma_m = \tilde\gamma'\sigma_m = 1$, we know that $||\gamma_m||$ and $||\tilde\gamma_m||$ are both bounded above. It follows that we can find a convergent subsequence indexed by $r$ such that $\gamma_r \rightarrow \gamma$. This, together with the fact that $|| \gamma_r - c_r \cdot \tilde \gamma_r|| \rightarrow 0$ and $c_r \rightarrow 1$ implies that $\tilde\gamma_r \rightarrow \gamma$ as well. Thus, we see that Assumption 4 can be violated only if we can find a sequence of distinct vertices $\gamma_r$ and $\tilde\gamma_r$ in $V_{\mathcal{B^*}}(X_r, \sigma_r)$ such that $\gamma_r - \tilde\gamma_r \rightarrow 0 $. 

The fact that $\gamma_r - \tilde\gamma_r \rightarrow 0 $ further implies that there exist a sequence of distinct vertices $\vartheta_s$ and $\tilde{v}_s$ in $V_{\mathcal{B^*}}(X_s, \upsilon)$ such that $\vartheta_s - \tilde{\vartheta}_s \rightarrow 0$. To see this, recall that we can write $\gamma_r = \lambda_{B_r}(X_r, \sigma_r) \gamma_{B_r}(X_r, \upsilon)$, where $\gamma_{B_r}(X, \upsilon) = A_{B_r}(X, \upsilon)^{-1} e_1$ and $\lambda_B(\cdot,\cdot)$ is a scalar which we showed to be bounded both above and away from zero in Lemma \ref{lem: lambda bounded}. Since the set of possible values for $B_r$ is finite, we can extract a subsequence $r_1$ on which $B_{r_1}$ is constant. We can likewise extract a further subsequence $r_2$ on which $\tilde{B}_{r_2}$ is constant, where $\tilde{B}_r$ is defined analogously to $B_r$, i.e. $\tilde\gamma_r = \lambda_{\tilde B_r}(X_r, \sigma_r) \gamma_{\tilde B_r}(X_r, \upsilon)$. Since the values of the $\lambda(\cdot)$ functions are bounded both above and away from zero, we can extract a further subsequence $s$ along which $\lambda_{B_s}(X_s, \sigma_s) \rightarrow \lambda^* >0$ and $\lambda_{\tilde{B}_s}(X_s, \sigma_s) \rightarrow \tilde\lambda^* >0$. Since $\gamma_s \rightarrow \gamma $ and $\lambda_{B_s}(X_s, \sigma_s) \rightarrow \lambda^*$, it follows that $\vartheta_s =  \gamma_{B_s}(X_s, \upsilon) \rightarrow \frac{1}{\lambda^*} \gamma$. Likewise, we have that $\tilde{\vartheta}_s =  \gamma_{\tilde B_s}(X_s, \upsilon) \rightarrow  \frac{1}{\tilde\lambda^*}\gamma$. However, by construction $\vartheta_s'\upsilon = \tilde{\vartheta}_s'\upsilon = 1$, which implies $$1 = \lim_{s\rightarrow \infty} \vartheta_s' \upsilon =  \frac{1}{\lambda^*} \gamma' \upsilon = \lim_{s\rightarrow \infty} \tilde{\vartheta}_s \upsilon =  \frac{1}{\tilde{\lambda}^*} \gamma' \upsilon,$$
\noindent and hence $\lambda^* = \tilde{\lambda}^*$. It follows that $\vartheta_s - \tilde{\vartheta}_s \rightarrow 0$.

However, by construction $\vartheta_s = A_B(X_s, \upsilon)^{-1} e_1$ and $\tilde\vartheta_s = A_{\tilde{B}}(X_s,\upsilon)^{-1} e_1$ with $\vartheta_s \neq \tilde\vartheta_s$. It follows that $|| A_B(X_s, \upsilon)^{-1} e_1 -  A_{\tilde{B}}(X_s, \upsilon)^{-1} e_1|| \rightarrow 0 $, which contradicts Assumption \ref{Assumption: Vertices Separated}. $\Box$

\section{Computation of $\mathcal{V}^{lo}_{n,0}$ and $\mathcal{V}^{up}_{n,0}$\label{sec: vlo vup computation}}

We now provide additional details on the computation of the truncation points $\mathcal{V}^{lo}_{n,0}$ and $\mathcal{V}^{up}_{n,0}$ for the conditional and hybrid tests. Equation (\ref{eq: V^lo, V^up}) gives formulas for $\mathcal{V}^{lo}_{n,0}$ and $\mathcal{V}^{up}_{n,0}$ that require taking a maximum/minimum over all of the dual vertices, which may be computationally challenging in practice. To facilitate computation, we provide two results which together allow for rapid calculation of these endpoints even when the number of dual vertices is large.

Our first result provides conditions under which $\mathcal{V}^{lo}_{n,0}$ and $\mathcal{V}^{up}_{n,0}$ can be calculated as the maximum/minimum over sets with at most $k$ elements. 

\begin{lemma} \label{lem: vlo vup primal}
	Suppose the primal problem (\ref{eq: Projection Linear Program}) has a solution $(\eta^*,\delta^*)$. Let $B \subset \{1,...,k\}$ denote the set of binding moments at $(\eta^*,\delta^*)$.\footnote{That is, $Y_{n,0,B}  - X_{n,0,B}\delta^* = \eta^* \cdot \hat\sigma_{n,0,B}$ and $Y_{n,0,-B}  - X_{n,0,-B}\delta^* < \eta^* \cdot \hat\sigma_{n,0,-B}$, where we use the notation $-B$ to denote rows not contained in $B$.} Let $W_{n,0} = (\widehat\sigma_{n,0},~X_{n,0})$ and let $M_B$ be the matrix so that $M_B W_{n,0}$ selects the rows of $W_{n,0}$ corresponding with the index set $B$. If $|B| = p+1$, $W_{n,0,B}$ is invertible (i.e., the primal solution is non-degenerate), and $e_1' W_{n,0,B}^{-1} \geq 0$, then the vector $\gamma$ with $M_B \gamma = (e_1' W_{n,0,B}^{-1})'$ and remaining elements equal to 0 is a solution to the dual problem. Moreover, for $L = (I - W_{n,0} W_{n,0,B}^{-1} M_B )$ and $\Delta = \widehat\Sigma_{n,0} \gamma / (\gamma' \widehat\Sigma_{n,0} \gamma)$, we have that 
	\begin{equation}\label{eq: V^lo primal}
		\mathcal{V}^{lo}_{n,0}= \max_{j:\left(L\Delta\right)_j<0}-\frac{\left(L S_{n,0,\gamma} \right)_j}{\left(L \Delta\right)_j} \,\,\, \text{ and } \,\,\, \mathcal{V}^{up}_{n,0}= \min_{j:\left(L\Delta\right)_j>0}-\frac{\left(L S_{n,0,\gamma} \right)_j}{\left(L \Delta\right)_j}
	\end{equation}
\noindent for $\mathcal{V}^{lo}_{n,0}, \mathcal{V}^{up}_{n,0}$ as defined in (\ref{eq: V^lo, V^up}).
\end{lemma}

%
%

\paragraph{Proof of Lemma \ref{lem: vlo vup primal}}
It is straightforward to verify that $\gamma$ satisfies the Karush-Kuhn-Tucker (KKT) conditions at $(\eta^*,\delta^*)$. The KKT conditions are necessary and sufficient for the solution to a linear program, and thus $\gamma$ is a solution to the dual problem. (In fact, if the primal is non-degenerate, then the dual is unique \citep[e.g.][Theorem 1(v)]{wachsmuth_licq_2013}, so $\gamma$ must be the unique dual solution, $\hat\gamma = \gamma$.) Observe that when $(\eta^*,\delta^*)$ is a solution to the primal problem with rows indexed by $B$ binding, then $(\eta^*,\delta^{*\prime})' = W_{n,0,B}^{-1} M_B Y_{n,0}$. Since the KKT conditions are necessary and sufficient, it follows that $\gamma^{\prime}y = \max_{\tilde\gamma \in V(X_{n,0}, \hat\sigma_{n,0})} \tilde\gamma' y$ if and only if $L y = y - W_{n,0} W_{n,0,B}^{-1} M_B y \leq 0$. But we argued in the proof to Lemma \ref{lem: can consider subset of vertices} that when $\hat\gamma = \gamma$, $\mathcal{V}^{lo}_{n,0}$ and $\mathcal{V}^{up}_{n,0}$ are respectively the minimum and maximum of the set $$\left\{\gamma^{\prime} y \,|\, y \text{ s.t. } \gamma^{\prime}y \geq \max_{\tilde\gamma \in V(X_{n,0}, \hat\sigma )} \tilde\gamma'y \text{ and } S(y,\gamma) = S_{n,0,\gamma} \right\} ,$$ which by the preceeding argument is equivalent to the set $$ \left\{ \gamma^{\prime}y \,|\, y \text{ s.t. } L y \leq 0 \text{ and } S(y,\gamma) = S_{n,0,\gamma}\right\}.$$ The result then follows from Lemma 5.1 in \citet{Leeetal2016}. $\Box$

Since the dual-simplex method naturally returns the solution $\eta^*$ and optimizer $\delta^*$, it is straightforward to verify that $W_{n,0,B}$ is invertible and $e_1'W_{n,0,B}^{-1} \geq 0$. If these conditions are met, then $\mathcal{V}^{lo}_{n,0}, \mathcal{V}^{up}_{n,0}$ can be calculated using (\ref{eq: V^lo primal}), which is computationally straightforward  since it involves a maximum/minimum over sets of at most $k$ elements. For cases where the conditions for Lemma \ref{lem: vlo vup primal} are not met, the following result provides a useful alternative method for computing $\mathcal{V}^{lo}_{n,0}, \mathcal{V}^{up}_{n,0}$. 

\begin{lemma} \label{lem: bisection for vlo vup}
	Suppose $\gamma$ is a solution to the dual problem and $\gamma' \widehat{\Sigma}_{n,0} \gamma > 0$. Then the values of $\mathcal{V}^{lo}_{n,0}$ and $\mathcal{V}^{up}_{n,0}$ associated with $\gamma$ correspond, respectively, to the minimum and maximum of the convex set
	$$C = \left\{c \,|\, c = \max_{\tilde\gamma \in V(X_{n,0},  \hat\sigma_{n,0} )}\tilde\gamma' \left(S_{n,0,\gamma}  + \frac{c}{\gamma' \widehat\Sigma_{n,0} \gamma} \widehat\Sigma_{n,0} \gamma   \right) \right\} .$$
\end{lemma}	

\paragraph{Proof of Lemma \ref{lem: bisection for vlo vup}}
Recall that the values of $\mathcal{V}^{lo}_{n,0}$ and $\mathcal{V}^{up}_{n,0}$ associated with $\gamma$ are the minimum and maximum of the set
$$\tilde C = \left\{\gamma' y \,|\, y \text{ s.t. } \gamma'y \geq \max_{\tilde\gamma \in V(X_{n,0}, \hat{\sigma}_{n,0} )} \tilde\gamma'y \text{ and } S(y,\gamma) = S_{n,0,\gamma} \right\} .$$
\noindent From the definition of $S(y, \gamma) = \left(I - \left(\gamma' \widehat{\Sigma}_{n,0} \gamma\right)^{-1} \widehat{\Sigma}_{n,0} \gamma \gamma'\right)y$, we have that $y = S(y, \gamma) + (\gamma'y) /\left(\gamma' \widehat{\Sigma}_{n,0} \gamma\right) \cdot  \widehat{\Sigma}_{n,0}  \gamma $, from which it follows that
$$\tilde C = \left\{\gamma' y \,|\, y \text{ s.t. } \gamma'y \geq \max_{\tilde\gamma \in V(X_{n,0}, \hat{\sigma}_{n,0} )} \tilde\gamma' \left( S_{n,0,\gamma}  + \frac{\gamma'y}{\gamma' \widehat\Sigma_{n,0} \gamma} \widehat\Sigma_{n,0} \gamma  \right) \text{ and } S(y,\gamma) = S_{n,0,\gamma} \right\} .$$

\noindent To establish that $\tilde{C} = C$, it thus suffices to show that $\{ \gamma'y \,|\,  S(y,\gamma) = S_{n,0,\gamma} \} = \mathbb{R}$, which follows from the assumption that $\gamma' \widehat{\Sigma}_{n,0} \gamma >0$ along with the fact that if $S(y,\gamma) = s$ then $S\left(y + a \cdot \widehat{\Sigma}_{n,0} \gamma,  \gamma \right) =s $ for any $a \in \reals$ (which follows immediately from the definition of $S(y,\gamma)$). Finally, convexity follows immediately from the form of $\tilde C$ and the fact that $\max_{\tilde\gamma \in V(X_{n,0}, \hat\sigma_{n,0} )} \tilde\gamma'y $ is convex in $y$.  
$\Box$

Lemma \ref{lem: bisection for vlo vup} implies that $\mathcal{V}^{lo}_{n,0}, \mathcal{V}^{up}_{n,0}$ can be calculated via a bisection method. The intuition for the algorithm is as follows. By construction, $\hat\eta_{n,0} \in C$. If there is some large value $M$ such that $M \not\in C$, then we know that $\mathcal{V}^{up}_{n,0}$ lies between $\hat\eta_{n,0}$ and $M$. We start by testing whether the midpoint between $\hat\eta_{n,0}$ and $M$ falls in the set $C$ by solving the linear program in the definition of $C$. If this point lies within $C$, then we can test the midpoint between the previously tested value and $M$, whereas if it does not, then we can test the midpoint between $\hat\eta_{n,0}$ and the previous midpoint. We can proceed in this way to narrow down the range in which $\mathcal{V}^{up}_{n,0}$ must fall. This tends to be computationally efficient, since the range in which $\mathcal{V}^{up}_{n,0}$ can lie is reduced by a factor of 2 in each step. Algorithm \ref{algo: bisection} below formally describes the algorithm used for bisection (and is implemented in our Matlab code). We recommend initializing the value of $M$ to some large value such that, for computational purposes, if $\mathcal{V}^{up}_{n,0} > M$ then it would suffice to set $\mathcal{V}^{up}_{n,0} = \infty$.\footnote{In our implementation, we set $M = \max \left(100,\hat{\eta}_{n,0} + 20 \sqrt{\gamma' \widehat \Sigma \gamma} \right)$, which guarantees that $M$ is at least 20 standard deviations above $\hat \eta_{n,0}$.} Note that the formulas in Lemma \ref{lem: bisection for vlo vup} require knowledge of a dual solution $\gamma$. Fortunately, the dual-simplex method returns a dual solution by default, and thus $\gamma$ can be obtained at no additional computational cost. 

We note that whenever the conditions of Lemma \ref{lem: vlo vup primal} are met, the dual solution is unique, since non-degeneracy in the primal implies uniqueness in the dual \citep[e.g.][Theorem 1(v)]{wachsmuth_licq_2013}. If the conditions of Lemma \ref{lem: vlo vup primal} are not met, then the dual may or may not be unique. A researcher interested in testing whether the dual is unique can use the algorithm suggested by \citet{appa_uniqueness_2002} to verify the uniqueness of a linear program. We note, however, that as described in Appendix \ref{appendix: nonunique dual}, uniqueness of the dual is not needed for the validity of the our tests in the finite-sample normal model. Tests based on the formulas given in Lemma \ref{lem: bisection for vlo vup} using an arbitrarily-chosen dual solution therefore remain valid in the finite-sample normal model. Our conditions for asymptotic size control do imply, however, that the dual will be unique with probability tending to one.

\begin{algorithm}
	\caption{Bisection Method for Calculating $V^{up}_{n,0}$}
	\label{algo: bisection}
	\begin{algorithmic}[1]
		
		\Procedure{computeVUP}{} 
		
		\If{CheckIfInC(M) } 
		\State $V^{up}_{n,0} \gets \infty$		
		\Else
		
		\State $lb \gets \hat\eta_{n,0}$
		\State $ub \gets M$

		\While{$ub - lb > TolV$} 
		\State mid $\gets \frac{1}{2}(lb + ub)$
		
		\If{  CheckIfInC(mid) }
		\State $lb \gets$ mid
		\Else
		\State $ub \gets$ mid
		\EndIf
		\EndWhile
		
		\State $V^{up}_{n,0} \gets \frac{1}{2} (lb + ub)$
		
		\EndIf
		\EndProcedure
	\end{algorithmic}	
	
	where we define the functions:
	\begin{algorithmic}[1] 		
		\Function{LPValue}{c}
		
		\State \Return $$ \begin{array}{c}
			\max_{\tilde\gamma}\tilde\gamma'\left(S_{n,0,\gamma} +\frac{\widehat\Sigma_{n,0}{\gamma}}{{\gamma}'\widehat\Sigma_{n,0} {\gamma}}c\right)\\
			\mbox{subject to }\tilde\gamma\ge0,\,\,W_{n,0}'\tilde\gamma=e_{1}
		\end{array}$$
		\EndFunction

		\Function{CheckIfInC}{c}
		\If{| $c - LPValue(c) | < TolLP$ } 
		\State \Return True
		\Else
		\State \Return False	
		\EndIf
		
		\EndFunction \\

	\end{algorithmic}		
	
\end{algorithm}

\clearpage

\section{Connections to LICQ \label{sec: LICQ}}

We now briefly discuss the connections and differences between Assumption \ref{Assumption: Nondegeneracy} and linear independence constraint qualification (LICQ) conditions that have been imposed in the literature. We refer the reader to \citet{kaido_constraint_2021} for detailed discussion of constraint qualifications in the moment inequality literature, and Section 3 of \citet{rambachan_more_2022} for additional results for our conditional test under LICQ. 

 We  focus on the special case where the target parameter is scalar $(\beta \in \mathbb{R})$ and enters the moments linearly, which simplifies exposition and facilitates comparisons to other papers that consider the LICQ or closely related assumptions in the linear case \citep[e.g.][]{cho_simple_2021, Garfarov2016, kaido_asymptotically_2014}. That is, we consider moments of the form $Y_i - X_{i,\beta} \beta - X_{i,\delta} \delta$, where $Y_i \in \reals^k$, $X_{i,\beta} \in \mathbb{R}^k$, $X_{i,\delta} \in \reals^{k \times p }$, and $(Y_i,X_{i,\delta},X_{i,\beta})$ doesn't depend on $\beta$ or $\delta$.  

To give a formal definition of LICQ, we introduce the following notation. Let $X_i = (X_{i,\beta}, X_{i,\delta})$ and $\tau = (\beta, \delta')'$, so that we can write the moments as $Y_i - X_i \tau$. Define $\mathbb{T} = \{ \tau \,|\, E_P[Y_i - X_i \tau] \leq 0 \}$ to be the set of values for $\tau$ such that the unconditional moments are satisfied, and define the set of support points in direction $p$ by $S(p) = \{ \tau \,|\, p'\tau = \sup_{\tilde\tau \in \mathbb{T} } p'\tilde\tau \}$. We will be most interested in the support points in the directions $e_1$ and $-e_1$, so that the optimization in the definition of $S(p)$ corresponds with the upper and lower bounds for $\beta$. We say that LICQ holds in the direction $p$ if for all $\tau^* \in S(p)$, the matrix $X_{B}$ has full row rank, where $X= E_P[X_i]$ and $B$ is the set of rows such that $E_P[Y_{i,B} - X_{i,B} \tau^*] = 0$.\footnote{LICQ is typically defined in terms of the Jacobian of the expectation of the moments with respect to $\tau$, but in our linear setting the Jacobian of $E_P[Y_i - X_i \tau ]$ is simply $-X$.}  

We now show that LICQ implies uniqueness in a ``population version'' of the dual problem for our test statistic. Specifically, for any $\sigma \in \mathbb{R}^k$ with $\sigma> 0$, let $$\eta(Y,X,\beta, \sigma) = \min_{\eta,\delta} \hspace{.1cm} \eta \text{ s.t. } Y - X_\beta \beta - X_\delta \delta \leq \sigma \cdot \eta.$$

\noindent We then have the following result for the dual problem to $\eta(Y,X,\beta, \sigma)$.

\begin{lemma}\label{lem: LICQ implies degeneracy}
Let $\beta^{ub} = \sup_{\tau \in \mathbb{T}} e_1'\tau$ and $\mu = E_P[Y_i]$. If LICQ holds in the direction $e_1$, then for any $\sigma>0$, $\eta(\mu, X, \beta^{ub}, \sigma)$ has a unique dual solution, i.e. there is a unique solution to $$\max_{\gamma \in V(X_\delta,\sigma)} \gamma' (\mu- X_\beta \beta^{ub}).$$
\end{lemma}

\noindent 
\paragraph{Proof of Lemma \ref{lem: LICQ implies degeneracy}}

We first show that $\eta(\mu,X,\beta^{ub},\sigma)=0$. Since $\beta^{ub} = \sup_{\tau \in \mathbb{T}} e_1'\tau$ by definition, we must have that $\eta(\mu,X,\beta^{ub},\sigma) \leq 0$. Towards contradiction, suppose that $\eta(\mu,X,\beta^{ub},\sigma) <0$. Then there exists $\delta^*$ such that $\mu - X_\beta \beta^{ub} - X_\delta \delta^* < 0$. But then for some $\epsilon>0$, $\mu - X_\beta (\beta^{ub}+\epsilon) - X_\delta \delta^* < 0$, which is a contradiction, since it implies that $\sup_{\tau \in \mathbb{T}} e_1' \tau > \beta$. 

We thus see that if $\delta^*$ is a solution for $\eta(\mu, X, \beta^{ub},\sigma)$, then $(\beta^{ub}, \delta^{*\prime})' \in S(e_1)$. Hence, LICQ implies that for $B$ the set of binding moments at $\delta^*$, we have that $X_B = (X_{\beta,B} \,,\, X_{\delta,B})$ has rank $|B|$.  It follows that $X_{\delta,B}$ has rank  $|B|-1$. However, observe that there can be no $\tilde \delta$ such that $X_{\delta,B} \tilde\delta > 0$, since if there were, then for $\epsilon >0$ sufficiently small we would have that $\mu_B - X_{\beta,B} \beta^{ub}  - X_{\delta, B} (\delta^* + \epsilon \tilde\delta) <0 $ while the remaining moments are still slack, and thus $\eta(\mu, X, \beta^{ub},\sigma) <0 $. Since $\sigma_B >0$, it follows that $W_B = (\sigma_B , X_{\delta,B})$ has rank $|B|$. Note that $W_B$ is the gradient of the binding constraints at the optimum to $\eta(\mu,X,\beta^{ub},\sigma)$. Since the gradient of the binding constraints has full-rank, Theorem 1(v) in \citet{wachsmuth_licq_2013} implies that $\eta(\mu,X,\beta^{ub},\sigma)$ has a unique Lagrangian, i.e. a unique dual solution.	
$\Box$

It is worth noting that uniqueness of $\max_{\gamma \in V(X_\delta,\sigma)} \gamma' (\mu- X_\beta \beta^{ub})$ can imply restrictions on the possible values of $\mu$ --- for example, if $X_\delta=0$ and $X_\beta = \sigma = \iota$, then it implies that $\mu$ has a unique maximal element. By comparison, Assumption \ref{Assumption: Nondegeneracy} implies that with probability approaching 1, the \textit{sample} dual problem (i.e., the dual to $\eta(Y_{n,0}, X_{n,0}, \beta_0, \hat\sigma_{n,0} )$) has a unique solution. When $X_\delta=0$ and $X_\beta = \sigma = \iota$, this is satisfied if $\Sigma$ is full-rank, regardless of the value of $\mu$. More generally, as shown in Section \ref{sec: sufficient conditions for nondegeneracy}, for a wide variety of settings Assumption  \ref{Assumption: Nondegeneracy} can be guaranteed to holds under restrictions on $X_{n,0}$ and $\Sigma$ only, without imposing restrictions on $\mu$.

\section{Simulation Details\label{sec: Monte Carlo Appendix}}

\subsection{Moment Inequality Specification}

We adopt the notation of Example 3 in the main text, so $J_{f,i,t}$ is the set of products marketed by firm $f$ in
market $i$ in period $t,$ and $\Delta \pi(J_{f,i,t},J^{\prime}_{f,i,t})$ is the difference in expected profits from marketing $J_{f,i,t}$ rather then $J^{\prime}_{f,i,t}$.  Following \cite{Wollmann}, and as discussed in the main text, the fixed cost to firm $f$ of marketing product $j$ at time $t$ is $\beta(\delta_{c,f}+\delta_g g_j)$ if the product was marketed last year ($j \in J_{f,i,t-1}$), and $\delta_{c,f}+\delta_g g_j$ otherwise.  Here $\delta_{c,f}$ is a per-product cost which is constant across products but may differ across firms, while $g_j$ is the gross weight rating of product $j$.

If we begin with the case where fixed costs are constant across firms ($\delta_{c,f}=\delta_c$ for all $f$) and again let $1\{\cdot\}$ denote the indicator function,  we obtain four conditional moment inequalities by adding and subtracting one product at a time from the set marketed.  For instance, similar to the Example 3, if firm $f$ markets product $j$ at both $t-1$ and $t$, then for
$$
m^1(\theta)_{j,f,i,t} \equiv -\left[\Delta \pi (J_{f,i,t},J_{f,i,t}\setminus j) - (\delta_c +\delta_g g_j)\beta\right] \times 1\left\{j\in J_{f,i,t}, j \in J_{f,i,t-1}\right\},
$$
we must have
$E\left[m^1(\theta)_{j,f,i,t}|V_{f,i,t}\right]\le 0$
for all variables $V_{f,i,t}$ in the firm's information set when time-$t$ production decisions were made, since otherwise the firm would have chosen not to market product $j$ in period $t.$  We can analogously obtain moments $m^2(\theta)_{j,f,i,t},..., m^4(\theta)_{j,f,i,t}$ corresponding with the cases where a firm markets product $j$ only at period $t$, only at period $t-1$, or in neither period.

We obtain two further conditional moment inequalities by considering the case where a firm markets a product of a given weight $g_j$ but not a higher or lower weight $g_{j'}$. For example, we obtain the moment 
	\begin{equation*}
			m^5_{j,f,i,t}(\theta) \equiv 
			\end{equation*}
			\begin{equation*}
			-\left( \frac{\sum_{j' \in J^-(j,f,i,t)} \left[\Delta \pi (J_{f,i,t}, (J_{f,i,t} \setminus j) \cup j') - \delta_g (g_j - g_{j'})\right]} {\# J^-(j,f,i,t)}\right)   \times 1\left\{j\in J_{f,i,t}, j \notin J_{f,i,t-1}\right\},		
	\end{equation*}
\noindent where $J^{-}(j,f,i,t)$ is the set of products not marketed by firm $f$ at time $t$ or $t-1$ with weight below $g_j$. We likewise construct a moment for heavier products that were not marketed. 

 As in Wollmann, there are nine firms ($F=9$).  To generate data we model the expected  and observed profits for firm $f$ from marketing product $j$ in market $i$ in period $t$, denoted by $\pi^*_{j,f,i,t}$ and
$\pi_{j,f,i,t }$ respectively, as 
$$
\pi^*_{j,f,i,t} =  \eta_{j,i,t} + \epsilon_{j,f,i,t}, \;\;  \hbox{and} \;\; \pi_{j,f,i,t} = \pi^*_{j,f,i,t} + \nu_{j,i,t} + \nu_{j,f,i,t} ,
$$
\noindent
where the $\nu$ terms are mean zero disturbances that arise from expectational and measurement error and the  $\eta$ and $\epsilon$ terms represent product-, market-, and firm-specific profit shifters known to the firm when marketing decisions are made. The distributions of these errors are calibrated to match moments in Wollmann's data, as described in the next section.\footnote{The terms $\eta_{j,i,t}$ and $\nu_{j,i,t}$ reflect product/market/time ``shocks'' that are known and unknown to the firms, respectively, when they make their decisions.  Shocks of this sort are an important aspect of Wollmann's setting.  Note that Wollmann also estimates (point-identified) demand and variable cost parameters in a first step, while for simplicity we treat the variable profits $\pi_{j,f,i,t}$ as known to the econometrician.}

As described below, each simulated dataset is a cross-section containing data on one period for 500 markets following the sequential process described above.  The moments used in our simulations are then averages (over markets $i$) of
\begin{equation}\label{eq: simulation moments}
\frac{1}{J}\sum_{j} \left(m^l_{j,f,i}(\theta) \otimes \tilde{Z}_{j,f,i}\right)',
\end{equation}
where we also average over all firms $f$ assumed to share the same fixed cost $\delta_{f,c}.$
Since we consider a single period for each market $i$ in cross-section, we suppress the time subscript.
We present results both for the case where $\tilde{Z}_{j,f,i}$ includes only a constant, and for the case where all moments are interacted with a constant and the first four moments are additionally interacted with the common profit-shifters $\eta$,
$$
\tilde{Z}_{j,f,i}= (1, \eta^+_{j,i},\eta^-_{j,i}),
$$
for $q^+=\max\{q,0\}$ and $q^-=-\min\{q,0\}$.
In the model with a single constant term, $\delta_{c,f}=\delta_c$ for all $f$, this generates 6 and 14 moment inequalities.
We also present results when the nine firms are divided into three groups each with a separate constant term, and when each firm has a separate constant term.  For each specification we 
consider the first four moments separately for the firm(s) associated with distinct parameters $\delta_{c,f}$, but average the last two moments across all firms as they
do not depend on the constant terms. This generates 14 and 38 moments for the three group classification, and 38 and 110 moments when each firm has a separate constant term. To estimate the conditional variance $\Sigma=\Omega,$ in each specification we define the value of the instrument $Z_i$ in market $i$ as the Jacobian of (\ref{eq: simulation moments}) with respect to the linear parameters $(\delta_g,\{\delta_{c,f}\})$.

\subsection{Data-generating Process Details}

\subsubsection{Competition and Firm Decisions \label{sec:firm choices}}
We now describe the data-generating process for a single market, suppressing the $i$ subscript for notational brevity.
We consider competition between $F$ firms, who in each period decide which set of products to offer. 
Firm $f$ estimates that marketing product $j$ in period $t$ will earn variable profits $\pi^*_{jft}$, and chooses to market the product if and only if the expected profits exceed the fixed costs. Thus, if a firm marketed product $j$ in period $t-1$, then the firm chooses to market $j$ in period $t$ if and only if
\begin{align*}
\pi^*_{jft} - \beta \theta_c -\beta \theta_g g_j \; >0.
\end{align*}
If the firm did not market the product $j$ in period $t-1$, then it chooses to add product $j$ if and only if 
\begin{align*}
\pi^*_{jft} - \theta_c -\theta_g g_j \; >0.
\end{align*}

\subsubsection{Distributional Assumptions}

We set $\pi^*_{jft} = \eta_{jt} + \epsilon_{jft}$, the sum of a product-level shock that is common to all firms and a firm-product idiosyncratic shock. We assume that $\eta_{jt} \sim \mathcal{N}(0, \sigma_{\eta}^2)$. If $j$ was not marketed in the previous period, then $\epsilon_{jft} \sim \mathcal{N} ( \beta \mu_f + \beta \theta_g g_j, \sigma_{\epsilon}^2)$; if the product was marketed previously, then $\epsilon_{jft} \sim \mathcal{N} (\mu_f + \theta_g g_j, \sigma_{\epsilon}^2)$. Note that the mean profitability of marketing a product depends on a firm-specific mean, $\mu_f$, which allows us to match the firm-level market shares observed in Wollmann's data. We also construct the mean of the $\epsilon_{jft}$ term to depend on the product's weight and whether it was marketed in the previous period in a way that guarantees that all simulated products will be offered with the same probability in our simulations.

While firms make their decisions using $\pi^*_{jft}$, we assume that the econometrician observes only $\pi_{jft} = \pi^*_{jft} + \nu_{jt} + \nu_{jft}$. The $\nu$ terms represent measurement or expectational errors. We assume that $\nu_{jt}$ and $\nu_{jft}$ are independently drawn from a normal distribution with mean $0$ and variance $\sigma_{\nu}^2$.

%
%
%
%
%
%
%
%
%
%
\subsection{Calibration}

We calibrate our parameters to estimates and moments reported in the November 2014 version of Wollmann. We set $F=9$ to match the number of firms in Wollmann's data, and $G = 22$ to match the number of unique values of GWR. We use $\theta_c = 129.73$, $\theta_g = -21.38$, and $\beta = 0.386$ to match the results from the estimates in Table VII in Wollmann.\footnote{Note that Wollmann denotes by $-\frac{1}{\lambda}$ what we have been calling $\beta$.} We set the values of $g$ to be 22 evenly spaced points between 12,700 and 54,277 to match the lowest and highest GWR figures reported in Table II, which gives the average GWR for different buyer types.

To calibrate the remaining parameters, we simulate data according to the process described above, and set the parameters to match moments of the simulated data to those in Wollmann's data. In order to simulate the data for the calibration, we first fix standard normal draws that are used to construct the $\eta$, $\epsilon$, and $\nu$ shocks. These standard normals draws are then scaled by the desired variance parameters in each simulation. Letting $J_{ft}$ denote the set of products offered by firm $f$ in period $t$, the simulations begin in state 0 with $J_{f0} = \emptyset$ for all firms. We then simulate $J_{ft}$ and $\pi^*$ going forward using the dynamics described above. We discard the first 1,000 periods as burnout so as to obtain draws from the stationary distribution, and calibrate the model using 27,000 subsequent periods. After discarding 1,000 draws, we obtain essentially identical results if we begin from the state where all products are in the market in rather than all products out of the market.

The remaining parameter values to calibrate are $ \{ \mu_f \}, \sigma_\eta, \sigma_\epsilon , \sigma_{\nu} $. The intuition for the calibration is as follows. The firm-specific means $\mu_f$ affect the number of products each firm offers, and so we calibrate these to match the market shares and total number of products offered in Wollmann's data. The $\sigma_\epsilon$ and $\sigma_\eta$ terms affect how often firms add and remove products, and so we calibrate these to match the variability of the number of products offered over time in Wollmann's data. Lastly, we calibrate $\sigma_\nu$, which governs the variance of the expectational/measurement error. We do not have direct measures of the variability of firm profits in Wollmann's data, but if markups are constant, then the variance in firm profits is one-to-one with the variance of quantity sold, and so we calibrate $\sigma_{\nu}$ to match the variability of quantities sold assuming mark-ups are fixed at 35\%.

Specifically, the calibration uses the following steps:

1) We first calibrate $(\sigma_\eta, \sigma_\epsilon )$ and the $\mu_f$ terms to match the market shares and variability of products offered in Wollmann. This calibration process involves an inner and outer loop, described below.

a) The inner loop for $\mu_f$. Given a guess for $(\sigma_\eta, \sigma_\epsilon )$, we calibrate $\mu_f$ to match the market share and average number of products in Wollmann's data. Market shares are taken from Table III in Wollmann. Wollmann does not provide the mean number of products offered by year, only the min and max, so we approximate it by taking the midpoint between the two extremes, which gives 48 total products per year on average.

b) In the outer loop, we calibrate $(\sigma_\eta, \sigma_\epsilon)$ to match a measure of the variability of the number of products offered in Wollmann's data. In particular, Table I in Wollmann lists 9-year averages for the total number of products offered for three 9-year periods (he has 27 years of data). We run 1,000 simulations of 27 periods, and for each 27-year period we calculate the average number of products offered within each 9-year subinterval, just as Wollmann does. We then calibrate $\sigma_{\eta}$ so that the average variance in the number of products offered across three consecutive 9 year periods matches that in Wollmann's data.

The simulated variance comes very close to the target variance whenever $\sigma_\eta = \sigma_\epsilon$, regardless of scaling. We therefore choose $\sigma_\eta =  \sigma_\epsilon = 30$, which gives that the variance of $\pi^*$ is roughly half of the variance of $\pi$. 

2) Lastly, we calibrate $\sigma_\nu$ to match a moment implied by the variability in quantity sold across time in Wollmann. If prices and markups are relatively constant, then the variance in quantities will be well-approximated by a constant times the variance in profits: $\var{\pi_{jft}} \approx \bar{p}^2 \bar{m}^2 \var{Q_{jft} }$, where $\bar{p}$ and $\bar{m}$ are the average prices and markups.\footnote{This is because if prices and costs are constant across firms, 
	\begin{align*}
	\pi_{jft} &= Q_{jft} (p - c)  \nonumber\\
	&= Q_{jft} \frac{p-c}{p} p  \nonumber\\
	&= Q_{jft} \times m \times p.
	\end{align*}
	Thus, $\var{\pi_{jft}} =m^2 p^2 \var{Q_{jft}}$ when $p$ and $c$ are constant, and this holds approximately with averages if the variance in $m$ and $p$ is small relative to that in $Q$. } For our calibration, we set $\bar{p}$ to be the average price in Wollmann's data (\$66,722), and set $\bar{m}$ equal to 0.35. As with the number of products offered, Wollmann does not report annual quantities, but rather the average for three 9-year periods. We thus use a procedure analogous to that described in step 1b) to match the variance of the 9-year averages of quantity sold. 

\subsubsection{Calibrated Parameters}

Tables \ref{tbl: mu_f} and \ref{tbl: variance params} show the calibrated values for the $\mu_f$ and variance parameters, respectively.

\begin{table}[hbtp]
	\caption{Calibrated $\mu_f$ Parameters}
	\label{tbl: mu_f}
	\begin{center}
		\begin{tabular}{ll}
			Firm&$\mu_f$\\
			\hline
			Chrysler&$74.31$\\
			Ford&$98.36$\\
			Daimler&$114.69$\\
			GM&$80.11$\\
			Hino&$67.71$\\
			International&$110.63$\\
			Isuzu&$80.15$\\
			Paccar&$114.63$\\
			Volvo&$94.17$\\
		\end{tabular}
	\end{center}
\end{table}

\begin{table}[hbtp]
	\caption{Calibrated Variance Parameters}
	\label{tbl: variance params}
	\begin{center}
		\begin{tabular}{ll}
			Parameter&Value\\
			\hline
			$\sigma_\eta$&$30.00$\\
			$\sigma_\epsilon$&$30.00$\\
			$\sigma_\nu$&$57.96$\\
		\end{tabular}
	\end{center}
\end{table}

\subsubsection{Sampling from the DGP}

Wollmann's data involves observations of sequential periods from the same market. If we were to construct moments at the product-period level in this setting, then the sequential nature of the model would induce serial correlation in the realizations of the moments. Although $\Sigma$ can be estimated in this setting, accounting for serial correlation substantially complicates covariance estimation. Since covariance estimation is not the focus of this paper, and \cite{Wollmann} performs inference assuming no serial correlation, we instead focus on a modified DGP corresponding to a cross-section of independent markets, a common setting in the industrial organization literature. To do this, we sample from the stationary distribution of the calibrated DGP described above as follows. We draw a 51,000 period sequential chain, and discard the first 1,000 observations as a burn-in period. For each simulated dataset, we then randomly subsample 500 periods from this chain. This cross-sectional set-up also allows us to consider specifications with more moments than in Wollmann.

\subsection{Implementation Details}

\subsubsection{Parameter Grids}

For procedures that require test inversion for the parameter of interest, we invert tests over a discretized parameter space.\footnote{For the LF and LFP approaches, we do not need to discretize the parameter space when the parameter of interest enters the moments linearly, since the endpoints of the confidence set can be calculated analytically using linear programming, as discussed in Section \ref{sec: Practical Implementation}.} For $\delta_g$ and the cost of the mean-weight truck, we use 1,001 gridpoints (plus estimates of the identified set bounds); for $\beta$, we use 100 gridpoints for our main simulations, and 1,000 gridpoints for timing comparisons.

\subsubsection{Implementation of LF and LFP tests}
To calculate the LFP critical values, we draw a fixed matrix $\Xi$ of standard normal draws of size $k \times 10,000$, and we use these for all of our calculations. Since the LF procedure is more computationally intensive, we calculate it using a matrix of size $k \times 1000$.  

In simulating the draws for the LF approach, in certain very rare cases we encountered computational issues in which the linear program for one of the draws did not converge. In these cases, we treat the draw as if it were infinity, which pushes the estimated critical value slightly higher. However, in all specifications this happens in no more than 0.01\% of cases (of approximately 50 million simulations), and is thus unlikely to have any substantial impact on our results. 

\subsubsection{Implementation of the sCC and sRCC tests}

We implement the sCC and sRCC tests using code provided by the authors. The refinement needed for the sRCC test is difficult to compute with many moments and many parameters. Thus, when our specification has both 100+ moments and 10+ parameters, we instead report the results of a test that rejects whenever the sRCC test rejects. In particular, the refinement to the sRCC test can matter only when there is one active moment ($\hat{r}=1$) and the test statistic falls between the $1-\alpha$ and $1-\alpha/2$ quantile of the $\chi^2$ distribution with 1 degree of freedom. For specifications with 100+ moments and 10+ parameters, we thus report the power of the test that rejects when either the sCC test rejects or the refinement could matter. The power and size of this test can thus be viewed as upper bounds on the power and size of the sRCC test, and its runtime is a lower bound on the runtime of the sRCC test. 

\subsubsection{Implementation of the AS and KMS tests}

We next describe the implementation of the AS and KMS tests, which uses the Matlab package developed by \citet{kaido_calibrated_2017}. The Matlab package is developed for the case where the moments are additively separable in the data and the parameters, i.e. when the moments take the form $E[m(D_i)] - g(\theta) \leq 0$, where $\theta$ is a vector of parameters and the target parameter takes the form $l'\theta$. Note that in our first two simulation designs, where the target parameter is $\delta_g$ or the cost of the mean-weight truck (and $\beta$ is known), the moments take the form $E[ Y_i | X_i ] - X_i \delta \leq 0$ and the target parameter is $l'\delta$. The moments thus take the form needed to use the Matlab package \textit{conditional} on $X_i$. The Matlab package, however, uses a bootstrap procedure that samples from the unconditional distribution of the data, which is unsuitable for our setting. To use the package in our setting with conditional moments, we adopt the following procedure. Given $Y_{n,0},X_{n,0},\widehat{\Sigma}_{n,0}$, we draw $Y_i^* \sim N(n^{-\frac{1}{2}}Y_{n,0}, \widehat{\Sigma}_{n,0})$ independently for $i=1,...,n$.\footnote{We re-center and re-scale the draws so that the sample mean of $Y_i^*$ is exactly $n^{-\frac{1}{2}}Y_{n,0}$ and the sample covariance is $\widehat{\Sigma}_{n,0}$.} We then provide the Matlab package with the data $(Y_i^*)_{i=1}^n$ and set $m(Y_i^*) = Y_i^*$ and $g(\theta) = X_{n,0} \theta $. This ensures that the bootstrap distribution of the sample mean of $Y_i^*$ (scaled by $\sqrt{n}$) within the Matlab package approximates the conditional distribution of $Y_{n,0} | X_{n,0}$.

We use the default tolerances in the Matlab package except we halve the default tolerance for the objective (i.e., we set EAM\_obj\_tol and EAM\_thetadistort to $0.005/2$). Tightening the objective tolerance appears to reduce numerical precision errors that can, for instance, lead the estimated bounds for the AS test to be tigher than for the KMS test. On the other hand, the tighter tolerances increase runtime and lead to some convergence issues. In the specification with the most moments and parameters, the KMS test fails to converge correctly in 6\% of the cases with the tigher tolerances. We discard all such draws and report size and excess length conditional on the algorithm converging correctly. We obtain qualitatively similar results using the default tolerances, which have fewer convergence issues but are less numerically precise.

\subsection{Additional Simulation Results}

\begin{figure}
	\centering
	
	\caption{Rejection probabilities for 5\% tests of $\theta_g$ \label{fig:theta_g power}}
	
	\subfloat[2 Parameters, 6 Moments]{\includegraphics[width=0.48\linewidth, height = .3 \textheight ]{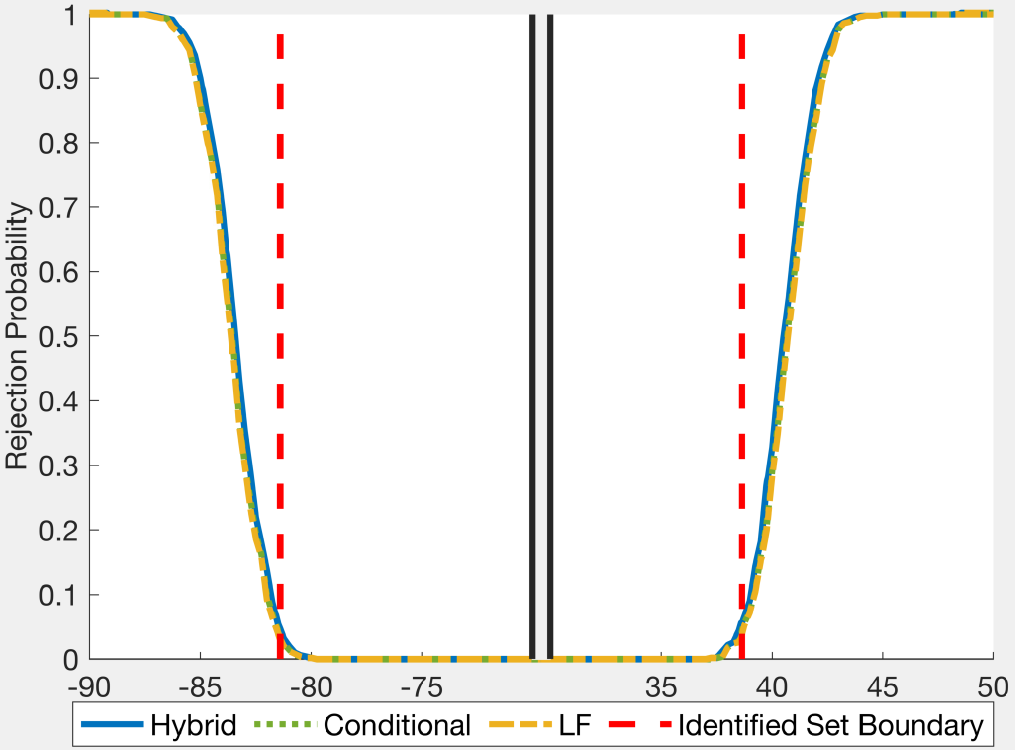} 
	}
	\hfill
	\subfloat[2 Parameters, 14 Moments]{\includegraphics[width=0.48\linewidth, height = .3 \textheight ]{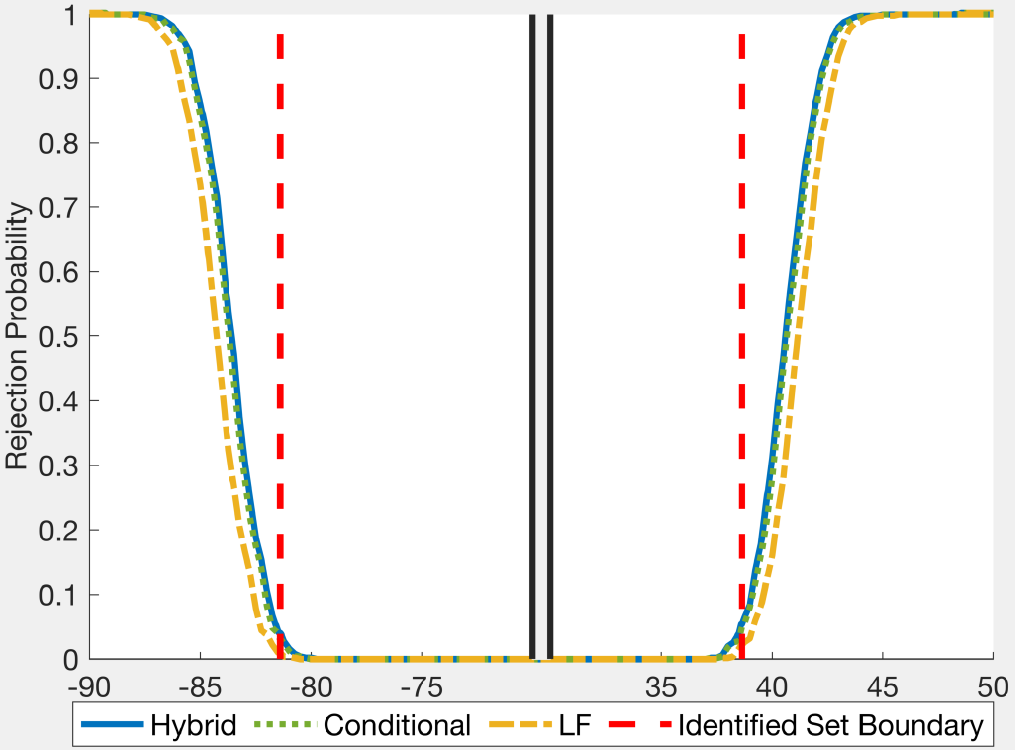} 
	}

	\subfloat[4 Parameters, 14 Moments]{\includegraphics[width=0.48\linewidth, height = .3 \textheight ]{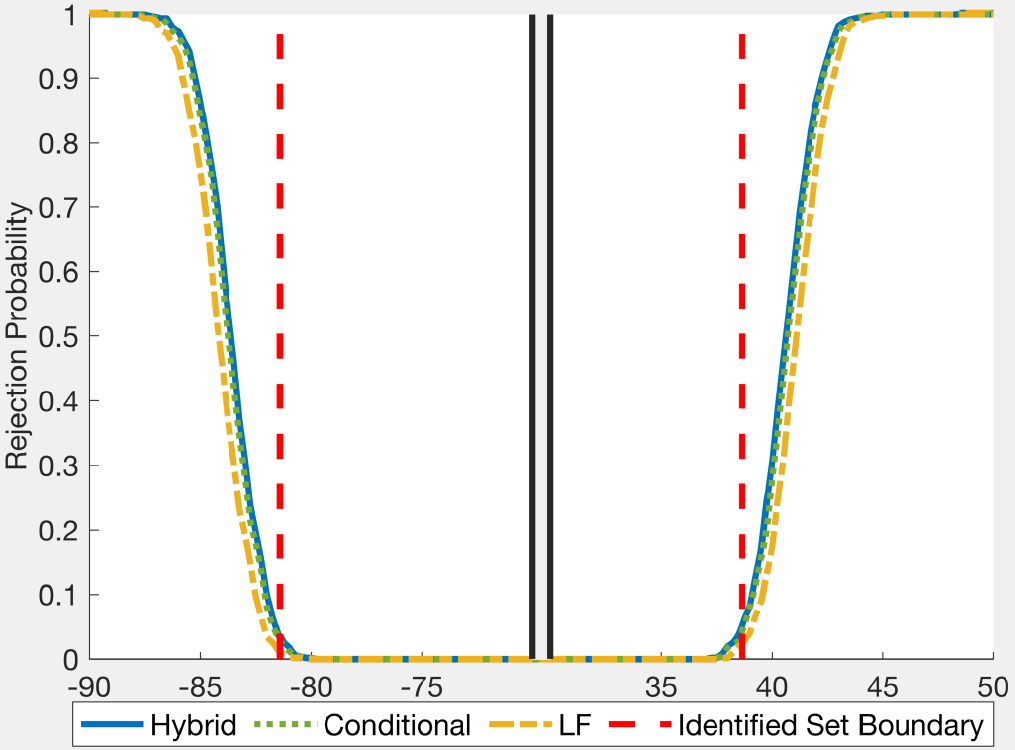} 
	}
	\hfill
	\subfloat[4 Parameters, 38 Moments]{\includegraphics[width=0.48\linewidth, height = .3 \textheight ]{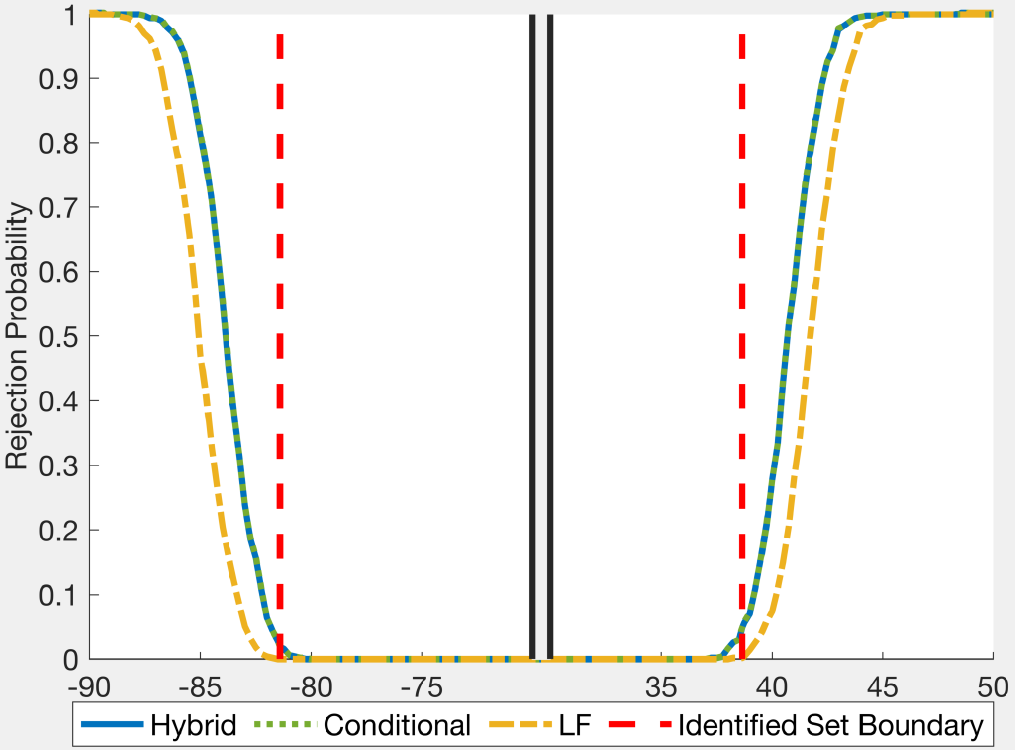} 
	}

	\subfloat[10 Parameters, 38 Moments]{\includegraphics[width=0.48\linewidth, height = .3 \textheight ]{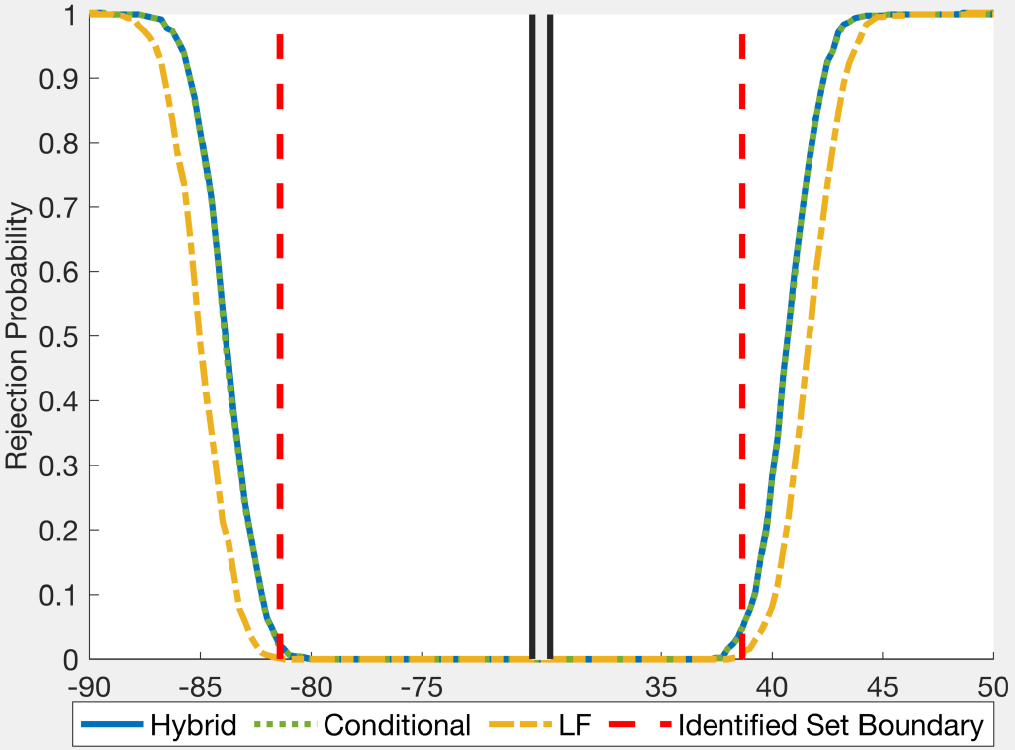} 
	}
	\hfill
	\subfloat[10 Parameters, 110 Moments]{\includegraphics[width=0.48\linewidth, height = .3 \textheight ]{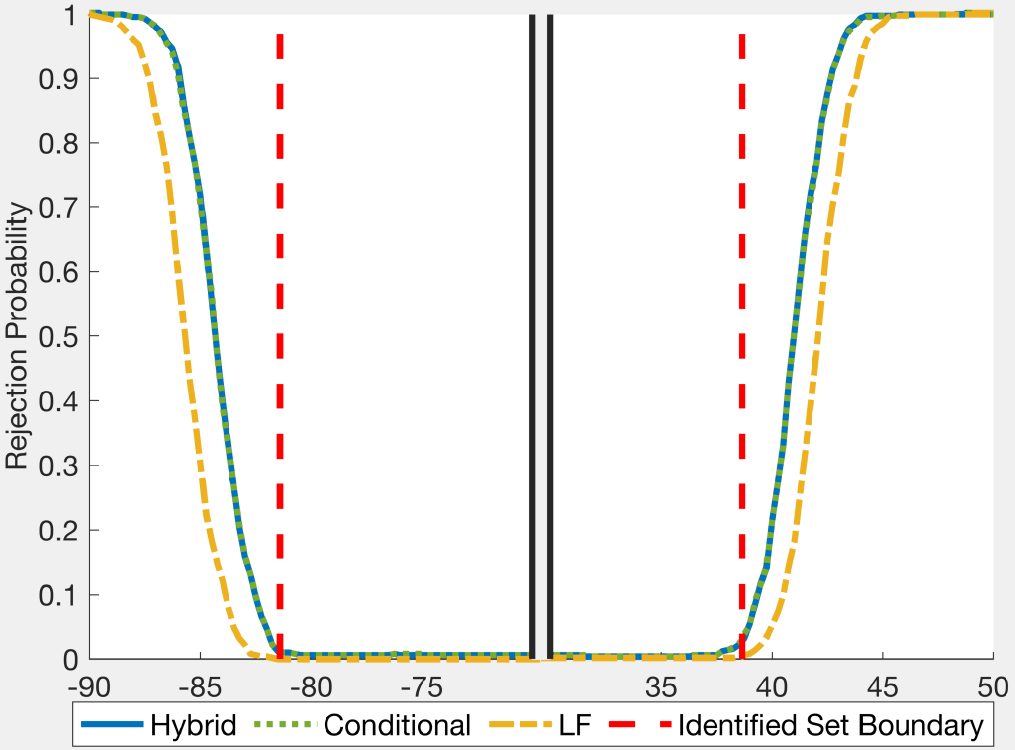} 
	}
\end{figure}

\begin{figure}
	\centering
	
	\caption{Rejection probabilities for 5\% tests of $\beta$\label{fig:beta power wollmann}}
	
	\subfloat[3 Parameters, 6 Moments]{\includegraphics[width=0.48\linewidth ]{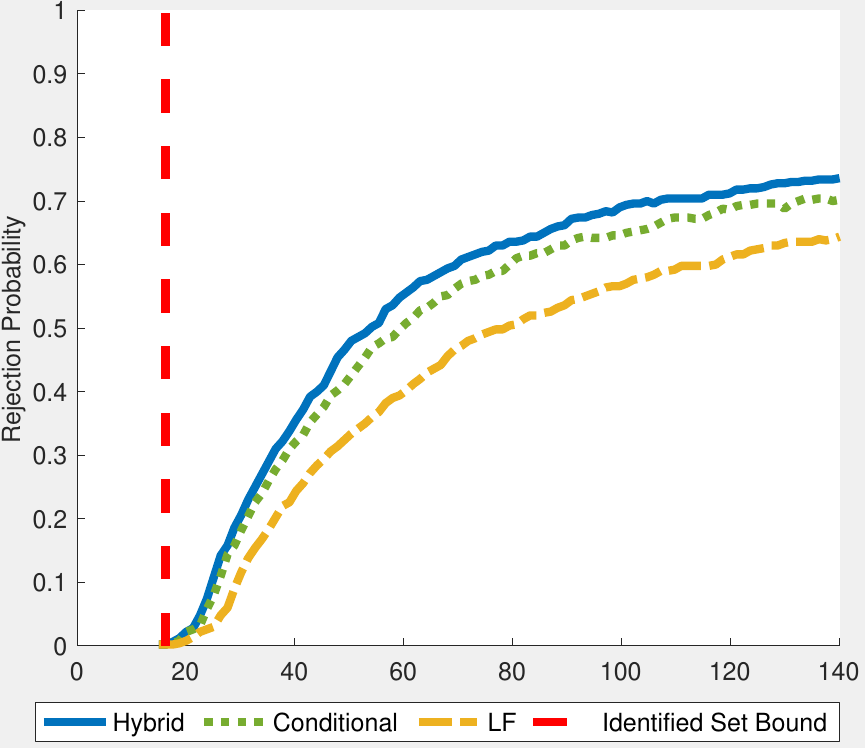} 
	}
	\hfill
	\subfloat[3 Parameters, 14 Moments]{\includegraphics[width=0.48\linewidth]{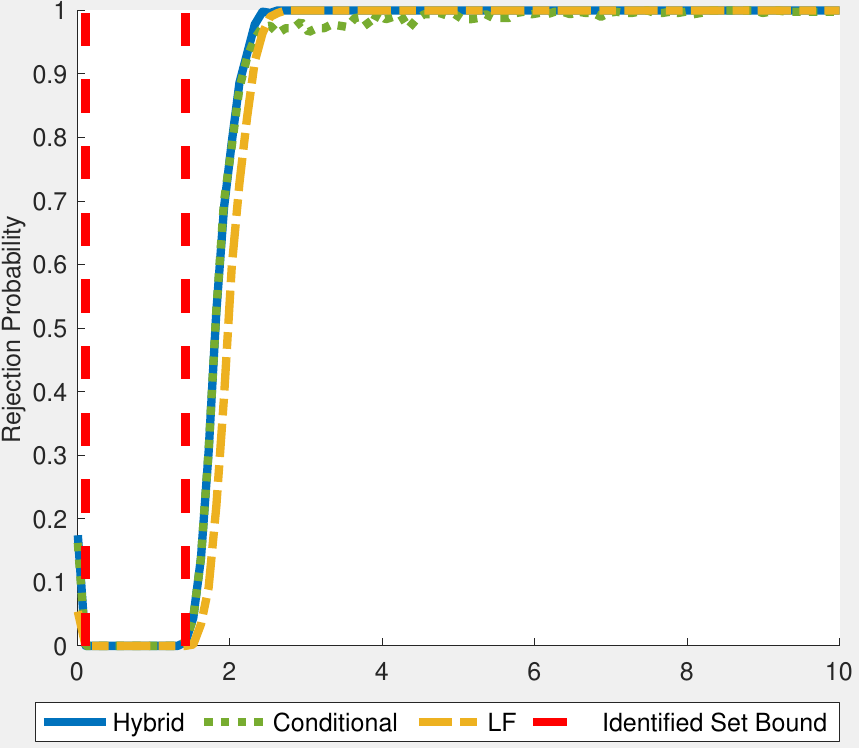} 
	}

	\subfloat[5 Parameters, 14 Moments]{\includegraphics[width=0.48\linewidth]{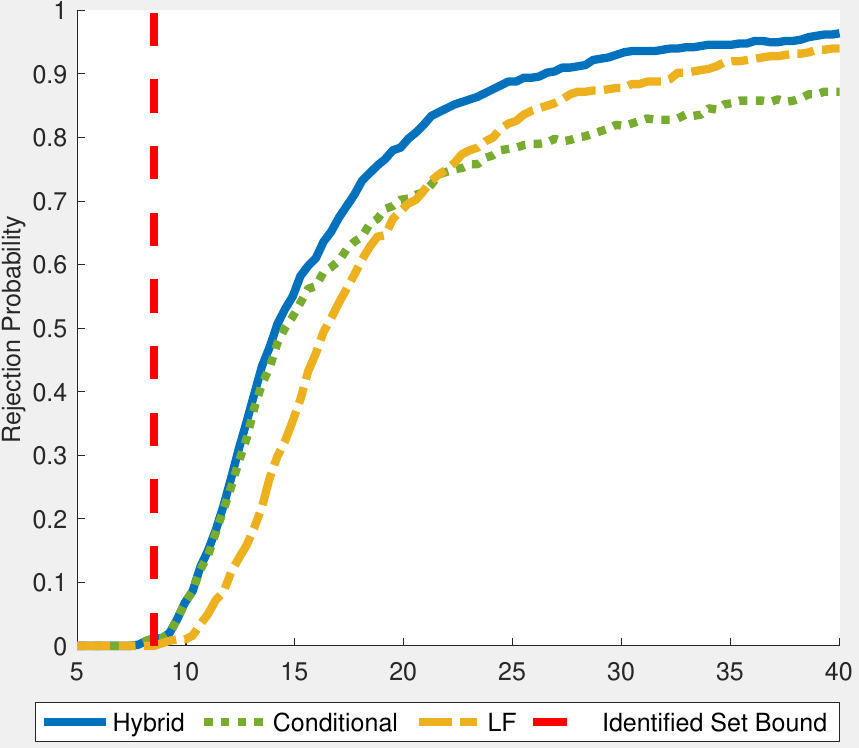} 
	}
	\hfill
	\subfloat[5 Parameters, 38 Moments]{\includegraphics[width=0.48\linewidth]{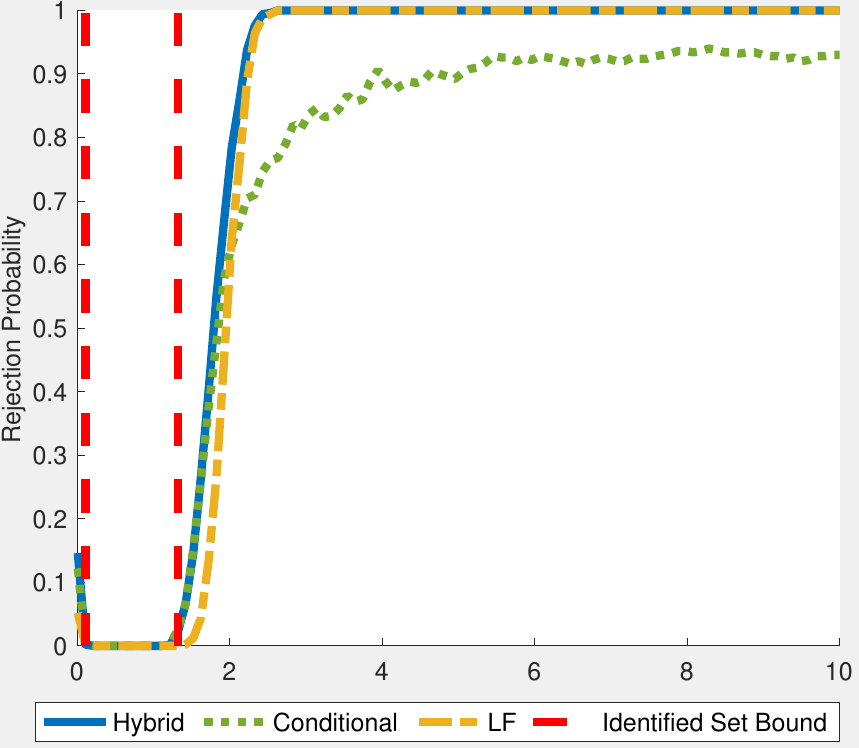} 
	}

	\subfloat[11 Parameters, 38 Moments]{\includegraphics[width=0.48\linewidth]{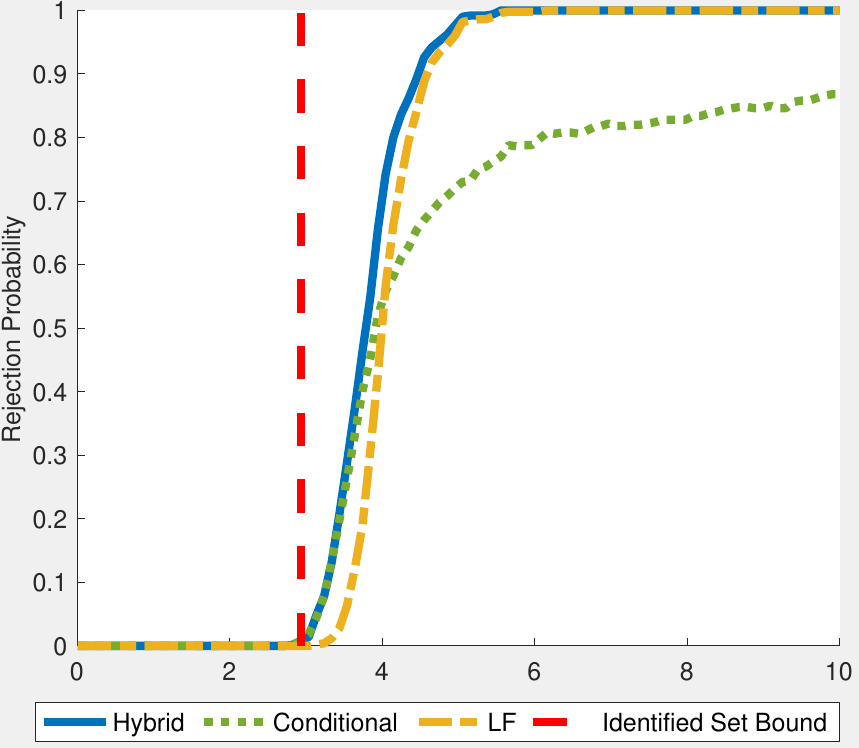} 
	}
	\hfill
	\subfloat[11 Parameters, 110 Moments]{\includegraphics[width=0.48\linewidth]{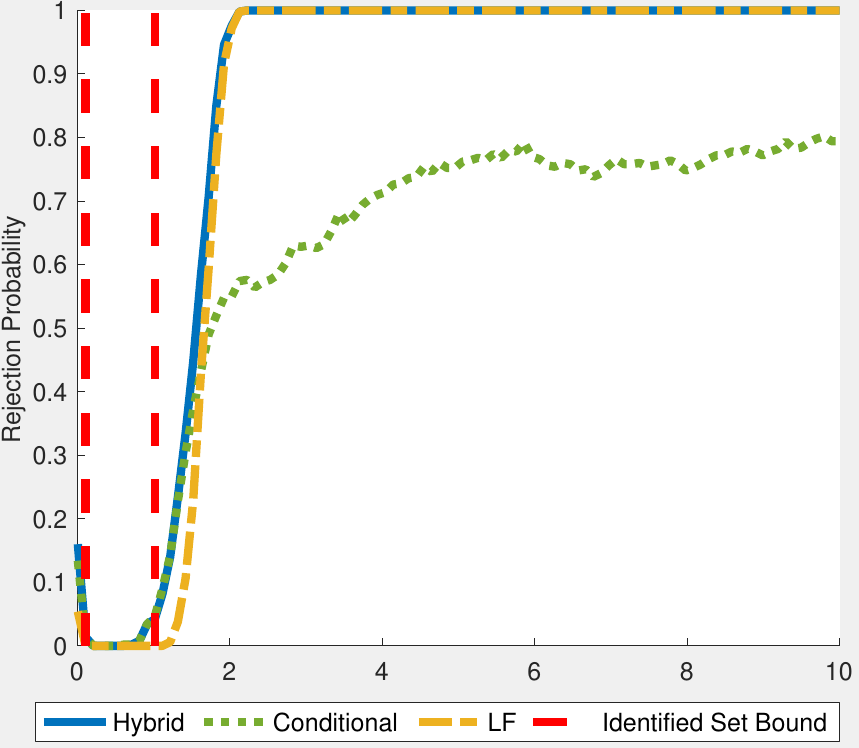} 
	}
\end{figure}

\begin{figure}
	\centering
	
	\caption{Rejection Probabilities for 5\% tests of Cost of Mean-Weight Truck: Comparisons to \citet{cox_simple_2020} and LFP tests \label{fig:meanweight power cox and shi}}
	
	\subfloat[2 Parameters, 6 Moments]{\includegraphics[width=0.48\linewidth, height = .3 \textheight ]{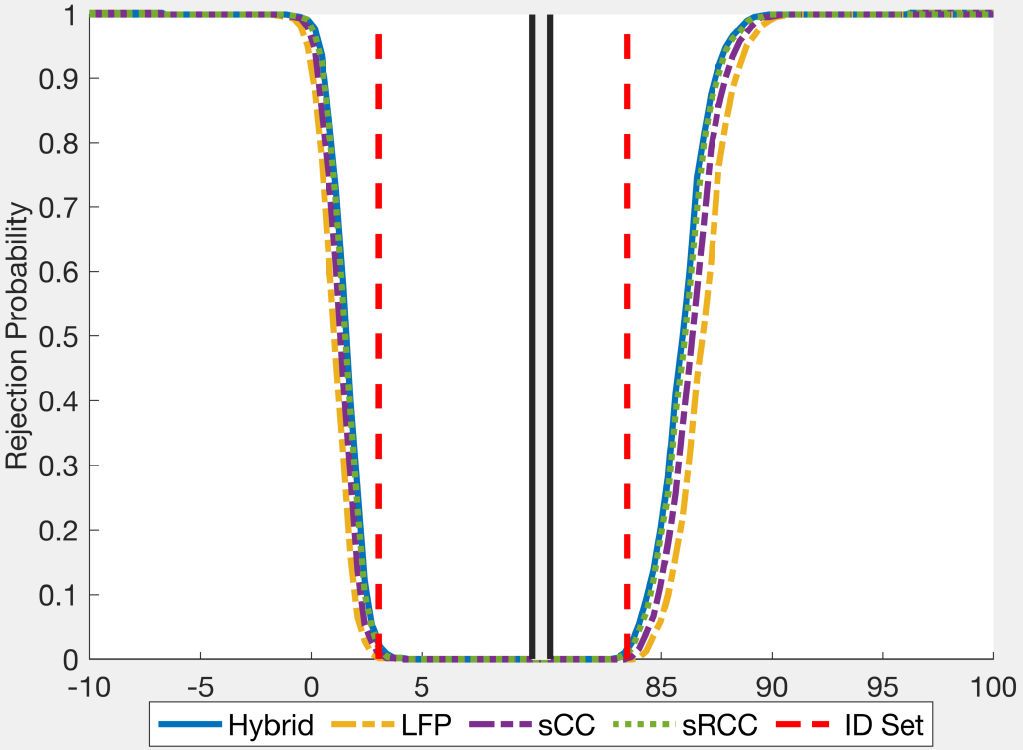} 
	}
	\hfill
	\subfloat[2 Parameters, 14 Moments]{\includegraphics[width=0.48\linewidth, height = .3 \textheight ]{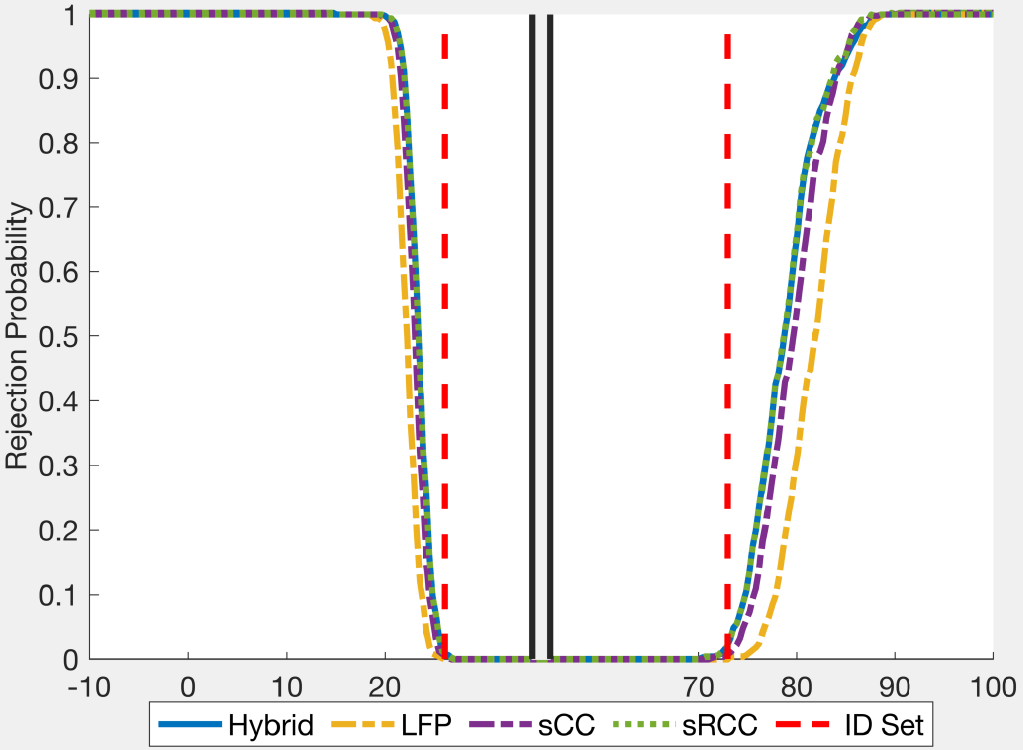} 
	}

	\subfloat[4 Parameters, 14 Moments]{\includegraphics[width=0.48\linewidth, height = .3 \textheight ]{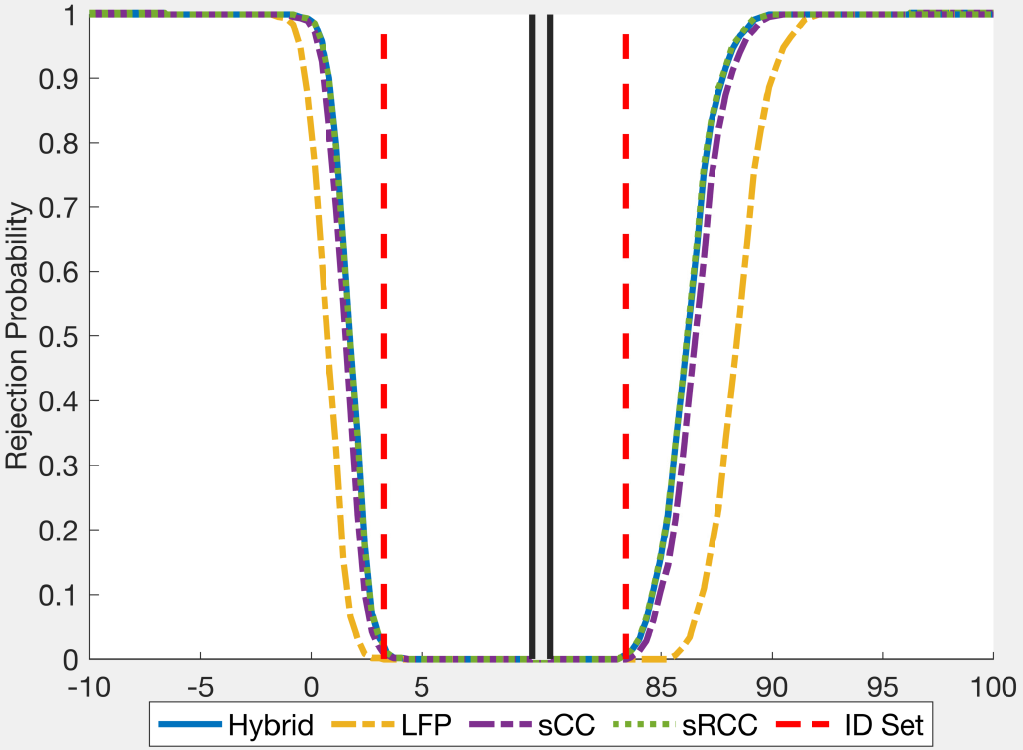} 
	}
	\hfill
	\subfloat[4 Parameters, 38 Moments]{\includegraphics[width=0.48\linewidth, height = .3 \textheight ]{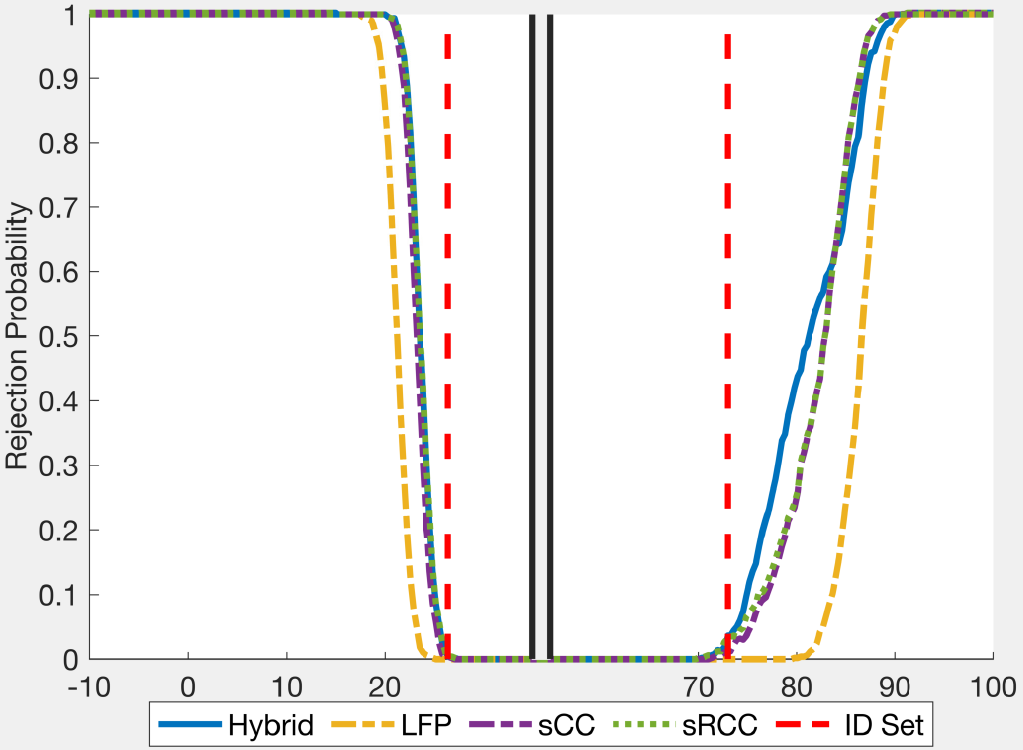} 
	}

	\subfloat[10 Parameters, 38 Moments]{\includegraphics[width=0.48\linewidth, height = .3 \textheight ]{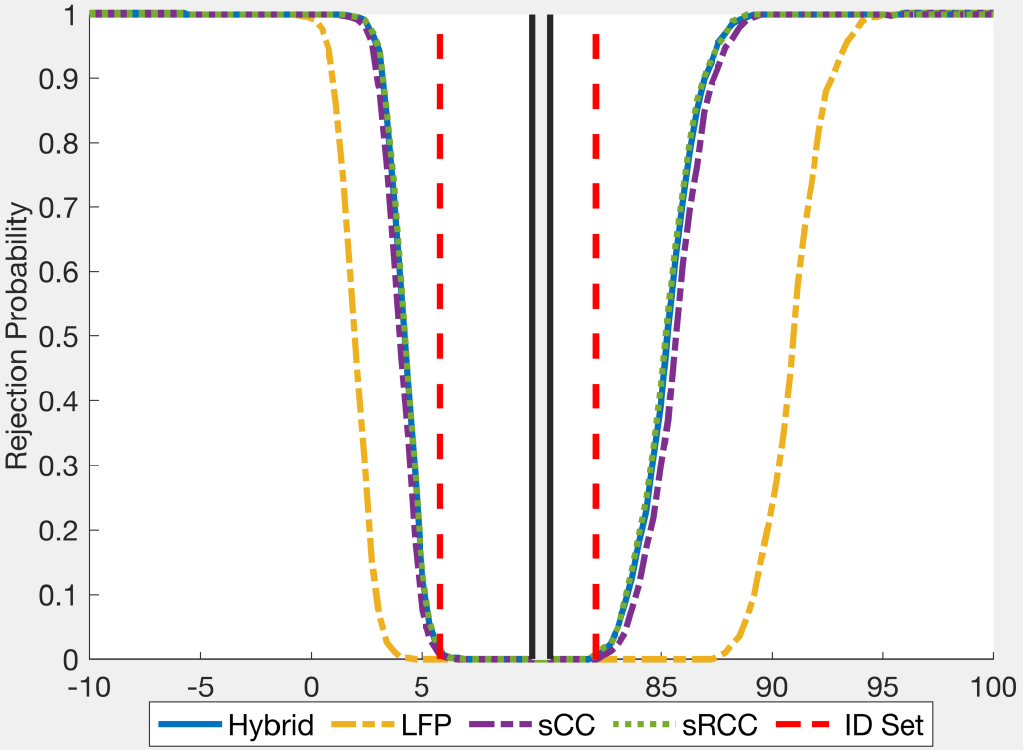} 
	}
	\hfill
	\subfloat[10 Parameters, 110 Moments]{\includegraphics[width=0.48\linewidth, height = .3 \textheight ]{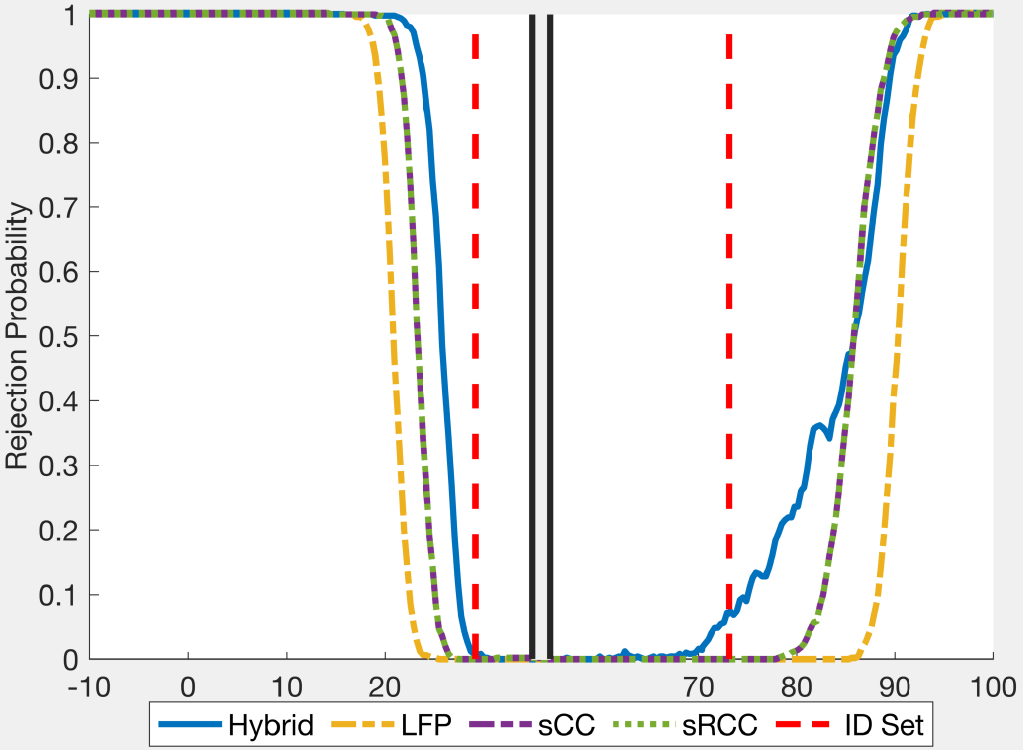} 
	}
\end{figure}

\begin{figure}
	\centering
	
	\caption{Rejection Probabilities for 5\% tests of $\theta_g$: Comparisons to \citet{cox_simple_2020} and LFP tests \label{fig:thetag power cox and shi}}
	
	\subfloat[2 Parameters, 6 Moments]{\includegraphics[width=0.48\linewidth, height = .3 \textheight ]{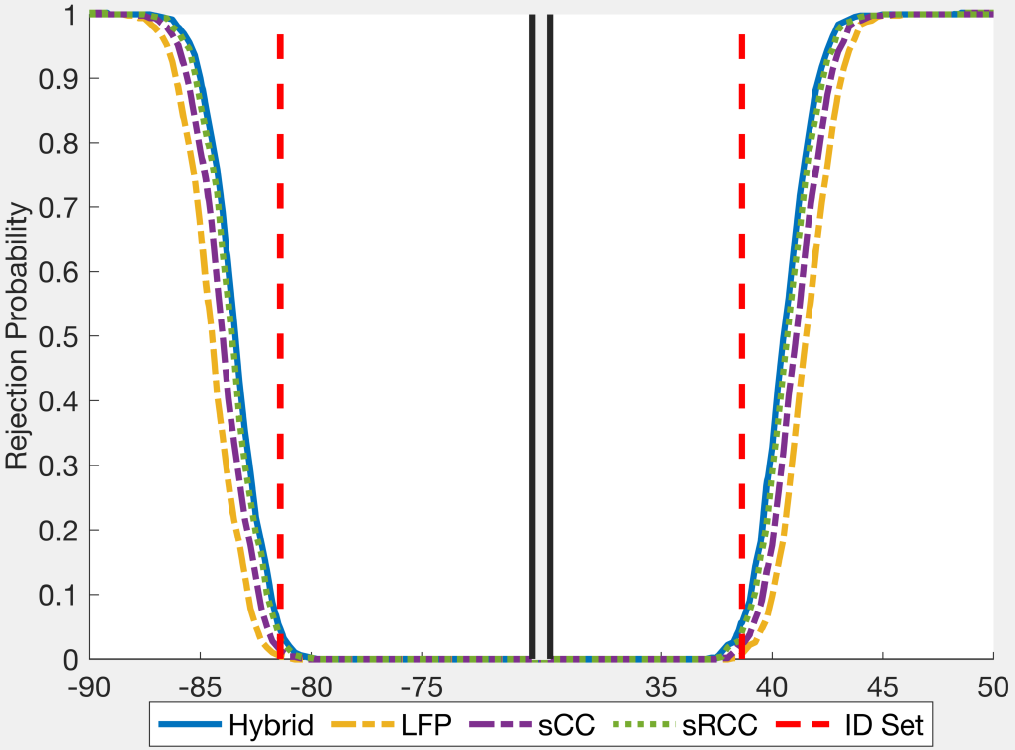} 
	}
	\hfill
	\subfloat[2 Parameters, 14 Moments]{\includegraphics[width=0.48\linewidth, height = .3 \textheight ]{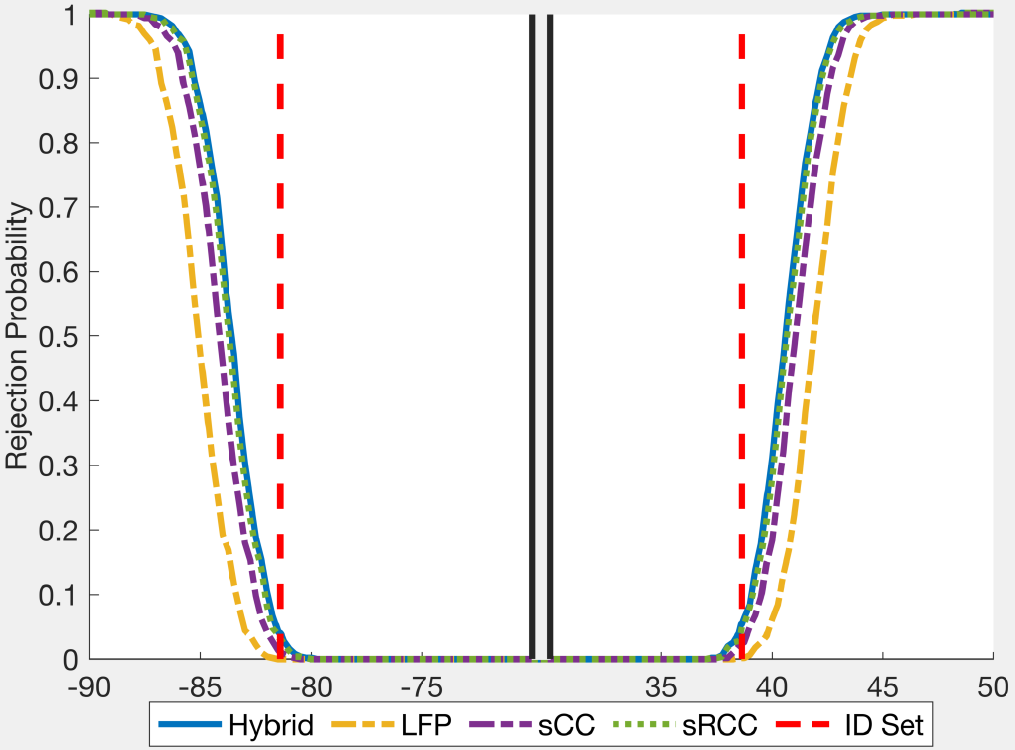} 
	}

	\subfloat[4 Parameters, 14 Moments]{\includegraphics[width=0.48\linewidth, height = .3 \textheight ]{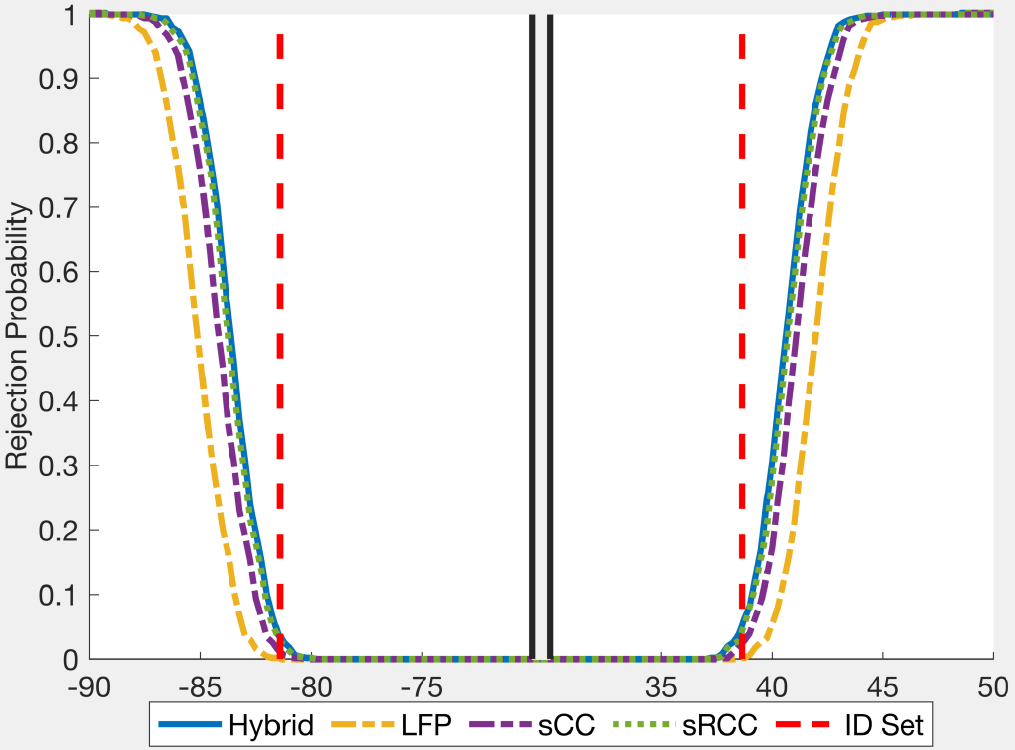} 
	}
	\hfill
	\subfloat[4 Parameters, 38 Moments]{\includegraphics[width=0.48\linewidth, height = .3 \textheight ]{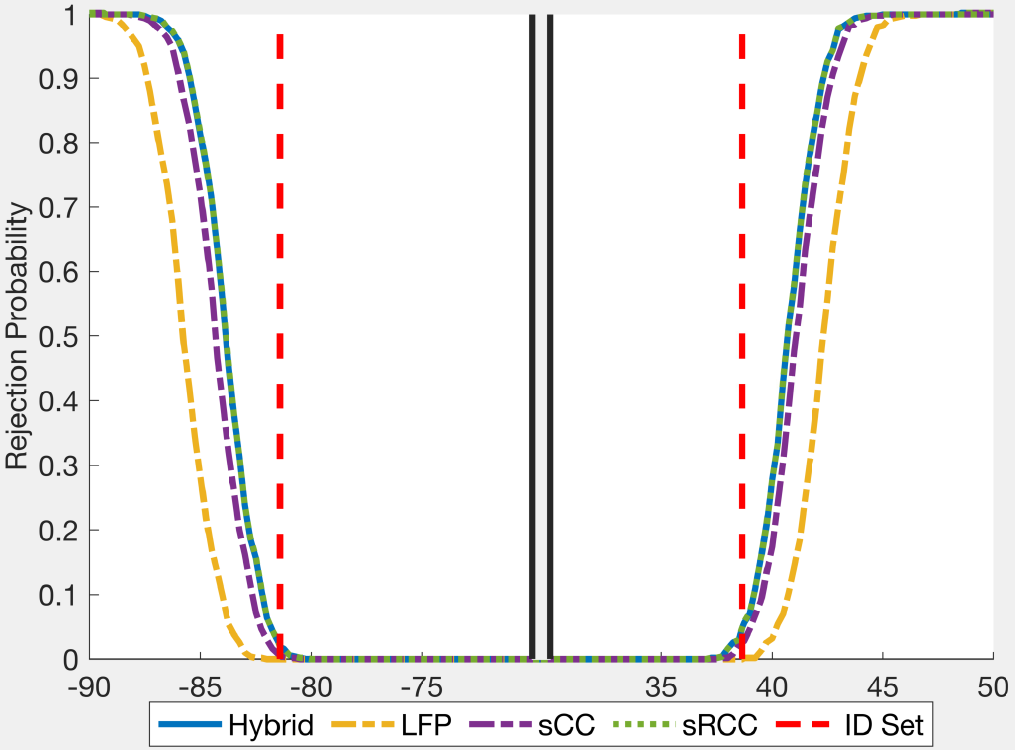} 
	}

	\subfloat[10 Parameters, 38 Moments]{\includegraphics[width=0.48\linewidth, height = .3 \textheight ]{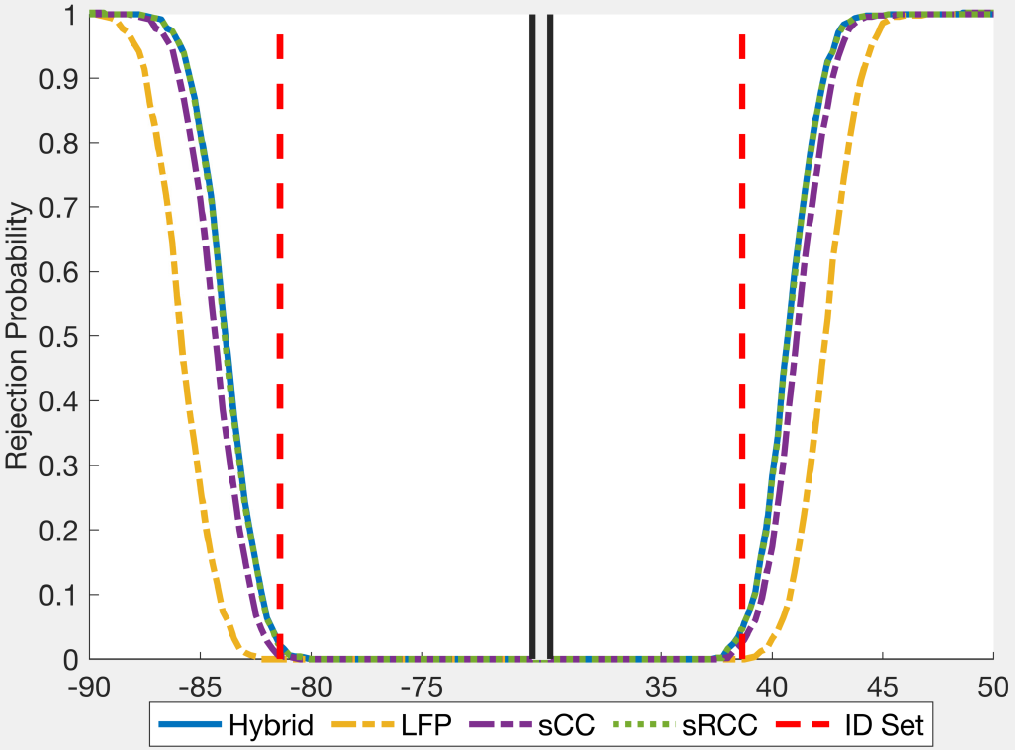} 
	}
	\hfill
	\subfloat[10 Parameters, 110 Moments]{\includegraphics[width=0.48\linewidth, height = .3 \textheight ]{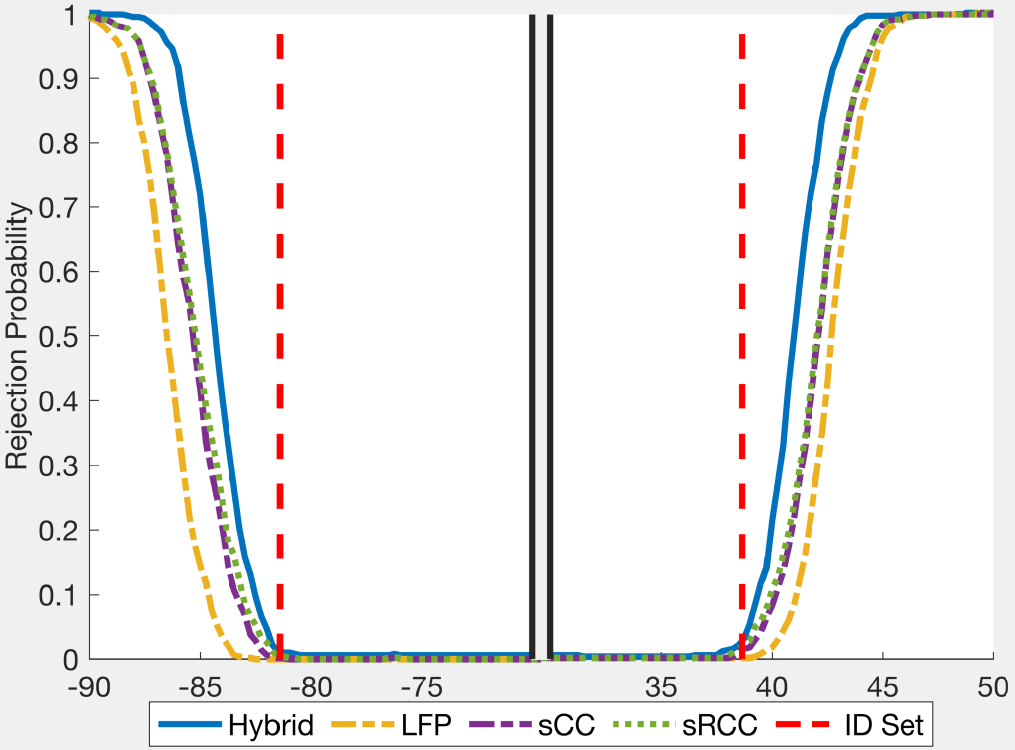} 
	}
\end{figure}

\begin{figure}
	\centering
	
	\caption{Rejection Probabilities for 5\% tests of $\beta$: Comparisons to \citet{cox_simple_2020} and LFP tests \label{fig:beta power cox and shi}}
	
	\subfloat[2 Parameters, 6 Moments]{\includegraphics[width=0.48\linewidth, height = .3 \textheight ]{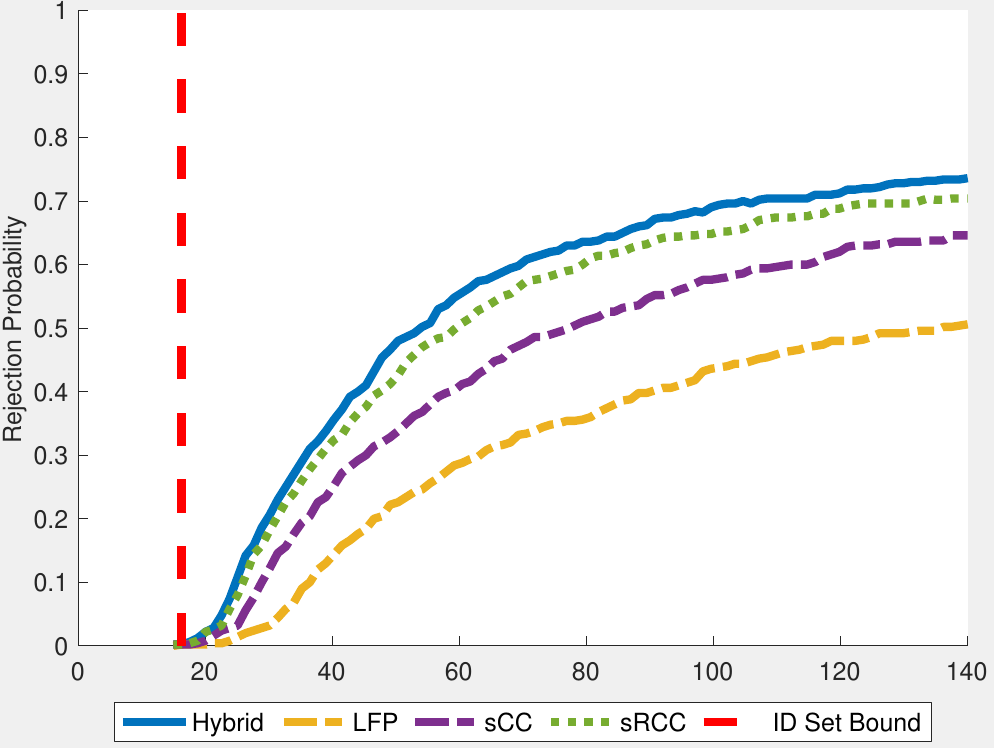} 
	}
	\hfill
	\subfloat[2 Parameters, 14 Moments]{\includegraphics[width=0.48\linewidth, height = .3 \textheight ]{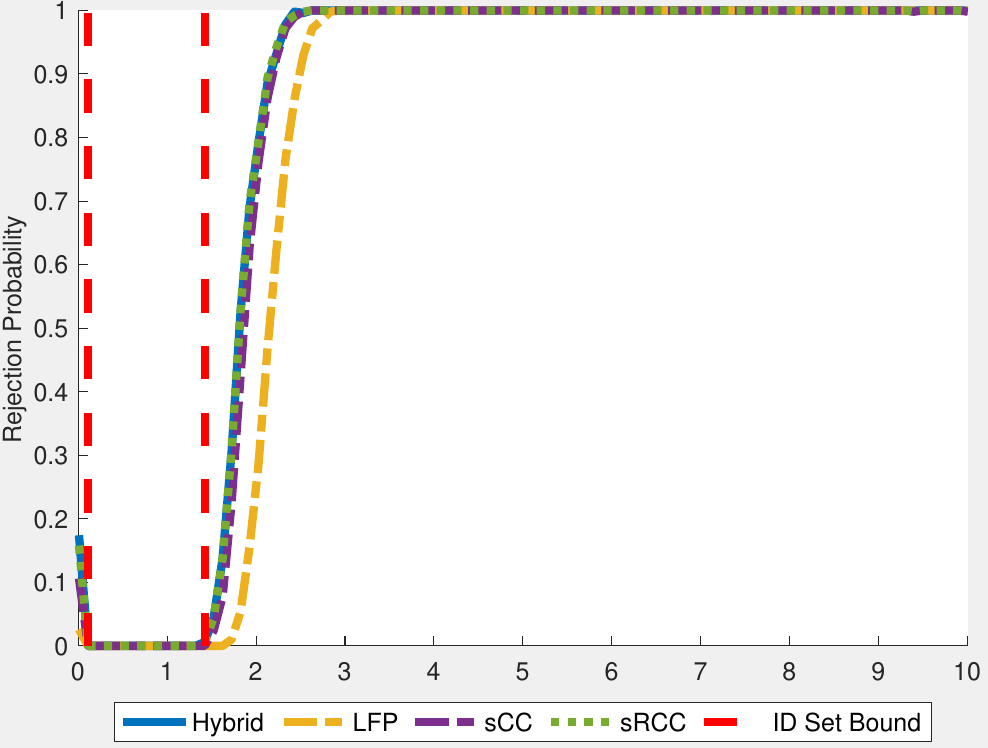} 
	}

	\subfloat[4 Parameters, 14 Moments]{\includegraphics[width=0.48\linewidth, height = .3 \textheight ]{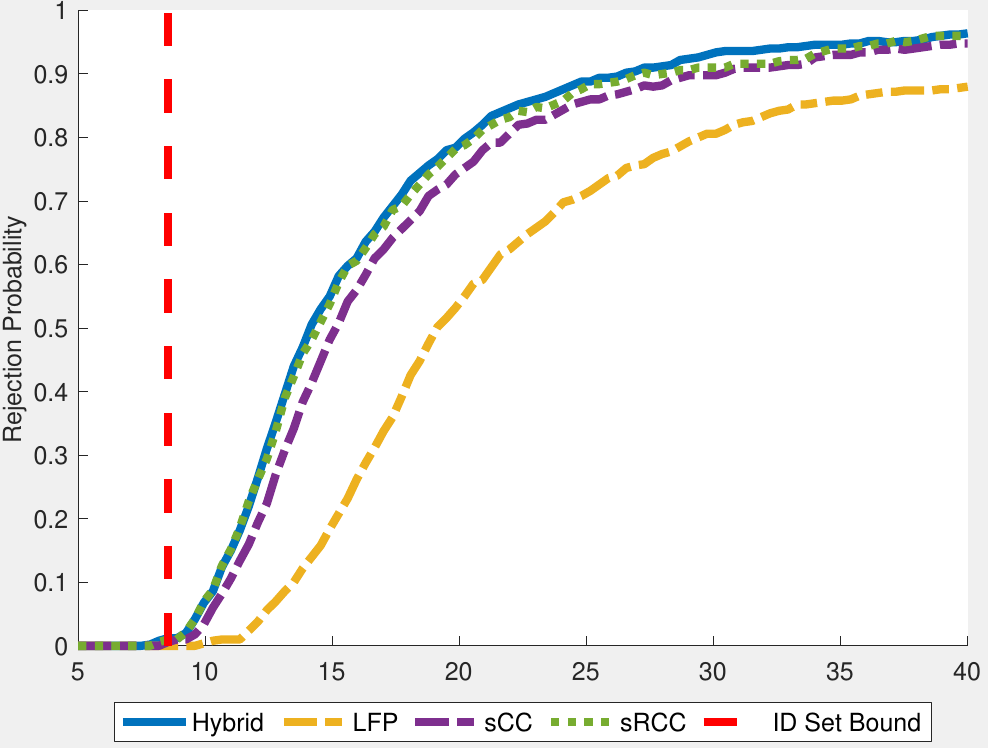} 
	}
	\hfill
	\subfloat[4 Parameters, 38 Moments]{\includegraphics[width=0.48\linewidth, height = .3 \textheight ]{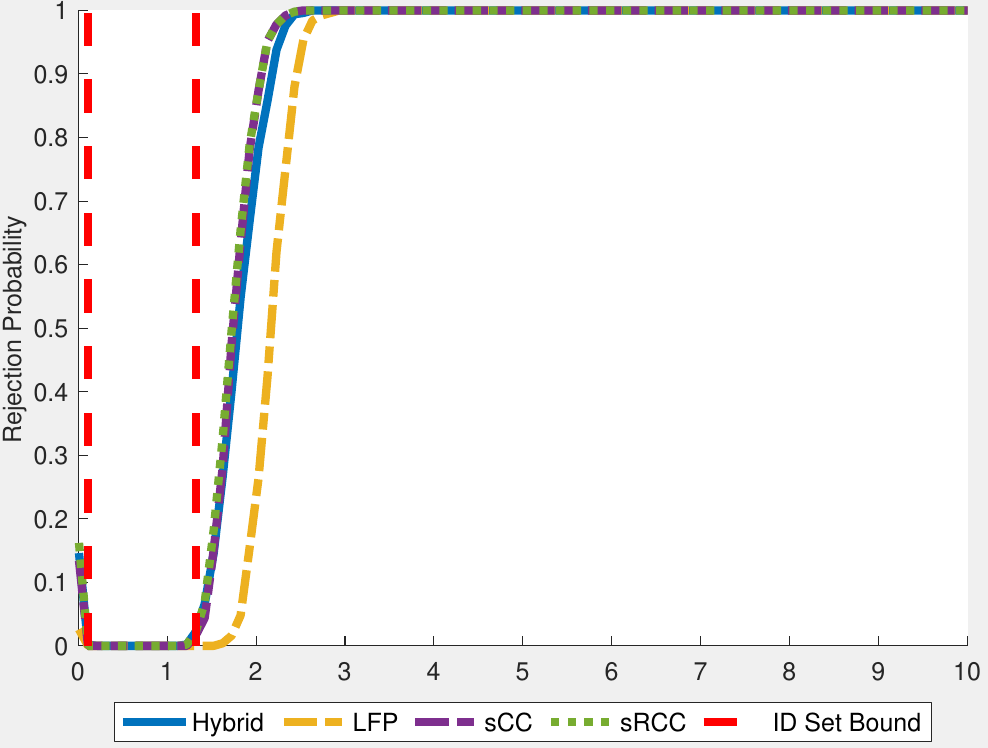} 
	}

	\subfloat[10 Parameters, 38 Moments]{\includegraphics[width=0.48\linewidth, height = .3 \textheight ]{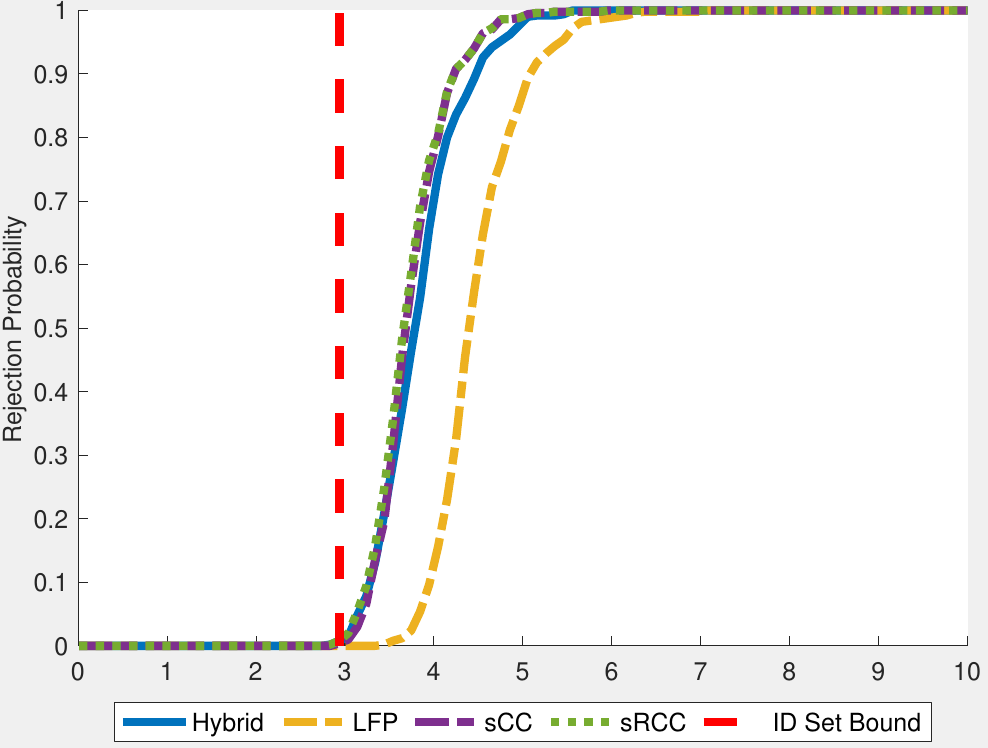} 
	}
	\hfill
	\subfloat[10 Parameters, 110 Moments]{\includegraphics[width=0.48\linewidth, height = .3 \textheight ]{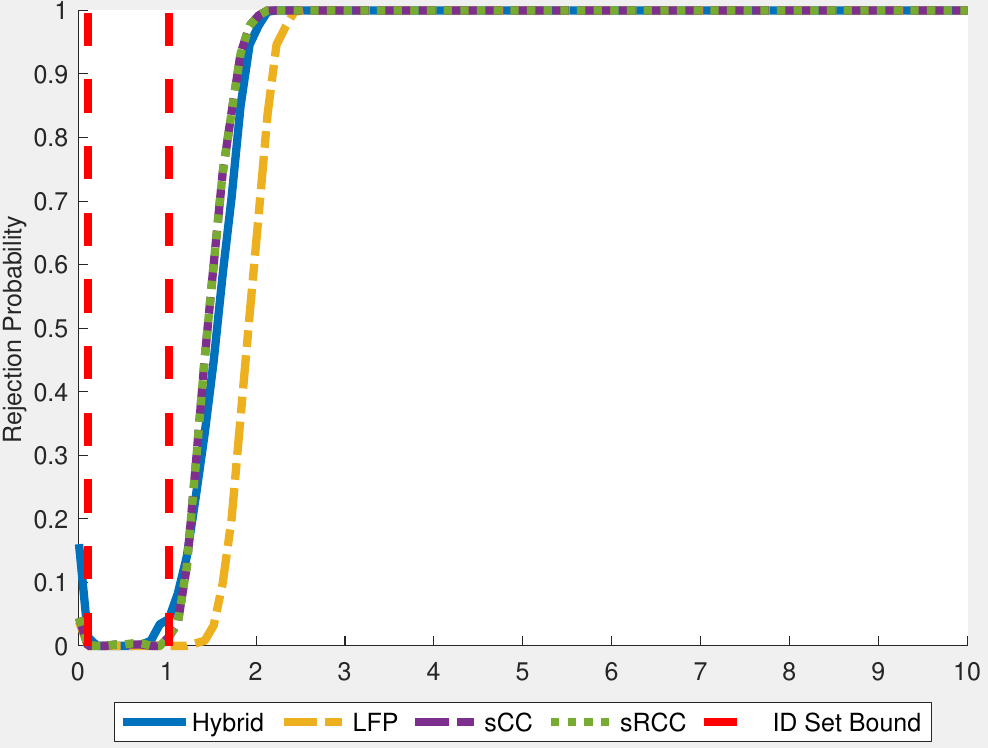} 
	}
\end{figure}

\begin{figure}
	\centering
	
	\caption{Rejection Probabilities for 5\% tests of Cost of Mean-Weight Truck: Comparisons to AS and KMS tests \label{fig:meanweight power as and kms}}
	
	\subfloat[2 Parameters, 6 Moments]{\includegraphics[width=0.48\linewidth, height = .3 \textheight ]{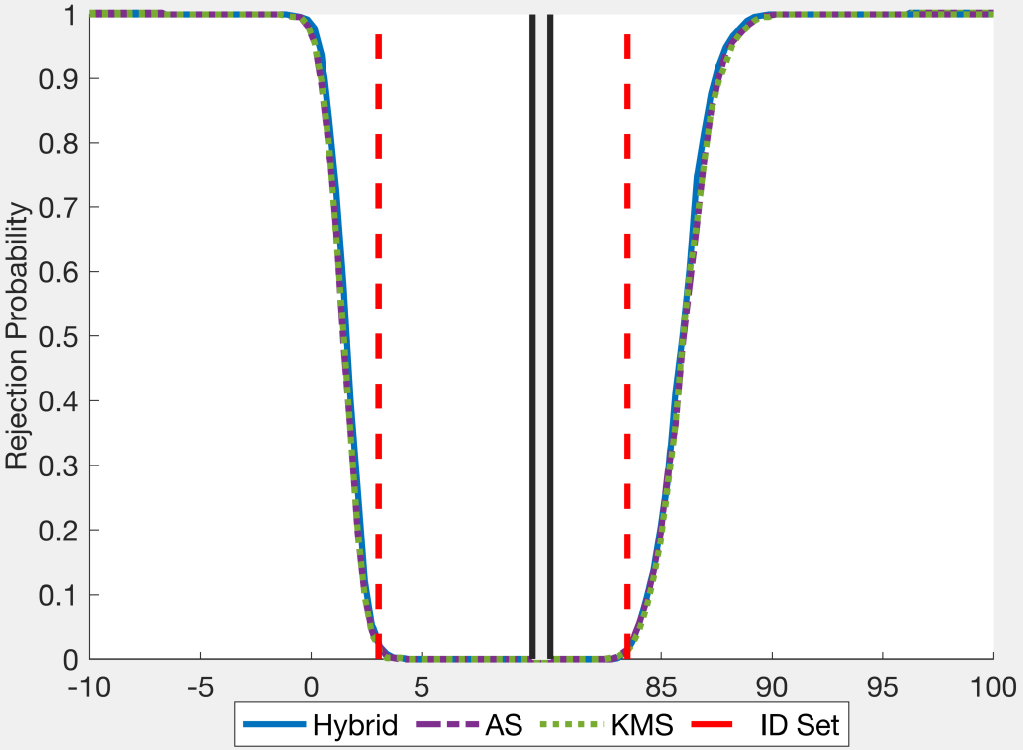} 
	}
	\hfill
	\subfloat[2 Parameters, 14 Moments]{\includegraphics[width=0.48\linewidth, height = .3 \textheight ]{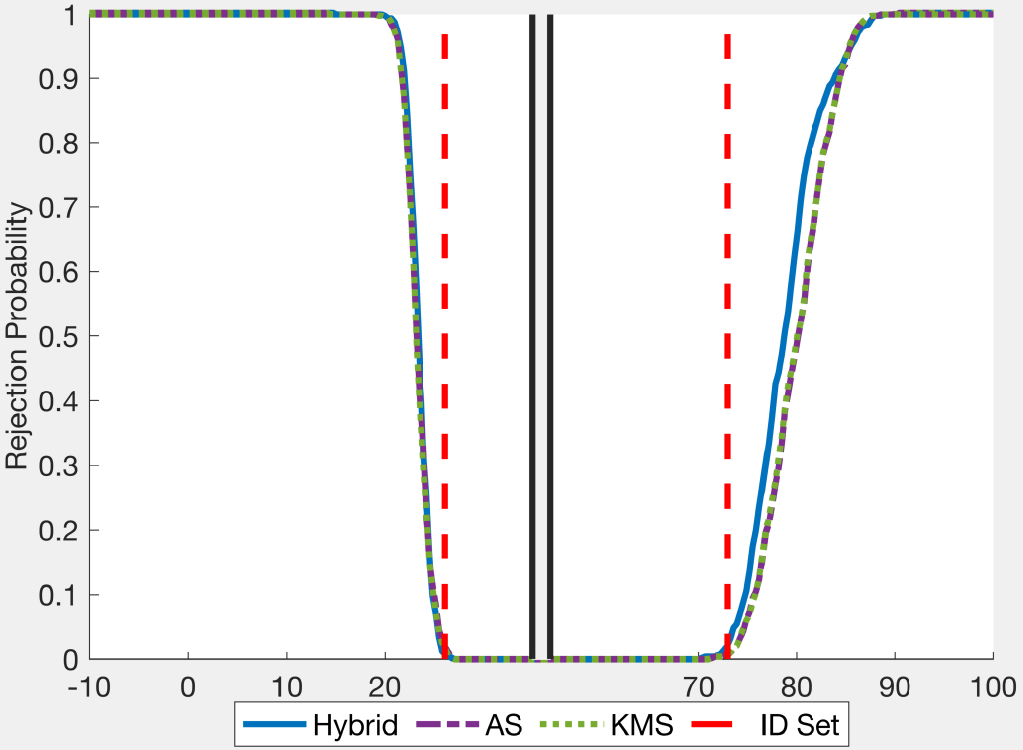} 
	}

	\subfloat[4 Parameters, 14 Moments]{\includegraphics[width=0.48\linewidth, height = .3 \textheight ]{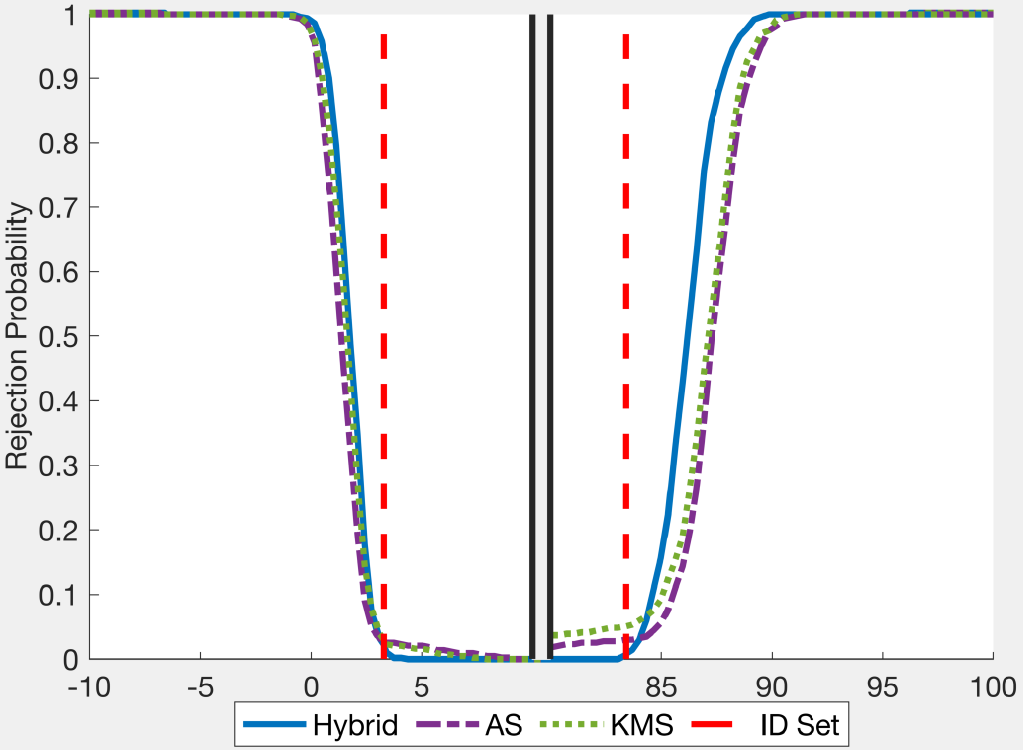} 
	}
	\hfill
	\subfloat[4 Parameters, 38 Moments]{\includegraphics[width=0.48\linewidth, height = .3 \textheight ]{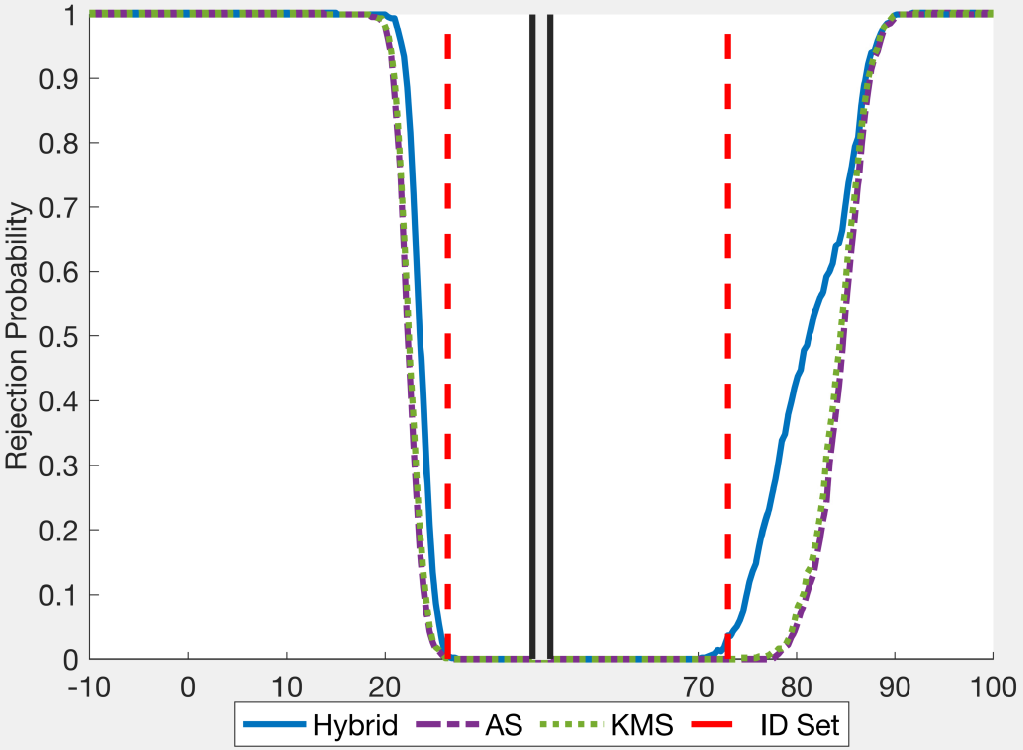} 
	}

\end{figure}

\begin{figure}
	\centering
	
	\caption{Rejection Probabilities for 5\% tests of $\theta_g$: Comparisons to AS and KMS tests \label{fig:thetag power as and kms}}
	
	\subfloat[2 Parameters, 6 Moments]{\includegraphics[width=0.48\linewidth, height = .3 \textheight ]{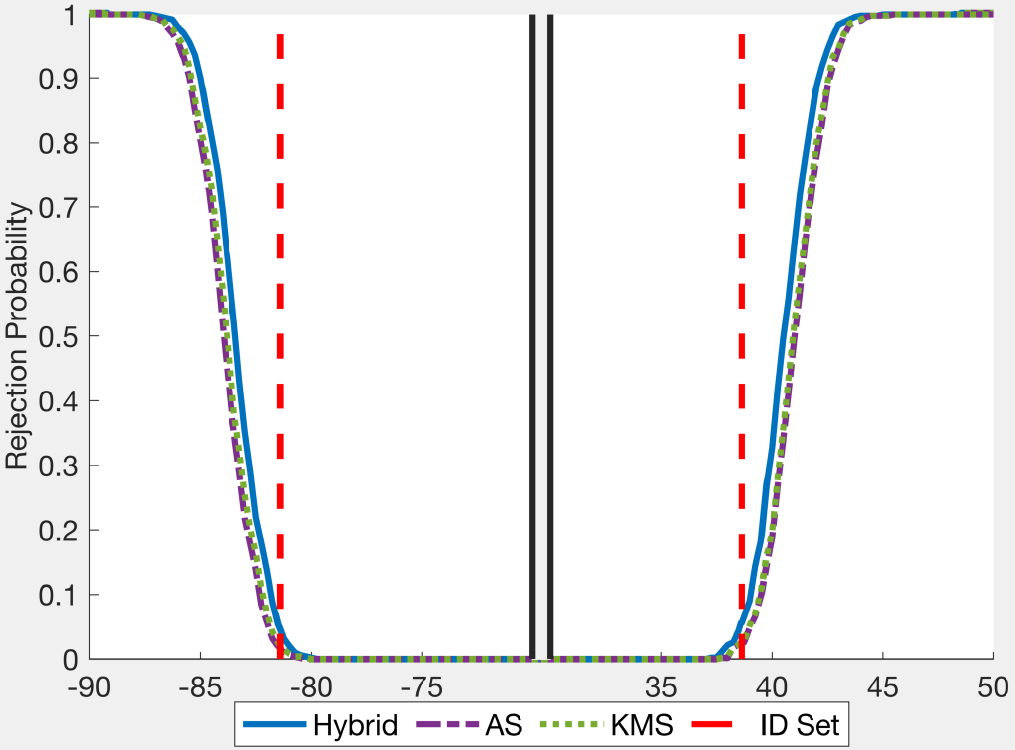} 
	}
	\hfill
	\subfloat[2 Parameters, 14 Moments]{\includegraphics[width=0.48\linewidth, height = .3 \textheight ]{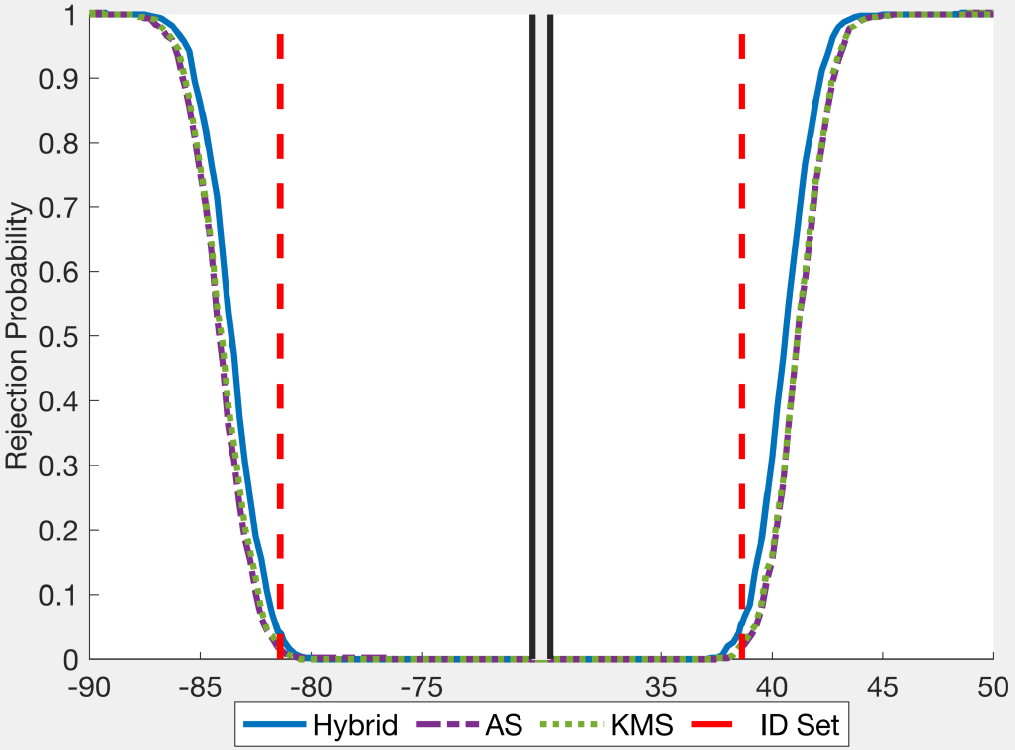} 
	}

	\subfloat[4 Parameters, 14 Moments]{\includegraphics[width=0.48\linewidth, height = .3 \textheight ]{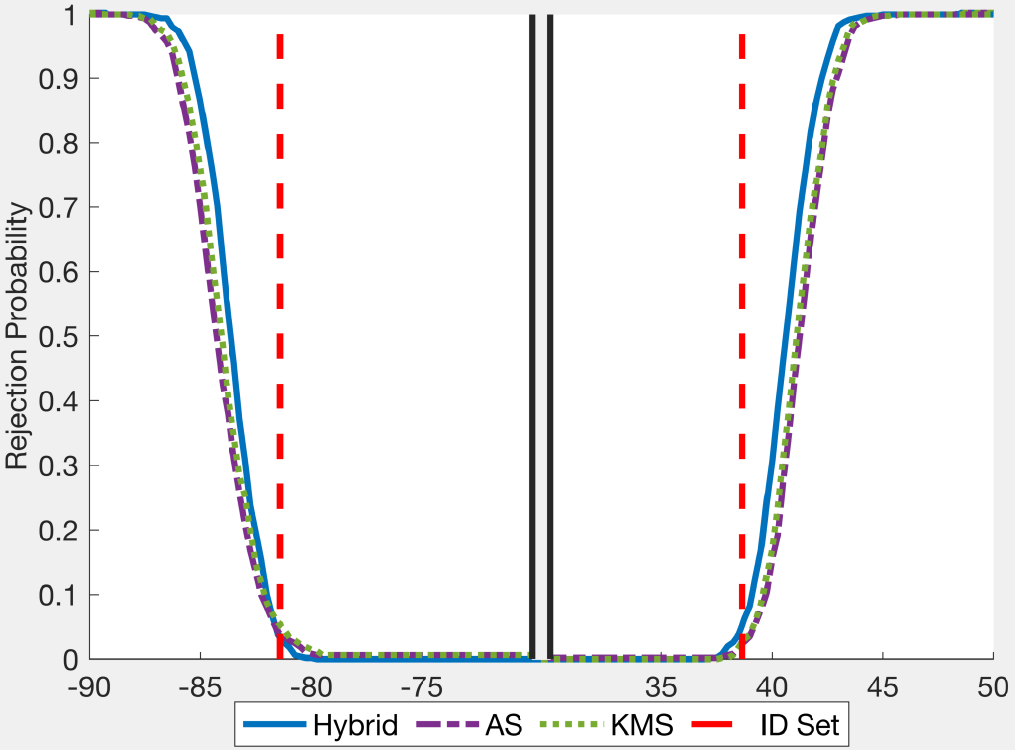} 
	}
	\hfill
	\subfloat[4 Parameters, 38 Moments]{\includegraphics[width=0.48\linewidth, height = .3 \textheight ]{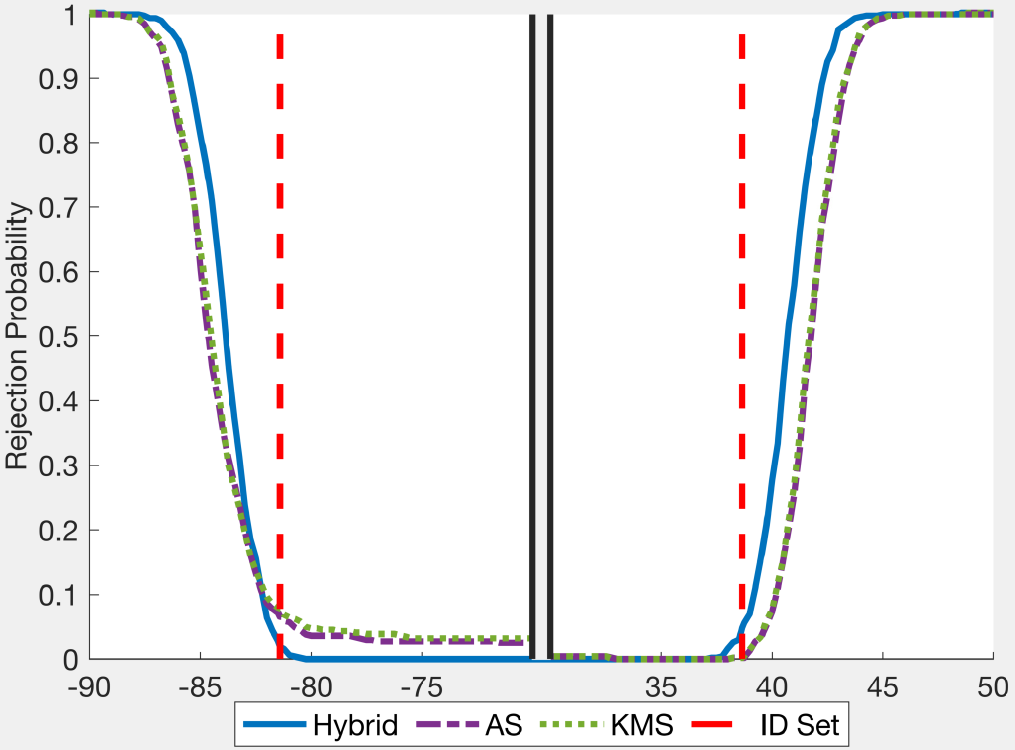} 
	}

\end{figure}

This appendix reports additional  simulation results to complement the results reported in Section \ref{sec: Monte Carlo} of the main text.  Figures \ref{fig:theta_g power}-\ref{fig:beta power wollmann} show comparisons analogous to Figure \ref{fig:mean weight power} except for the alternative parameters $\delta_g$ and $\beta$. Figures \ref{fig:meanweight power cox and shi}-\ref{fig:beta power cox and shi} show comparisons of the hybrid to the LFP, sCC, and sRCC tests, while Figures \ref{fig:meanweight power as and kms}-\ref{fig:thetag power as and kms} show comparisons to the AS and KMS tests. 

\clearpage

\bibliographystylesupp{agsm}
\bibliographysupp{Conditional_MI_Tests}

\end{document}